\documentclass[twocolumn]{aastex63}

\usepackage{graphicx}	
\usepackage{color}
\usepackage{amsmath}	
\usepackage{amssymb}	
\usepackage{mathrsfs}
\usepackage{mathrsfs}

\newcommand{\msun}{M_\odot}

\newcommand{\zsun}{Z_\odot}

\newcommand{\cc}{{\rm cm}^{-3}}

\newcommand{\msunyr}{M_\odot~{\rm yr}^{-1}}

\newcommand{\kpc}{{\rm kpc}}
\newcommand{\mpc}{{\rm Mpc}}
\newcommand{\gpc}{{\rm Gpc}}
\newcommand{\pc}{{\rm pc}}
\newcommand{\mum}{\mu {\rm m}}
\newcommand{\kms}{{\rm km~s}^{-1}}

\newcommand{\K}{{\rm K}}
\newcommand{\beq}{\begin{equation}}
\newcommand{\eeq}{\end{equation}}

\shorttitle{Early BH-Galaxy Coevolution}
\shortauthors{Inayoshi et al.}

\begin{document}

\title{Rapid Growth of Seed Black Holes during Early Bulge Formation}

\correspondingauthor{Kohei Inayoshi}
\email{inayoshi@pku.edu.cn}

\author{Kohei Inayoshi}
\affiliation{Kavli Institute for Astronomy and Astrophysics, Peking University, Beijing 100871, China}
\author{Riouhei Nakatani}
\affiliation{RIKEN Cluster for Pioneering Research, 2-1 Hirosawa, Wako-shi, Saitama 351-0198, Japan}
\author{Daisuke Toyouchi}
\affiliation{Research Center for the Early Universe (RESCEU), The University of Tokyo Hongo, 7-3-1, Bunkyo-ku Tokyo, 113-0033, Japan}
\author{Takashi Hosokawa}
\affiliation{Department of Physics, Kyoto University, Kyoto 606-8502, Japan}
\author{Rolf Kuiper}
\affiliation{Zentrum für Astronomie der Universität Heidelberg, Institut für Theoretische Astrophysik, Albert-Ueberle-Straße 2, 69120 Heidelberg, Germany}
\author{Masafusa Onoue}
\affiliation{Kavli Institute for Astronomy and Astrophysics, Peking University, Beijing 100871, China}
\affiliation{Kavli Institute for the Physics and Mathematics of the Universe (Kavli IPMU, WPI), The University of Tokyo, Chiba 277-8583, Japan}
\affiliation{Max-Planck-Institut f\"ur Astronomie, K\"onigstuhl 17, D-69117 Heidelberg, Germany}

\begin{abstract}
We study the early growth of massive seed black holes (BHs) via accretion 
in protogalactic nuclei where the stellar bulge component is assembled,
performing axisymmetric two-dimensional radiation hydrodynamical simulations.
We find that when a seed BH with $M_\bullet \sim 10^5~\msun$ is embedded 
in dense metal-poor gas ($Z=0.01~\zsun$) with a density of $\ga 100~\cc$ and bulge stars with a total mass of 
$M_\star \ga 100~M_\bullet$, a massive gaseous disk feeds the BH efficiently at rates of
$\ga 0.3-1~\msunyr$ and the BH mass increases nearly tenfold within $\sim 2$ Myr.
This rapid accretion phase lasts until a good fraction of the gas bounded within the bulge accretes onto the BH,
although the feeding rate is regulated owing to strong outflows driven by ionizing radiation
emitted from the accreting BH.
The transient growing mode can be triggered for seed BHs 
formed in massive dark-matter halos with masses of $\ga 10^9~\msun$ at $z\sim 15-20$
(the virial temperature is $T_{\rm vir}\simeq 10^5~\K$).
The host halos are heavier and rarer than those of typical first galaxies, but are more likely to end up in
quasar hosts by $z\simeq 6$.
This mechanism naturally yields a mass ratio of $M_\bullet/M_\star >0.01$ higher than the value 
seen in the local universe and the existence of such overmassive BHs provides us with a unique opportunity 
to detect highly accreting seed BHs at $z\sim 15$ 
with AB magnitude of $m_{\rm AB} \sim26 - 29$ mag at $2~\mum$ (rest-frame 10 eV) 
by the upcoming observations by the {\it James Webb Space Telescope} and {\it Nancy Grace Roman Space Telescope}.
\end{abstract}


\section{Introduction}
\label{sec:intro}

Supermassive black holes (SMBHs) are one of the most essential objects that comprise galaxies
and their coevolution with host galaxies are suggested by the empirical relations between 
BH masses and galaxy properties seen in the local universe \citep[e.g.,][]{Magorrian_1998,Ferrarese_2000,Kormendy_Ho_2013}. 
Nevertheless, the origin of those SMBHs remains one of the most intriguing and longest-standing 
unsolved puzzles in astrophysics. 
In the past decades, the discoveries of very luminous quasars in the high-redshift universe ($z>6$)
have revealed that SMBHs with masses greater than $\sim 10^9~\msun$ formed within the first billion 
years after the Big Bang \citep[e.g.,][]{Fan_2006,Willott_2010,Mortlock_2011,Wu_2015,Banados_2018,
Wang_2021,Yang_2021}.
Beyond the rarest and most massive SMBHs that represent the tip of the iceberg of 
the high-redshift BH population, a recent wide-field survey with Subaru Hyper Suprime-Cam (HSC) 
brings the total number of $z>6$ quasars to nearly 200 and enables us to construct their luminosity 
function down to the faint-end regime \citep{Matsuoka_2016,Matsuoka_2018}.
The BH mass and host galaxy's properties of those low-luminosity quasars at $z>6$ have also 
been extensively investigated in order to construct the BH mass distribution at $z>6$ and understand 
the early coevolution between BHs and galaxies \citep[e.g.,][]{Onoue_2019,Izumi_2019}.

The existence of those high-redshift quasars requires their quick assembly mechanisms 
\citep{Volonteri_2012,Haiman_2013,Inayoshi_ARAA_2020};
for instance rapid gas collapse into the nuclei of early protogalaxies \citep{Volonteri_Rees_2005,IHO_2016,Toyouchi_2021} and
the formation of massive heavy seed BHs through primordial star formation \citep{O01,Bromm_Loeb_2003,
Lodato_Natarajan_2006,Shang_2010,Latif_2013,Regan_2014,IOT_2014,Becerra_2015,Chon_2018,Wise_2019,Regan_2020,Sassano_2021} and 
runaway stellar mergers in dense regions \citep{Devecchi_2009,Sakurai_2017,Tagawa_2020,Chon_Omukai_2020}.
In most previous studies, those efficient BH assembly processes are considered to take place 
in atomically-cooling dark-matter (DM) halos with virial temperature of $T_{\rm vir} \ga 10^4~\K$,
(corresponding to DM halo masses of $M_{\rm h}\ga 10^{7}~\msun$ at $z\sim 10-15$), 
where typical first-galaxies would form \citep{Bromm_Yoshida_2011}.
However, a majority of cosmological simulations have suggested that the subsequent growth of seed
BHs formed in typical first galaxies with shallow gravitational potential would be strongly quenched 
by various feedback processes including supernova (SN) explosions of forming massive stars and 
radiative/mechanical outputs associated with BH accretion \citep[e.g.,][]{Dubois_2013,Prieto_2016,
Angles-Alcazar_2017,Habouzit_2017,Latif_2018}.

Recently, semi-analytical studies by \cite{Lupi_2021} and \cite{Li_2021} have proposed that 
BH seeding processes preferentially occur in highly-biased, overdense regions of the universe
where quasars form by $z\simeq 6$.
The quasar progenitor halos are substantially heavier than those hosting first galaxies and 
their average mass exceeds $M_{\rm h}\sim 10^9~\msun$ by $z\sim 20$ via quick assembly 
after the birth of BH seeds with $M_\bullet \sim 10^{4-5}~\msun$.
Large-scale cosmological simulations demonstrate that cold gas streams feed the centers
of such massive halos and sustain high mass accretion rates 
until the mass of the galaxy reaches $\sim 10^{12}~\msun$ 
(\citealt{Li_2007,Sijacki_2009,DiMatteo_2012,Zhu_2020}; see also \citealt{Dekel_2006}).
However, it is worth emphasizing that due to numerical limitations, cosmological simulations with a spatial 
resolution of $\sim O({\rm kpc})$ treat feedback effects with subgrid models.
Therefore, it is essential to directly resolve physical processes in the nuclear region and investigate 
the growth phases of seed BHs hosted in massive halos that will end up as high-redshift quasars.

The nature of mass accretion onto BHs in galactic nuclei has been explored by analytical and numerical work.
In the recent decade, radiation hydrodynamical (RHD) simulations have shown that rapid accreting flows 
can feed the BH at super-Eddington rates as long as a sufficient amount of gas already exists or is efficiently 
supplied to the vicinity of the BH horizon scale of $r_{\rm nuc}\simeq 10^{2-3}~r_{\rm Sch}$
\citep{Ohsuga_2005,Jiang_2014,Sadowski_2015,McKinney_2015}, where $r_{\rm Sch}\equiv 2GM_\bullet/c^2$
is the Schwarzschild radius of the BH, where $G$ is the gravitational constant and $c$ is the speed of light.
Note that the nuclear size for a seed BH is as small as $r_{\rm nuc} \simeq 10^{-5}~\pc~ (M_\bullet/10^5~\msun)$.
However, ionizing radiation emitted from the accreting BH heats the gas inflowing from the BH gravitational 
influence radius of $\sim 1-10~\pc$, within which the BH gravitational energy dominates
the gas thermal energy. 
Thus, the mass accretion rate is generally self-regulated
below the Eddington value \citep{Ciotti_Ostriker_2001,Milosavljevic_2009,Johnson_2011,PR_2011,PR_2012,
Jeon_2012,Park_2017,Smith_2017}.

Accordingly, RHD simulations that cover the BH influence radii and resolve sub-parsec scales 
showed that when the BH is embedded in a dense gas cloud so that the gas inflowing rate from 
the BH influence radius substantially exceeds the Eddington value, the global quasi-steady structure of 
rapid inflows is maintained \citep*{IHO_2016}.
In this case, since the emergent radiation flux is reduced by photon trapping and 
dust absorption in the flow, the gas inflow is not prevented by radiative feedback but 
leads to collapse of the ionized region surrounding the BH \citep[see also][]{Park_2014,Sakurai_2016,Yajima_2017,Park_2020}.
Moreover, anisotropic radiation produced from the accreting BH toward the polar regions moderates
the negative feedback effect \citep{Sugimura_2017,Takeo_2018} and mechanical feedback associated 
with strong outflows completely evacuates the polar regions but does not affect the gas dynamics
of disk accretion \citep{Takeo_2020}. 
Recent three-dimensional RHD simulations have shown that massive gas supply can be sustained
when the dusty disk becomes sufficiently optically thick to ionizing radiation \citep{Toyouchi_2021}.
Applying the conditions to seed BHs in high-redshift protogalaxies, they found that BHs formed 
in massive DM halos with $M_{\rm h}\ga 10^9~\msun$ in the early universe can experience 
a rapidly growing phase.
We note that the halo conditions nicely agree with those suggested by previous studies in the context 
of seed formation \citep{Lupi_2021, Li_2021} and subsequent BH growth \citep[e.g.,][]{DiMatteo_2012}.

However, those RHD simulation studies that focus on the intermediate physical scales within the gravitational 
influence radius of a seed BH ($\la 1-10~{\rm pc}$) have not taken into account the existence of its host galaxy, 
specifically stellar bulge components.
A previous study by \cite{Park_2016} pioneered the role of stellar bulge gravity in promoting mass accretion onto seed BHs,
using spherically symmetric one-dimensional RHD simulations that assume metal-free gas with a uniform initial distribution.
They found that when the bulge mass is greater than a critical mass of $\sim 10^6~\msun$, the bulge gravity
leads to gas accumulation and increases the BH fueling rate.

In this paper, we investigate the dynamics of accreting and outflowing gas in the intermediate 
missing region surrounding a massive seed BH in protogalaxies, performing a series of 
axisymmetric two-dimensional (2D) RHD simulations with a sufficiently large computation domain covering
$\sim 0.1-100~\pc$ that enables us to capture the multi-scale physics properly. 
We consider a massive metal-polluted cloud with metallicity of $Z=0.01~\zsun$ concentrated
inside a massive DM halo, adopting initial conditions of gas motivated by cosmological simulations.
We consider bulge stars formed in the protogalaxies that cause two relevant effects on BH feeding.
First, since the bulge mass is considered to be heavier than the BH mass (for instance, 
$M_\star/M_\bullet \simeq 10^3$ in the local universe), a larger amount of gas can 
be accumulated from larger scales of $\sim O({\rm kpc})$ due to additional stellar gravity 
and can efficiently be delivered into the BH gravitational sphere of influence. 
This would lead to rapid growth of BHs, as seen in the previous study.
On the other hand, since the bulge consists of a number of stars (presumably young stars in protogalaxies), 
photoionization and heating caused by stellar radiation affect the thermal properties of the surrounding gas
and would prevent the gas from feeding the BH.
In our study, we quantify the critical conditions required for rapid growth of BHs induced by the galactic bulges.

One of the most interesting questions is how and when the BH-galaxy correlations have been established.
To answer this, we apply our RHD simulation results to seed formation in the cosmological framework of galaxy formation
and translate the conditions required for rapid BH accretion in terms of the properties of the host protogalaxies and halos.
We find that this BH-growth mechanism naturally yields a mass ratio of $M_\bullet/M_\star >0.01$ at $z\ga 10$,
which is significantly higher than the canonical value seen in the local universe \citep{Kormendy_Ho_2013} but is 
suggested from the observations of the brightest quasars at $z>6$ \citep{Wang_2010, Wang_2013,Pensabene_2020}.
Furthermore, the highly accreting seed BHs with $M_\bullet \simeq 10^{5-6}~\msun$ produce intense radiation 
with luminosities of $\sim 10^{44-45}~{\rm erg~s}^{-1}$. 
Therefore, the existence of such {\it overmassive} BHs provides with us a unique opportunity to detect 
seed BHs in the very early universe with upcoming deep near-infrared observations by e.g., 
James Webb Space Telescope (JWST) and Nancy Grace Roman Space Telescope (RST).
The result would shed the light on the unresolved coevolutionary process between BHs and galaxies in the early universe.

The rest of the paper is organized as follows. 
In Section \ref{sec:method}, we first describe the numerical method and settings of our 2D RHD simulations. 
In Section \ref{sec:result}, we present the simulation results and their dependence on physical parameters
that characterize the bulge and halo properties.
Estimating the radiative luminosity from the highly accreting BHs, we quantify the detectability of 
the signature in upcoming observations by JWST and RST.
In Section \ref{sec:analytic}, based on the simulation results, we provide a theoretical explanation for 
the conditions required to promote rapid accretion onto BH seeds in protogalaxies.
In Section \ref{sec:discussion}, we discuss the conditions for rapid BH accretion in the framework of the hierarchical structure formation,
the $M_\bullet/M_\star$ ratio expected in the early bulge formation stage, and implications for 
the measurements of quasar lifetime.
Finally, we summarize our findings in Section \ref{sec:summary}.

\section{Method}\label{sec:method}

\subsection{Basic equations}

We solve the axisymmetric two-dimensional hydrodynamical equations using the open source code 
{\tt PLUTO} \citep{Mignone_2007}, which has been modified to study massive star formation and 
the evolution of protoplanetary disks \citep[e.g.,][]{Kuiper_2010,Kuiper_2011, Nakatani_2018a, Nakatani_2018b}.
The detailed description of the code and its applications to many other astrophysical systems are shown in
\cite{Kuiper_2020}.
In particular, we update the specific version of the code to investigate the physics of BH feeding and feedback
in the high-redshift protogalactic nuclei.

The basic equations we solve are as follows: the equation of continuity,
\begin{equation}
\frac{\partial \rho}{\partial t}+ \nabla \cdot (\rho \mbox{\boldmath $v$})=0,
\end{equation}
and the equation of motion,
\begin{align}
\frac{\partial (\rho v_r)}{\partial t} + \nabla \cdot(\rho v_r \mbox{\boldmath $v$}) = &
-\frac{\partial p}{\partial r} +\rho \frac{v_\theta^2 + v_\phi^2}{r} -\rho \frac{\partial \Phi}{\partial r}
\nonumber\\
& + (\nabla \cdot \mbox{\boldmath $\sigma$})_r + f_{{\rm rad},r},
\end{align}
\begin{align}
\frac{\partial (\rho v_\theta)}{\partial t} + \nabla \cdot(\rho v_\theta \mbox{\boldmath $v$})= &
-\frac{1}{r}\frac{\partial p}{\partial \theta}
-\rho \frac{v_\theta v_\phi}{r} + \rho \frac{v_\phi^2}{r}\cot \theta \nonumber\\
&+ (\nabla \cdot \mbox{\boldmath $\sigma$})_\theta
+ f_{\rm rad,\theta},
\end{align}
\begin{equation}
\frac{\partial (\rho v_\phi)}{\partial t} + \nabla^l \cdot(\rho v_\phi  \mbox{\boldmath $v$})= 
(\nabla \cdot \mbox{\boldmath $\sigma$})_\phi
\end{equation}
where $\rho$ is the mass density, {\boldmath $v$} is the velocity, $p$ is the gas pressure,
$\Phi$ is the gravitational potential, {\boldmath $\sigma$} is the stress tensor due to viscosity,
and $f_{{\rm rad}, r}$ and $f_{{\rm rad}, \theta}$ are the radial and polar components of the radiation force.

We solve the energy equation of
\begin{equation}
\frac{\partial E}{\partial t} + \nabla \cdot (H \mbox{\boldmath $v$})=-\rho v_r \frac{\partial \Phi}{\partial r} 
+ (\mbox{\boldmath $\sigma$} \cdot \nabla) \mbox{\boldmath $v$}
 + \Gamma -\Lambda,
\end{equation}
where $E = (1/2)\rho \mbox{\boldmath $v$}^2 + \rho e$ and $H = E+p$ are the total energy density and enthalpy
per unit volume, and $e$ is the internal energy per unit mass.
The equation of state for ideal gas is assumed as 
\begin{equation}
p=(\gamma-1)\rho e,
\end{equation}
where the adiabatic exponent $\gamma$ depends on the chemical abundances and gas temperature 
\cite[e.g.,][]{Omukai_Nishi_1998}.
The first and second terms on the right-hand-side present work done by the gravitational force 
and viscous heating.
The last two terms are volumetric radiative cooling and heating (in units of ${\rm erg~s^{-1}~cm^{-3}}$).

The gravitational potential is set to
\begin{equation}
\Phi = -\frac{GM_\bullet}{r}
+ \Phi_\star (r)+ \Phi_{\rm DM}(r),
\end{equation}
where $M_\bullet$ is the BH mass, 
$\Phi_\star$ and $\Phi_{\rm DM}$ are the gravitational potential owing to
bulge stars and a DM halo, respectively.
We here do not consider the self-gravity of gas.
For the bulge stellar potential, we assume that the stellar-mass density profile follows a Hernquist profile 
and thus the potential is given by
\begin{equation}
\Phi_\star (r) = -\frac{GM_\star}{r+r_{\rm c}},
\end{equation}
\citep{Hernquist_1990}, where $M_\star$ is the total bulge mass and $r_{\rm c}$ is the core radius.
Note that the bulge mass increases with time at a given star formation rate (see \S\ref{sec:BC}).
With the stellar mass distribution, the velocity dispersion of stars at the half-mass radius of $R_{\rm e}=(1+\sqrt{2})r_{\rm c}$
is analytically calculated as $\sigma_{\rm e} \simeq 0.23 \sqrt{GM_\star/r_{\rm c}}$
(see Appendix \ref{sec:appA})\footnote{\cite{Park_2016} calculated the bulge size so that the average stellar mass density within $r_{\rm c}$ 
is comparable to that of the Milky Way. Namely, they adopt $r_{\rm c}= 50~\pc (M_\star/10^7~\msun)^{1/3}$, which 
yields $M_\star \simeq 2.5\times 10^{11}~\msun (\sigma_{\rm e}/200~\kms)^3$ and 
underestimates the velocity dispersion of bulge stars compared to that observed in the local universe.}
Assuming that the bulge stellar component in high-redshift protogalaxies obeys
the $M_\star$-$\sigma_{\rm e}$ relation \citep{Kormendy_Ho_2013}, the core radius is set to
\begin{equation}\label{eq:rbulge}
r_{\rm c} \simeq  6.5~{\rm pc} \left(\frac{M_\star}{10^{7}~M_\odot}\right)^{0.47}.
\end{equation}
Note that the observed slope of the $R_{\rm e}$-$M_\star$ relation is known to depend on the stellar mass, galaxy type, and redshift
\citep[e.g.,][]{Shen_2003, Mowla_2019, Li_size_2021,Kawinwanichakij_2021}.
For the DM potential, we consider a Navarro, Frenk \& White (NFW) potential \citep{NFW_1997, Mo_1998}, 
approximately written as 
\begin{equation}
\Phi_{\rm DM} (r) = -\frac{2k_{\rm B}T_{\rm vir}}{\mu m_{\rm p}}~
\frac{\ln(1+r/r_{\rm s})}{r/r_{\rm s}}~f(c_{\rm N}),
\end{equation}
where $k_{\rm B}$ is the Boltzmann constant, $\mu$ is the mean molecular weight of gas,
$T_{\rm vir}$ and $r_{\rm vir}$ are the virial temperature and virial radius of the halo, 
$r_{\rm s} (\equiv r_{\rm vir}/c_{\rm N})$ is the characteristic radius of the NFW density profile,
$c_{\rm N}$ is the concentration factor, and $f(c_{\rm N}) = c_{\rm N}/[\ln(1+c_{\rm N})-c_{\rm N}/(1+c_{\rm N})]$.
We model the dependence of the mean concentration on halo mass $M_{\rm h}$ and redshift $z$: 
$c_{\rm N} \simeq 1.56~(M_{\rm h}/10^9~\msun)^{-0.13}[(1+z)/21]^{-1}$ \citep{Bullock_2001}.

In our axisymmetric simulations without magneto-hydrodynamical (MHD) effects,
angular momentum transport in the accreting flow is given by imposing explicit viscosity.
The viscous stress tensor is given by
\begin{equation}
\sigma_{ij}=\rho \tilde{\nu} \left[ 
\left(\frac{\partial v_j}{\partial x_i} + \frac{\partial v_i}{\partial x_j}\right)
-\frac{2}{3} (\nabla \cdot \mbox{\boldmath $v$} )\delta_{ij}
\right],
\end{equation}
where $\tilde{\nu}$ is the shear viscosity and the bulk viscosity is neglected.
To mimic angular momentum transport associated with MHD turbulence driven by 
the magneto-rotational instability (MRI) in a sufficiently ionized disk \citep[e.g.,][]{Balbus_Hawley_1998,
Stone_Pringle_2001,McKinney_&_Gammie_2004,Bai_2011,Narayan_2012},
we assume the azimuthal components of the shear tensor are non-zero and, in spherical polar coordinates, are given by
\begin{equation}
\sigma_{r\phi}=\rho {\tilde{\nu}} \frac{\partial}{\partial r}\left(\frac{v_\phi}{r}\right),
\end{equation}
\begin{equation}
\sigma_{\theta \phi}=\rho {\tilde{\nu}} \frac{\sin \theta}{r}\frac{\partial}{\partial \theta}\left(\frac{v_\phi}{\sin \theta}\right)
\end{equation}
\citep[e.g.,][]{Stone_1999,Fernandez_2013,Inayoshi_2019}.
The strength of anomalous shear viscosity is calculated with the $\alpha$-prescription \citep{SS_1973},
\begin{equation}
{\tilde{\nu}} = \alpha \frac{c_{\rm s}^2}{\Omega_{\rm K}} \cdot \exp \left(-\frac{|z|}{H}\right),
\label{eq:alpha_visc}
\end{equation}
where $\alpha$ is the viscous parameter, $c_{\rm s}$ is the sound speed, 
$\Omega_{\rm K}\equiv (GM_\bullet/r^3)^{1/2}$, and
$H (\equiv c_{\rm s}/ \Omega_{\rm K})$ is the disk scale height.
Note that the exponential factor imposes that the viscous process is active near the midplane.
The strength of viscosity is set to 
\begin{equation}\label{eq:alpha}
\alpha = \alpha_0 + \alpha_{\rm max} \exp\left[-\left(\frac{\rho_{\rm crit}}{\rho}\right)^2\right].
\end{equation}
The first term corresponds to the strength of MRI turbulence 
and the value is set to $\alpha_0=0.01$ \citep[e.g.,][]{Zhu_Stone_2018,Takasao_2018}.
The second term characterizes the torque caused by non-axisymmetric structure
(e.g., spiral arms) excited in a marginally unstable disk against its self-gravity.
The density threshold, above which viscosity turns on, is assumed to be 
{$\rho_{\rm crit} \equiv \Omega_{\rm K}^2/(\pi G) \simeq 2.2\times 10^{-21}
(M_\bullet/10^5~\msun) (r/10~{\rm pc})^{-3}~{\rm g~cm}^{-3}$.
This choice is motivated by the following reasons.
The local gravitational instability of a rotating disk is described by Toomre's $Q$ parameter \citep{Toomre_1964}
defined by
\begin{equation}
Q\equiv \frac{c_s \kappa_\Omega}{\pi G\Sigma},
\end{equation}
where $\kappa_\Omega^2 =4\Omega^2 + d\Omega^2/d\ln r $ is the epicyclic frequency and
$\Sigma$ is the disk surface density.
For a geometrically-thin cold disk around a point mass, the $Q$-value is approximated to
$Q\simeq \Omega_{\rm K}^2/(2\pi G \rho) = 0.5 (\rho_{\rm crit}/\rho)$.
Thus, the second term of the right-hand-side of Eq.~(\ref{eq:alpha}) is written as 
$\propto \exp(-4Q^2)$, which is a commonly used parameterization of
the effective viscosity adopted in semi-analytical models of a self-gravitating disk;
$\propto \exp(-Q^A/B)$ where $A>0$ and $B>0$ (\citealt{Zhu_2009,Takahashi_2013}; see also \citealt{Kratter_2016}).
The value of $\alpha_{\rm max}$ depends on the level of non-axisymmetric structures in a disk.
A previous 3D RHD simulation study of a dusty circum-nuclear disk around a BH shows 
that the mass inflow velocity is as high as a substantial fraction of the free-fall velocity
due to strong torque caused by spiral-arms in the disk, indicating $\alpha_{\rm max} \simeq O(1)$ \citep{Toyouchi_2021}.
We here adopt $\alpha_{\rm max} =2$ \citep{Hirano_2014,Fukushima_2020}.
The critical $Q$-value for the onset of gravitational torque caused by spiral arms in a disk is 
considered to be $Q\sim 1$, but the exact value depends on various properties of the disk 
(e.g., cooling, heating, and disk irradiation).
We note that in our viscous model, the second term in Eq.~(\ref{eq:alpha}) becomes larger than $\alpha_0$
when $Q\leq 1.15$.

We consider cooling processes associated with chemical reactions and radiative processes of 
metal-polluted gas, which is composed of 11 chemical species (H, H$_2$, e$^-$, H$^+$, H$^{+}_{2}$, 
H$^-$, He, He$^+$, He$^{++}$, C$^+$, and O) and dust grains. 
We solve the chemical reaction network of primordial gas among the 9 species (hydrogen and helium) and adopt the same 
reaction rate coefficients compiled in \cite{Li_2021}.
The cooling rates by collisional excitation and ionization, radiative recombination
of H, He, He$^+$ atoms, their free-free emission \citep{Glover_Jappsen_2007}, 
and the fine-structure lines of C$^+$ and O \citep{Hollenbach_McKee_1989} are 
calculated in the optically-thin limit.
The cooling rate of H$_2$ rovibrational transitions is included \citep{Galli_Palla_1998},
but this effect is negligible under strong H$_2$-dissociating photons produced from bulge stars.
We assume that the amount of heavy elements in gas-phase and dust grains is proportional to 
the relative metallicity $Z/\zsun$ and give the dust-to-gas mass ratio by $0.01~Z/\zsun$.

We implement photoionization of H, He, and He$^+$ by intense BH radiation and associated photoheating 
by solving radiation transfer.
In addition, we consider H$_2$-dissociating radiation in the Lyman-Werner (LW) band and X-ray radiation background 
produced from bulge stars surrounding the accreting BH and its gravitational influence radius.
As for X-ray irradiation, the secondary ionization effect caused by energetic primary electrons
is considered \citep[][references therein]{IO11}. 
More detailed treatments of BH and stellar radiation are described in \S\ref{sec:RT}.

\subsection{Radiation transfer}\label{sec:RT}

We quantify the radiative heating rate $\Gamma$, ionization rate $\zeta$, and radiation pressure force
by solving the radiation transfer (RT) of both direct and diffusion components. 
We adopt a hybrid RT scheme \citep[e.g.,][]{Kuiper_2020}: the direct component emitted from the central BH is solved 
by means of the multi-frequency ray-tracing method, while the diffusive component owing to thermal (re-)emission by 
dust grains is solved by a flux-limited diffusion (FLD) approximation.

The direct component of the flux associated with BH accretion is the primary source of radiative feedback that affects 
the properties of accreting flows.
The radiation flux can analytically be expressed as
\begin{equation}
F_{\nu}(r,\theta, t) = \frac{\mathscr{L}_{\bullet, \nu}(\theta,t)}{4\pi r^2}~e^{-(\tau_{\rm g,\nu} + \tau_{\rm d,\nu})},
\end{equation}
where $\nu$ is a frequency of radiation, $F_\nu (r,\theta, t)$ is the radial component of the specific radiative flux 
at a distance of $r$ from the center toward a direction of $\theta$ radian from the pole, and
$\mathscr{L}_{\bullet, \nu}/(4\pi r^2)$ is the unattenuated specific radiative flux that includes the anisotropic degree (see \S\ref{sec:BHmodel}).
The optical depth of gas and dust to a photon at a frequency of $\nu$ are calculated as 
\begin{align}
\tau _{\rm g,\nu} &= \int_{r_{\rm in}}^r  \sum_i n_i(\hat{r},\theta,t) \cdot \sigma_{i,\nu} ~d\hat{r},\\
\tau _{\rm d,\nu} &= \int_{r_{\rm in}}^r  \rho (\hat{r},\theta,t) \cdot \kappa_{\rm d, \nu} ~d\hat{r},
\end{align}
respectively, where $n_i$ is the number density of the $i$-th component ($i=$H, He, and He$^+$),
$\sigma_{\nu,i}$ is the bound-free cross section \citep{Verner_1996,Yan_1998}, and 
$\kappa_{\rm d,\nu}$ is the dust opacity.
In our simulations, we use the opacity table taken from \cite{Draine_Lee_1984} in the RT calculation 
for non-ionizing radiation, while the attenuation level of ionizing (EUV) radiation is calculated by using one single opacity value 
of $\kappa_{\rm d,0}\simeq 2.8~[Z/(0.01~\zsun)]~{\rm cm^2~g^{-1}}$ (per gas mass).
The radiation flux absorbed by each component is used for calculating the heating and ionization rates ($\Gamma$ and $\zeta$) 
owing to EUV irradiation and the dust heating rate $S_\bullet$ caused by all the photons emitted from the accreting BH.
Those rates are estimated so that the number of photons emitted along any line of sight 
equals the number of photoionization in that direction over any time interval 
\citep[e.g.,][]{Whalen_Norman_2006,IHO_2016,Nakatani_2018a,Nakatani_2018b}.
We note that the component of diffusive EUV radiation produced by radiative recombination of gas is not considered
because UV feedback onto accretion flows is dominated by the direct component emitted from the central bright object
\citep[e.g.,][]{Hosokawa_2011}.
However, in order to calculate the size of an ionized region precisely, we need to take into account that recombinations directly 
to the ground state lead to immediate ionization of a hydrogen atom in the vicinity.
To treat this effect, we adopt the ``on-the-spot" approximation using the case-B recombination rate coefficient.

The diffusion component of infrared (IR) thermal radiation produced by dust grains is treated with a gray-approximated 
FLD method by solving the moment equation
\begin{equation}
\frac{\partial E_{\rm rad}}{\partial t} + \nabla \cdot \mbox{\boldmath $F$}_{\rm IR} =
\rho \kappa_{\rm P} c(a_RT_{\rm d}^4 - E_{\rm rad}) + S_\bullet,
\end{equation}
where $E_{\rm rad}$ is the radiation energy density, $T_{\rm d}$ is the dust temperature,
$\kappa_{\rm P}$ is the Planck-mean dust opacity, $a_{\rm R}$ is the radiation constant,
and $\mbox{\boldmath $F$}_{\rm IR}$ is the IR radiation flux (re-)emitted from dust and 
is approximated as
\begin{equation}
\mbox{\boldmath $F$}_{\rm IR} = -\frac{\lambda c}{\rho \kappa_{\rm R}}
\nabla E_{\rm rad},
\end{equation}
where $\kappa_{\rm R}$ is the Rosseland-mean dust opacity and $\lambda$ is
the flux limiter \citep{Levermore_1981}.
We solve the radiation transfer equations with an implicit solver
imposing a zero-gradient boundary condition for the radiation energy.
The dust temperature is calculated separately from the radiation temperature by using 
the two-temperature linearization approach \citep[see more details in][]{Kuiper_2020}.

With the direct component of radiation incident from the BH, we calculate the outward radiation force
through electron scattering, bound-free transitions, and attenuation by dust grains.
The diffusive component exerts the IR radiation force to both the radial and polar directions.

\subsection{BH radiation model}\label{sec:BHmodel}

We inject photons from the unresolved central region (i.e., the sink cell) to the computational domain,
supposing that a compact accretion disk forms around the nuclear BH.
Following \cite{Sazonov_2004}, we consider a broad-band radiation spectral energy distribution (SED)
obtained from the averaged quasar samples
\begin{align}
L_{\bullet, \nu} = \begin{cases}
L_0 \left( \dfrac{\nu}{\nu_0}\right)^{-0.6} & (\nu_{\rm min} \leq \nu < \nu_0), \vspace{10pt}\\
L_0 \left( \dfrac{\nu}{\nu_0}\right)^{-1.5} & (\nu_0 \leq \nu \leq \nu_{\rm max}),
\end{cases}
\label{eq:specin}
\end{align}
where $h\nu_0 = 10~{\rm eV}$, $h\nu_{\rm min} = 1~{\rm eV}$, and 
$h\nu_{\rm max} = 1~{\rm keV}$.
With this spectral shape, the bolometric (frequency integrated) luminosity is calculated as 
$L_{\bullet} =\int_{\nu _{\rm min}}^{\nu _{\rm max}} L_{\bullet, \nu} d\nu 
\simeq 3.3~ \nu_0 L_0$
and the total EUV luminosity ($h\nu \geq h\nu_{\rm Ly}\equiv 13.6$ eV) is  
$L_{\rm EUV} \simeq 1.5~ \nu_0 L_0$.
The normalization of the luminosity is determined as a function of the mass accretion rate onto the BH.
We note that the spectral power-law indices we adopt are consistent with those obtained from 
the stacked UV (rest-frame wavelengths $600$-$2500~{\rm \AA}$) spectrum of 53 luminous quasars 
at $z\simeq 2.4$ with a correction for the intervening Lyman forest and Lyman continuum absorption
\citep{Lusso_2015}; namely, the continuum spectrum is approximated by a broken power law with indices of
$\simeq -0.61\pm 0.01$ at $\nu < \nu_0$ and $\simeq -1.7\pm 0.61$ at $\nu \geq \nu_0$, respectively.

We set a model for radiation luminosity emitted from the accreting BH as
\begin{align}
\frac{L_{\bullet}}{L_{\rm Edd}} = \begin{cases}
\dot{m} & (\dot{m}<2), \vspace{10pt}\\
2\left[ 1+\ln \left(\dfrac{\dot{m}}{2} \right) \right] & (\dot{m}\geq 2), 
\end{cases}
\label{eq:ledd}
\end{align}
where $L_{\rm Edd}$ is the Eddington luminosity, $\dot{m}$ is the BH mass accretion rate
normalized by the Eddington rate $\dot{M}_{\rm Edd}~[\equiv L_{\rm Edd}/(\eta_0c^2)]$,
and $\eta_0=0.1$ is the radiative efficiency in the sub-Eddington and mildly-Eddington regime \citep{SS_1973}.
The radiative luminosity at $\dot{m}>2$ is obtained from \cite{Watarai_2000}
in a Slim-disk model \citep{Abramowicz_1988}.
We note that the two expressions in Eq.~(\ref{eq:ledd}) are broadly consistent with 
the results obtained with general-relativistic, (magneto-) RHD simulations 
(\citealt{Jiang_2014, Sadowski_2015}; see also a model comparison in
\citealt{Inayoshi_ARAA_2020}).

Furthermore, we assume isotropic/anisotropic radiation fields depending on the 
bolometric luminosity emitted from the accreting BH.
When the radiative luminosity is lower than a critical value of $L_{\rm crit}$,
we inject the isotropic radiation flux from the center.
On the other hand, when the BH is rapidly growing and $L_\bullet > L_{\rm crit}$, 
the radiation flux with $L_{\rm crit}$ is considered to be isotropic but the excess luminosity 
$\Delta L_\bullet (\equiv L_\bullet - L_{\rm crit}$) is distributed into the radiation flux
anisotropically as $\Delta L_\bullet (\mathcal{N}+1) |\cos^{\mathcal{N}} \theta |$,
where $\mathcal{N}$ characterizes the anisotropic degree and is set to $\mathcal{N}=4$ 
in our simulations \citep[see also][]{Ohsuga_2005, Takeo_2018}.
Therefore, the anisotropic degree of the flux is characterized as
$\mathscr{L}_\bullet (\equiv \int \mathscr{L}_{\bullet, \nu}d\nu)=  L_{\rm crit}+ \mathcal{F}(\theta)\Delta L_\bullet$
with a function of $\mathcal{F}(\theta)$ defined by
\begin{align}
\mathcal{F}(\theta) = \begin{cases}
(\mathcal{N}+1) |\cos ^{\mathcal{N}} \theta| & (\Delta L_\bullet > 0), \vspace{5pt}\\
1.0 & ({\rm otherwise}).
\end{cases}
\label{eq:aniso}
\end{align}
It is worth mentioning that this anisotropic model injects radiation to the equatorial region 
($\theta \simeq \pi/2$) and the non-zero radiation flux through the equator affects the thermal 
properties of an accretion disk via IR re-emission from hot dust grains.
In addition, RHD simulations performed by \cite{Ohsuga_2005} suggest that a super-Eddington 
accreting disk transports a significant fraction of radiation energy through the equator via diffusion 
and the luminosity is comparable to the Eddington value (although most radiation is collimated 
toward the poles).
Therefore, we adopt $L_{\rm crit}=L_{\rm Edd}$.

\vspace{2mm}
\subsection{Stellar irradiation}

Bulge stars are not only sources of external gravitational potential
but also radiation sources associated with star formation activity.
We here consider LW radiation and X-rays, which play an important role in determining 
the thermal and chemical properties of gas around the nuclear region.
The LW and X-ray radiation is mainly produced by short-lived massive stars 
and X-ray binaries, and their luminosities are proportional to the star formation rate (SFR).
We assume that the bulge is composed of metal-poor stars with $Z=0.02~\zsun$ whose mass $m_\star$ 
follows a Salpeter initial mass function ($1 \leq m_\star/\msun \leq 100$), yielding the number flux of LW photons
$Q_{\rm LW} \simeq 5.5\times 10^{53}~{\rm s}^{-1}~({\rm SFR}/\msunyr)$.
Note that the number flux of LW radiation is lowered only by $\simeq 15\%$ over 
$5\times 10^{-4}\leq Z/\zsun \leq 0.2$ \citep{Schaerer_2003}.
Taking the mean energy of LW bands ($h\nu_{\rm LW}=12.4~{\rm eV}$), the luminosity is calculated as
$L_{\rm \star,LW}=h\nu_{\rm LW}Q_{\rm LW}\simeq 1.1\times 10^{43}~{\rm erg~s}^{-1}({\rm SFR}/\msunyr)$.
We also assume that the X-ray luminosity correlates with the SFR as seen in local star forming galaxies:
$L_{\rm \star,X}\simeq 6.7\times 10^{39}~{\rm erg~s}^{-1}({\rm SFR}/\msunyr)$ in the $2-10$ keV band
\citep[e.g.,][]{Grimm_2003,Mineo_2014}.

We treat the stellar radiation field in the optically-thin limit
(note that the radiation flux produced from the central accreting BH dominates the stellar radiation flux
at the vicinity of the BH within $1~{\rm pc}$).
Assuming a spherically symmetric distribution of bulge stars, the radiation energy density at a distance of $r$ 
from the center is given by
\begin{align}
cE_{\star,i}(r) = \int_{0}^{r} \epsilon_i(\hat{r})  \frac{\hat{r}^2}{r^2} d\hat{r}
+ \int_{r}^{\infty} \epsilon_i(\hat{r})  \frac{\hat{r}}{r} \ln \sqrt{\frac{\hat{r}+r}{\hat{r}-r}}  d\hat{r},
\label{eq:Estar}
\end{align}
where the first and second terms in the right hand side represent contributions from inside and outside 
the distance of $r$, $\epsilon_i(r)$ is the radiation emissivity at each cell
\begin{align}
\epsilon_i(r) 
= \frac{L_{\star,i}}{2\pi} \cdot \frac{r_{\rm c}}{r(r+r_{\rm c})^3}
\hspace{10pt}{\rm for}~i={\rm LW~and~X},
\end{align}
where the functional form is given by the stellar mass distribution, and
$\rho_\star(r) = \Delta \Phi_\star(r)/(4\pi G)$.
Integrating Eq. (\ref{eq:Estar}), we obtain 
\begin{equation}
cE_{\star,i}(r) =  \frac{L_{\star,i}}{4\pi r_{\rm c}^2}~\Xi(r/r_{\rm c}),
\end{equation}
where 
\begin{align}
\Xi(x) \equiv 
\frac{1}{(1+x)^2} + 
\int_{x}^\infty \frac{\ln(\frac{x_0+x}{x_0-x})}{(1+x_0)^3} \frac{dx_0}{x}.
\end{align}
The function of $\Xi(x)$ is approximated as $O(1)$ at $x<1$ and decreases as $\propto x^{-2}$ at $x\gg 1$.

In this paper, we consider H$_2$ photodissociation by LW radiation (${\rm H}_2 + \gamma_{\rm LW} \rightarrow 2~{\rm H}$),
H$^-$ photodetachment (${\rm H}^- + \gamma_{\rm 1eV} \rightarrow {\rm H}+{\rm e}^-$) by photons with $h\nu \simeq 1~{\rm eV}$,
and ionization/heating of H and He by X-rays with $h\nu = 2$ keV.
Those reaction rates are calculated in the same way as in \cite{Inayoshi_Tanaka_2015}.
For the H$^-$ photodetachment rate, we set the minimum rate coefficient at a range of the effective temperature of bulge stars
($10^4~\K \leq T_{\rm eff} \leq 10^5~\K$)\footnote{
While the H$_2$ photodissociation rate hardly depends on $T_{\rm eff}$, the H$^-$ photodetachment rate 
increases with lower $T_{\rm eff}$ because photons with $\simeq 1$ eV ($\ll h\nu_{\rm LW}$) contributes to the rate.
With a realistic stellar spectrum for metal-poor galaxies \citep{Inoue_2011}, the H$^-$ photodetachment rate boosts by
a factor of $\sim 20$ from that for $T_{\rm eff}=10^5~\K$.}.
We also take into account secondary ionization by energetic electrons produced by primary X-ray ionization
\citep{Shull_1985}.

\subsection{Boundary and initial conditions}\label{sec:BC}

We set a computational domain of $r_{\rm min}\leq r \leq r_{\rm max}$ and $\epsilon \leq \theta \leq \pi- \epsilon$,
where $\epsilon =0.001$ radian is set to avoid numerical singularity at poles.
We adopt logarithmically spaced grids in the radial direction and uniformly spaced grids in the polar direction.
The number of the grid points is set to ($N_r, N_\theta)=(250,240)$. 
For all the simulations we conduct, the minimum and maximum radius are set to $r_{\rm min}=0.1~\pc$
and $r_{\rm max}=100~\pc$.

We adopt the outflow boundary conditions at the innermost and outermost grid, where zero gradients across 
the boundaries are imposed on physical quantities in order to avoid spurious reflection of wave energy at the boundaries. 
We also impose $v_r \leq 0$ at the inner boundary (i.e., inflowing gas from ghost cells is prohibited). 
At the poles, the reflective condition is imposed on the polar component of the velocity $v_\theta$.
We set a density floor of $n_{\rm min} = 1~\cc$ and a maximum radial velocity of $v_{\rm max} = 150~\kms$.
This treatment allows us to avoid cells with a high Mach number ($\gg 100$) in the outflowing regions near the inner 
and polar boundary.

\begin{deluxetable*}{lcccccc}
\tablenum{1}
\tablecaption{Models considered}
\tablewidth{0pt}
\tablehead{
\colhead{Model} & \colhead{$M_\star$ ($\msun$)} & 
\colhead{$T_{\rm vir}$ ($\K$)} & \colhead{$\epsilon_\star$} & \colhead{$n_{\rm c}$ ($\cc$)} &
\colhead{transition} & 
 \colhead{References}}
\startdata
{\bf Massive Halos}$\dag$ &&&&\\
B7T5N4    &  $10^{7~~}$ & $10^5$  & 0.05 & $10^{4~~}$ &Y& \S\ref{sec:nc}\\
B7T5N3.5 &  $10^{7~~}$ & $10^5$ & 0.05 &  $10^{3.5}$ &Y& \S\ref{sec:nc}\\
B7T5N3~(fiducial) &  $10^{7~~}$ &    $10^5$ &  0.05 & $ 10^{3~~}$ &Y& \S\ref{sec:fid}~~~\\
B7T5N2.5 &  $10^{7~~}$ & $10^5$ & 0.05 & $ 10^{2.5}$ &Y& \S\ref{sec:nc}\\
B7T5N2 &  $10^{7~~}$ &    $10^5$ & 0.05 & $ 10^{2~~}$ &N& \S\ref{sec:nc}\\
B0T5N3 &  $0~~~~$    &       $10^5$ &  0.05 &    $10^{3~~}$ &N& \S\ref{sec:Mb}\\
B5T5N3 &  $10^{5~~}$ &    $10^5$ & 0.05 & $10^{3~~}$ &N& \S\ref{sec:Mb}\\
B6T5N3 &  $10^{6~~}$ &    $10^5$  & 0.05 & $10^{3~~}$ &N& \S\ref{sec:Mb}\\
B6.5T5N3 &  $10^{6.5}$ &    $10^5$  & 0.05 & $10^{3~~}$ &N& \S\ref{sec:Mb}\\
B7T5N3-highSFE &  $10^{7~~}$ & $10^5$ & 0.5 & $10^{3~~}$ &Y&  \S\ref{sec:highSFE}
\vspace{3pt}\\
{\bf Normal Halos}$\ddag$ &&&&\\
B7T4N4    &  $10^{7~~}$ & $10^4$  & 0.05 & $10^{4~~}$ &N& \S\ref{sec:Tv}\\
B7T4N3 &  $10^{7~~}$ &    $10^4$ &  0.05 & $ 10^{3~~}$ &N& \S\ref{sec:Tv}\\
B6T4N3 &  $10^{6~~}$ &    $10^4$ &  0.05 & $ 10^{3~~}$ &N& \S\ref{sec:Tv}\\
\enddata
\tablecomments{Simulation runs and input parameters. Column (1) model ID, (2) initial bulge mass, 
(3) virial temperature of the host DM halo, (4) star-formation efficiency, (5) initial gas density at the center,
and (6) Y(/N) indicates the cases where a transition to rapid accretion does (not) occur by the simulation termination 
at $t=4.0$ Myr (high-SFE case) and $t=3.5$ Myr (otherwise).
Movies of these simulation are available \href{https://www.youtube.com/playlist?list=PLqDrKvxmu0l4fyJ74u5ZYyQ8PmwuNZA32}{here}.
\\
$\dag$: a massive dark-matter halo with $M_{\rm h}=2\times 10^9~\msun$ at $z=15$.\\
$\ddag$: a normal atomic-cooling halo with $M_{\rm h}= 10^8~\msun$ at $z=10$.}
\label{tab:model}
\end{deluxetable*}

As initial conditions, we consider a neutral gas cloud with a temperature of $T=10^4~\K$
and a spherically symmetric density profile with a flat core with the central density of $n_{\rm c}$ and an envelope following $\propto r^{-2}$,
\begin{align}
n(r) = \frac{n_{\rm c}}{1+(r/r_0)^2},
\label{eq:nin}
\end{align}
where the core radius $r_0$ is set so that the density profile becomes consistent with that 
of high-redshift protogalaxies obtained by cosmological simulations at $r\gg r_0$.
Estimating the gas density at the outer boundary of the halo by the minimum-energy truncated 
isothermal sphere model as $n_{\rm vir}\simeq 0.07~\cc[(1+z)/21]^3$ \citep{Shapiro_1999,Iliev_2001}, 
the density profile outside 
the core is approximated as $n(r)\simeq f_n n_{\rm vir}(r/r_{\rm vir})^{-2}$ or equivalently as
\begin{align}
n(r) 
&\simeq 2.5\times 10^{2}~f_n T_{\rm vir,4} \left(\frac{r}{10~\pc}\right)^{-2}~\cc,
\end{align}
and thus the core radius is approximated as 
\begin{align}
r_0 \simeq 5 ~f_n^{1/2} T_{\rm vir,4}^{1/2} n_{\rm c,3}^{-1/2}~\pc,
\end{align}
where $T_{\rm vir,4} = T_{\rm vir}/(10^4~\K)$, $n_{\rm c,3} = n_{\rm c}/(10^3~\cc)$, 
and $f_n\sim O(1)$ is a numerical factor.
We note that the density profile with $f_n \simeq 4$ agrees with those of atomic cooling haloes 
with $T_{\rm vir}\simeq 10^4~\K$ obtained by cosmological simulations without star formation
and stellar feedback \citep{Wise_2008,Shang_2010,Regan_2014}.
After a seed BH forms at the center of the halo, the density normalization becomes consistent with 
$f_n\simeq 1$.

In addition, star formation in the halo is modeled by assuming a conversion efficiency from gas 
into stars $\epsilon_\star$.
In our fiducial case, we adopt $\epsilon_\star = 0.05$ \citep[e.g.,][]{Visbal_2015}, which is motivated by
abundance matching and the observed UV luminosity function of galaxies at $z\simeq 6$ \citep{Bouwens_2015}. 
Note that the efficiency is calculated as the average value over time and scales in galaxies.
Alternatively, we suppose that a star-formation episode with a high value of $\epsilon_\star$ lasts 
within a few Myrs in the nuclear region before SN explosions of massive stars begin to occur
and regulate the star formation efficiency in the bulge.
According to numerical simulations that study star cluster formation from a giant molecular cloud with
a size of $10-100~\pc$, the star formation efficiency (SFE) becomes as high as $\epsilon_\star \ga 0.2-0.3$
when the initial gas surface density is higher than $\sim 10^3~\msun \pc^{-2}$ \citep{Fukushima_2020,Fukushima_2021}.
Referring to those observational and theoretical studies, 
we assume $f_n=f_0 (1-\epsilon_\star)$, where $f_0=1$ and $\epsilon_\star =0.05$ (fiducial case)
and $f_0=4$ and $\epsilon_\star =0.5$ (high SFE case).
Following the definition of the star formation efficiency $\epsilon_\star$, the SFR is 
approximated as 
\begin{align}\label{eq:SFRmodel}
{\rm SFR} &= \frac{4\pi \epsilon_\star}{1-\epsilon_\star} \rho(r) r^2V_{\rm c},\nonumber\\
&\simeq 0.11~\epsilon_\star f_0~ T_{\rm vir,4}^{3/2}~\msunyr,
\end{align}
where $V_{\rm c}$ is the halo circular velocity.
This SFR is used to estimate the emissivity of stellar irradiation (LW and X-rays) and to calculate 
the bulge growth in mass.
Note that in most cases, the gas mass within $r\simeq r_{\rm c}$ for the given initial condition is lower than 
the total mass of newly forming stars\footnote{Although we do not consider the self-gravity of gas in our simulations,
it would dominate the bulge gravity at the intermediate scale of $\sim O(10~\pc)$ only in the early stage of 
the bulge formation where $M_\star \la10^6~\msun$. 
In this case, however, radiative feedback associated with BH accretion blows the gas outward
and then the gas self-gravity becomes less important eventually (see \S\ref{sec:Mb}).}.
Therefore, our star-formation model implicitly assumes that the bulge growth is not led by in-situ star formation 
but efficient migration of stars formed at larger radii with morphological evolution owing to stellar relaxation
(see also Appendix \ref{sec:appA}).

The initial velocity field is set to $(v_r, v_\theta, v_\phi) = (0,0,j_{\rm in}/R)$,
where $R=r\sin \theta$ is the cylindrical radius.
We assume that the initial specific angular momentum $j_{\rm in}(R)$ is proportional to 
the Keplerian velocity measured with the enclosed gas mass $\mathcal{M}$ within a distance of $R$.
Namely, we adopt $j_{\rm in}(R) = j_0 \sqrt{G \mathcal{M} R}$ and set $j_0=0.3$.
The model assumption is motivated by the fact that the dynamics of a collapsing gas obeys 
a self-similar solution and the rotational velocity in the accreting envelope where $\rho \propto r^{-2}$
is a good fraction of the Keplerian velocity in agreement with hydrodynamical simulations of 
collapsing gas in a DM halo \citep{Abel_2002,Yoshida_2008,IOT_2014}.

It is worth noting that if a uniform distribution of $j_{\rm in}$ is assumed, mass accretion within the well-defined 
centrifugal radius ($\equiv j_{\rm in}^2/G \mathcal{M}$) is prevented without efficient viscous angular momentum transport.
In this case, gas accretion through an isothermal, geometrically thin disk is stunted unless the centrifugal radius is smaller than 
$\la 1\%$ of the BH gravitational influence radius \citep{Sugimura_2018}.
As shown in \S\ref{sec:result}, however, the initial condition adopted in our simulations leads to a wide range of 
angular momentum of the inflowing gas and thus allows mass accumulation at various radii.
Thus, the disk turns out dense enough to be gravitationally unstable and additional viscosity given by 
the second term of Eq.~(\ref{eq:alpha}) is activated. 
As a result, the accretion flow overcomes the angular momentum barrier and feeds the central BH at super-Eddington rates
(see \S\ref{sec:fid}).

An alternative way to alleviate the angular momentum problem is to consider the acceleration of the BH
that would be expected in the presence of a dense (nuclear) star cluster.
Gravitational scattering with the stars results in a jitter at the location of the BH, which leads to a cancellation of 
the angular momentum of the infalling gas onto the BH in its own rest-frame.
\cite{Alexander_Natarajan_2014} showed that this mechanism permits extended periods of super-Eddington accretion
until the BH grows sufficiently to outweigh the star cluster \citep[see also][]{Natarajan_2021}.
The application to off-centered jittering BHs is left for future investigation, while in our simulations the location of the BH is
fixed at the center but anomalous viscosity is responsible for angular momentum transport
even without its cancellation.

\begin{figure*}
\begin{center}
{\includegraphics[width=150mm]{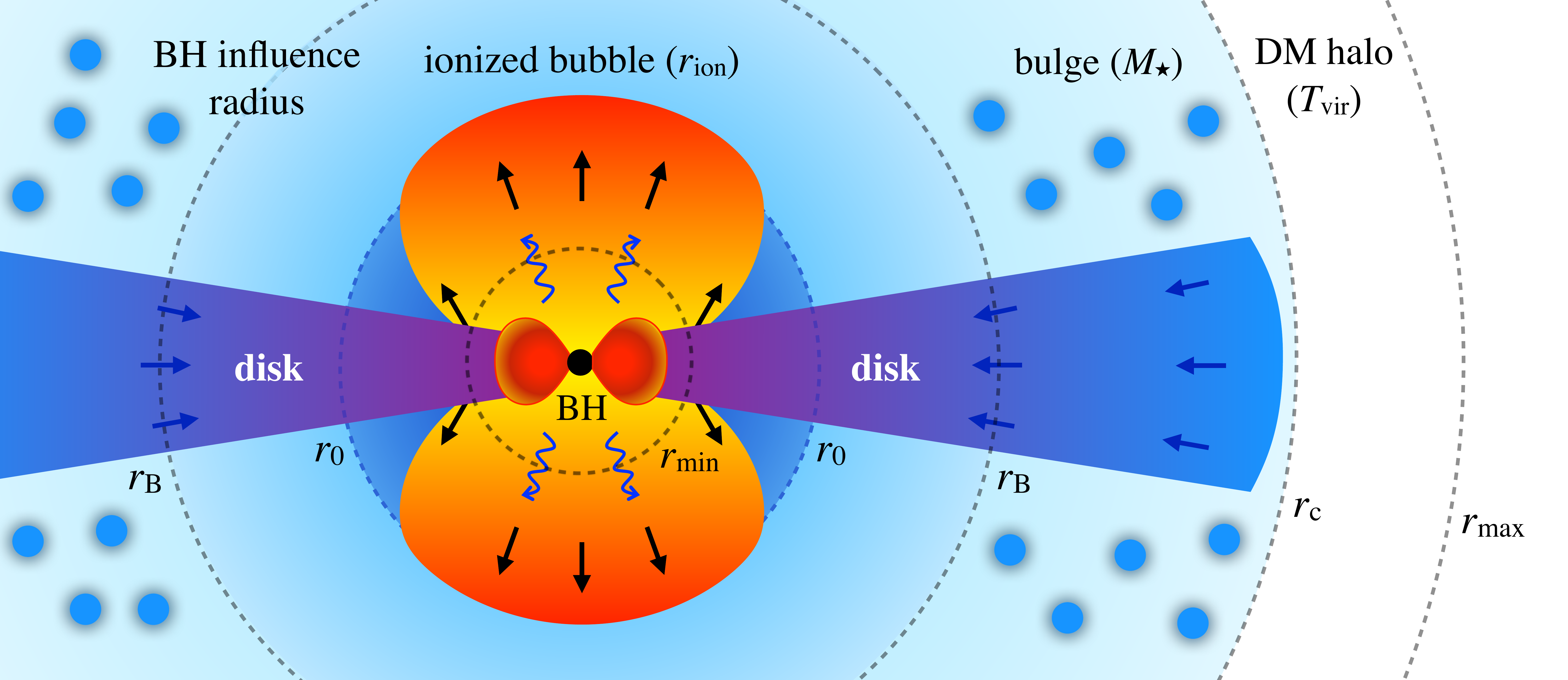}}
\caption{A schematic picture of accretion flows onto a massive BH embedded in a protogalactic bulge with
a mass of $M_\star$ and DM halo with a virial temperature of $T_{\rm vir}$.
There are four characteristic scales: the core radius of the initial density distribution ($r_0$),
the BH gravitational influence radius ($r_{\rm B}$), the bulge core size ($r_{\rm c}$), and 
the size of ionized regions created by the radiating BH ($r_{\rm ion}$).
The computational domain covers $r_{\rm min}\leq r \leq r_{\rm max}$, 
where those characteristic radii are well resolved.
}
\label{fig:cartoon}
\end{center}
\end{figure*}

\subsection{Cases considered}

In Table~\ref{tab:model}, we summarize the models we consider in this paper. 
We consider two types of DM halo with different virial temperatures of $T_{\rm vir} \simeq 10^5~\K$
($M_{\rm h}=2\times 10^9~\msun$ and $z=15$) and $T_{\rm vir} \simeq 10^4~\K$
($M_{\rm h}=10^8~\msun$ and $z=10$). 
The virial temperature of the halo is calculated from the given values of $M_{\rm h}$ and $z$
\citep[see Eq.~81 in][]{Iliev_2001} by setting $\mu =0.6$ for $T_{\rm vir}\geq 10^4~\K$
(note that the mean molecular weight is calculated self-consistently 
by solving non-equilibrium thermochemistry in our simulations).
The two cases are referred to ``Massive Halo" model and ``Normal Halo" model, respectively.
In the former case, we consider a relatively massive DM halo because high-redshift 
quasar host galaxies form in rare, overdense regions of the universe at $z> 6$ and their progenitor
halos would be likely as massive as our choice at higher redshifts when seed BHs form \citep{Li_2021}.
The latter case corresponds to that of typical first galaxies where star formation is triggered by hydrogen 
atomic cooling \citep{Bromm_Yoshida_2011}.

We perform the 13 simulations shown in Table~\ref{tab:model} with different bulge mass 
($M_\star = 0$, $10^5$, $10^6$, $10^{6.5}$, and $10^7~\msun$), initial gas density at the center 
($n_{\rm c}=10^2$, $10^{2.5}$, $10^{3}$, $10^{3.5}$, and $10^4~\cc$), and halo virial temperature 
($T_{\rm vir}=10^4$ and $10^5~\K$).
In all the simulations, we adopt the initial BH mass to $M_{\bullet, 0}=10^5~\msun$.
The SFE and the initial density normalization are assumed to be $\epsilon_\star = 0.05$ 
and $f_0=1$, respectively, for most of the cases.
To see the impact of their parameter choice, we additionally consider a case (B7T5N3-highSFE),
where a higher SFE of $\epsilon_\star = 0.5$ and $f_0=4$ are set.

In Fig.~\ref{fig:cartoon}, we summarize the characteristic physical scales that determine the properties 
of accretion flows onto a seed BH embedded in a protogalaxy.
In our case, the typical value of the BH gravitational influence radius (the so-called Bondi radius) is given by 
\begin{equation}
r_{\rm B} \equiv \frac{GM_\bullet }{c_{\rm s}^2} \simeq  6.4~\left(\frac{M_\bullet}{10^5~\msun}\right) 
\left(\frac{T}{10^4~\K}\right)^{-1}~\pc,
\end{equation}
for neutral gas ($\mu=1.22$) and $r_{\rm B,ion} \simeq 0.32~{\rm pc}~(M_\bullet/10^5~\msun)(T/10^5~\K)^{-1}$ 
for ionized gas ($\mu=0.6$).
We note that both the scales are well resolved in our simulations (i.e., $r_{\rm min}<r_{\rm B,ion}<r_{\rm B}< r_{\rm max}$).
This indicates the distance from the BH within which the gravitational energy dominates over the thermal energy
of the gas and thus gas accretion begins to occur unless BH feedback plays an important role.
In reality, however, when the BH grows via mass accretion, the accreting flow releases its gravitational 
energy as radiation, which heats the surrounding gas and forms an ionized bubble with a size of $r_{\rm ion}$.
In the intermediate region, stars form a galactic bulge component with a total mass of $M_\star$ and 
a half-mass radius of $R_{\rm e}=(1+\sqrt{2})r_{\rm c} \simeq 16~\pc ~(M_\star/10^7~\msun)^{0.47}$.
For both the ``Massive" and ``Normal" halo cases, the halo virial radius is $r_{\rm vir}\simeq 1.6~\kpc$
and $0.9~\kpc$, respectively.
Therefore, our computational domain covers the interior of the host DM halo ($r_{\rm max}\simeq 0.1~r_{\rm vir}$).

\vspace{2mm}
\section{Results}\label{sec:result}

\subsection{Fiducial case}\label{sec:fid}

\begin{figure}
\begin{center}
{\includegraphics[width=85mm]{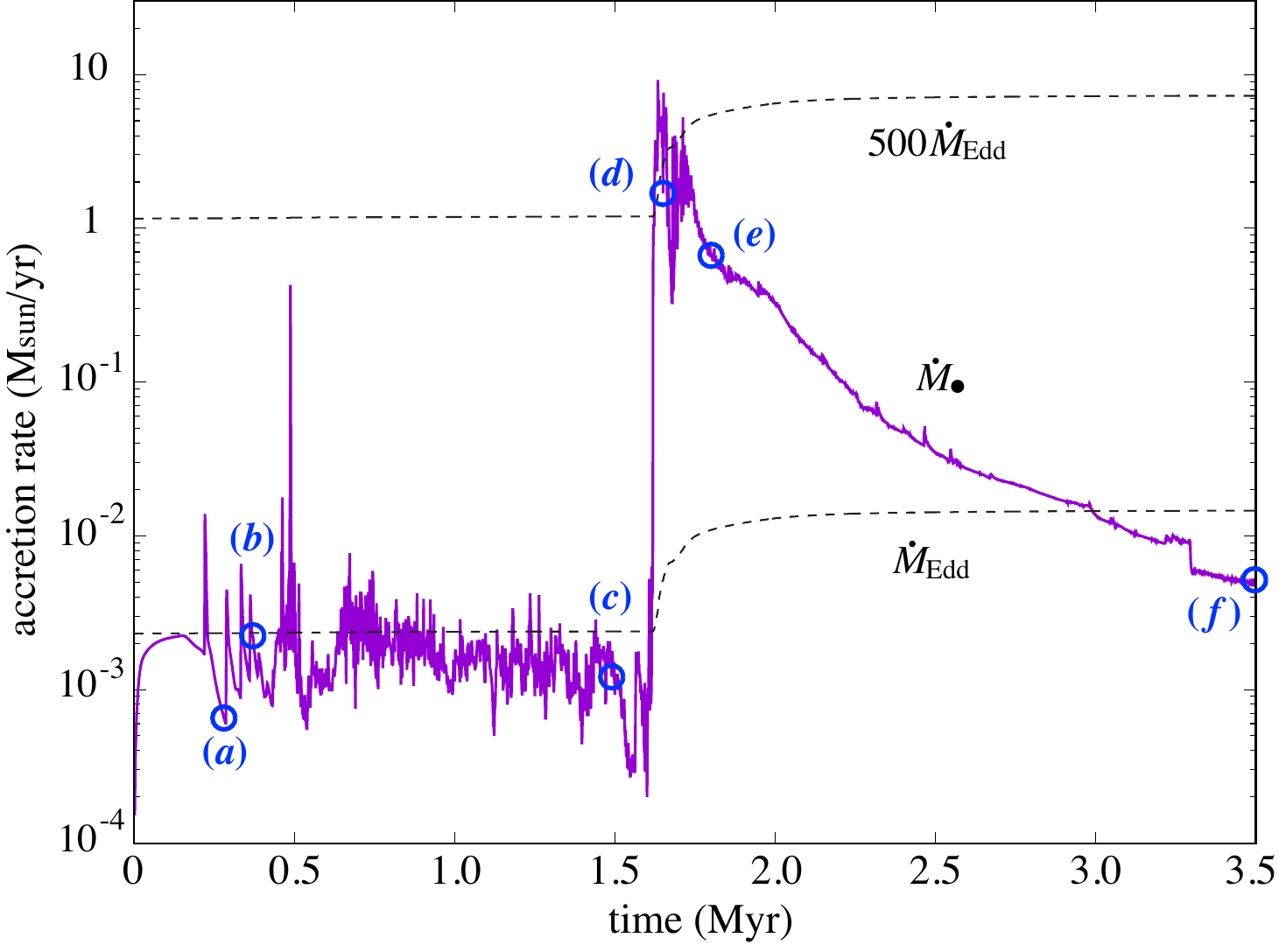}}
\caption{Time evolution of the gas accretion rate onto a massive BH with an initial mass of $M_\bullet = 10^5~\msun$
embedded in a dense gas cloud with the central density of $n_{\rm c}=10^3~\cc$ hosted in a stellar bulge
with $M_\star = 10^7~\msun$ and a massive DM halo with a virial temperature of
$T_{\rm vir}\simeq 10^5~\msun$ (model: B7T5N3).
The dashed curves show the Eddington accretion rate $\dot{M}_{\rm Edd}$ and the critical rate
for the onset of rapid mass accretion ($\sim 500~\dot{M}_{\rm Edd}$) obtained in previous RHD simulations \citep{IHO_2016}.
Open circles mark the six epochs at which we show the density and temperature distributions in Fig.~\ref{fig:rho_dist}
and radial profiles of the physical quantities in Fig.~\ref{fig:rad_prof}. 
}
\label{fig:time_Mdot}
\end{center}
\end{figure}

\begin{figure*}
\begin{center}
{\includegraphics[width=140mm]{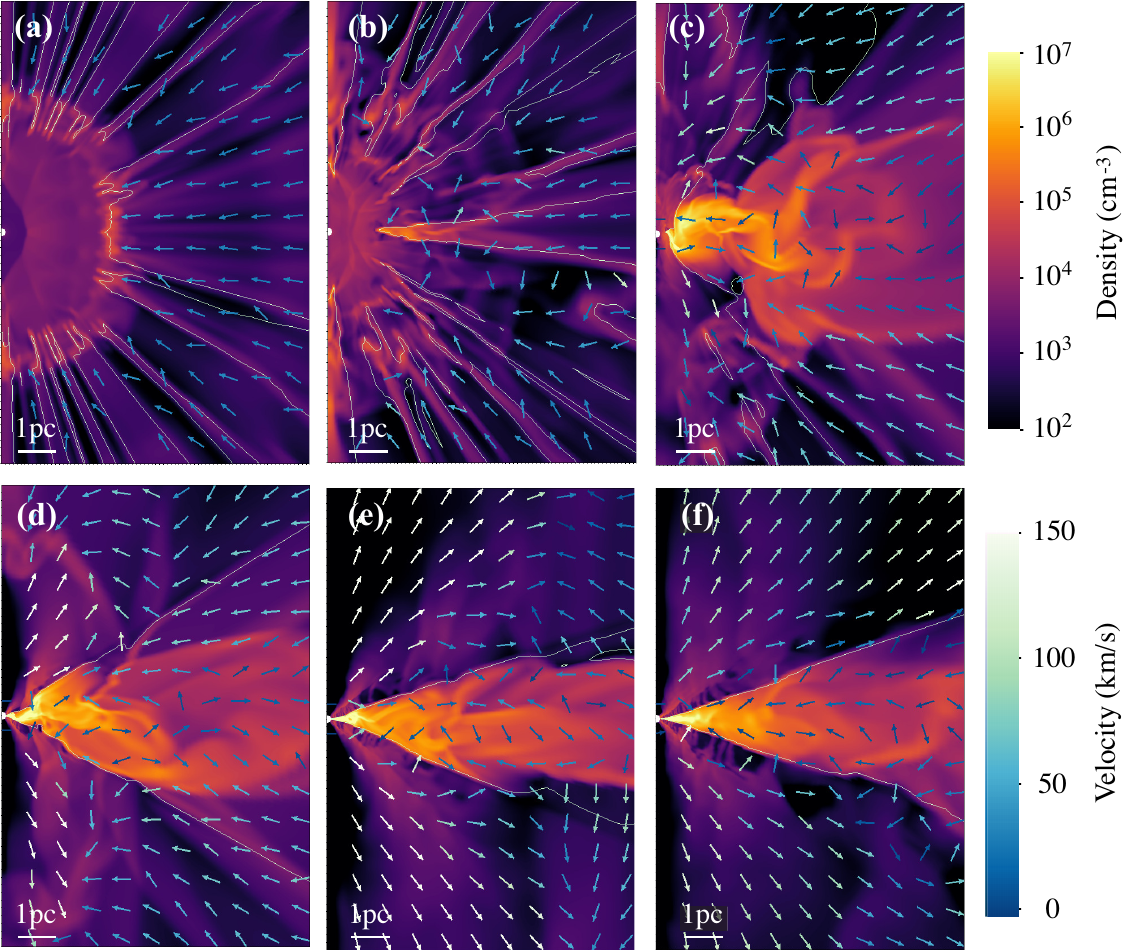}}
\caption{Distribution of the gas density for the fiducial case at the six elapsed times 
denoted by open circles in Fig.~\ref{fig:time_Mdot}.
In each panel, the location of the ionization front where the neutral fraction is $x_{\rm HI} = 0.95$ (thin contours)
and the velocity vectors are overlaid. In the first two panels of (a) and (b), the velocity vectors with $|\mbox{\boldmath $v$}|\geq 30~\kms$ are shown.
At the early stage ($t<1.6$ Myr; phases $a$-$c$), the intense inflows of neutral gas lead to 
collapse of the ionized region and form a dense gaseous accretion disk within the BH influence radius.
In the late stage ($t>1.6$ Myr; phases $d$-$f$), the disk feeds the BH at rates of $\gg \dot{M}_{\rm Edd}$
and the bipolar outflows driven by radiation decreases the gas supplying rate from larger radii.
Movies of this simulation are available \href{https://www.youtube.com/playlist?list=PLqDrKvxmu0l4fyJ74u5ZYyQ8PmwuNZA32}{here}.
}
    \label{fig:rho_dist}
    \end{center}
\end{figure*}

We first discuss the fiducial case where a BH with $M_\bullet = 10^5~\msun$ is embedded 
in a dense gas cloud with the central density of $n_{\rm c}=10^3~\cc$ hosted in a stellar bulge
with an initial mass of $M_\star = 10^7~\msun$ hosted in a massive DM halo with 
$M_{\rm h}=2\times 10^9~\msun$ at $z=15$ (the corresponding virial temperature 
is $T_{\rm vir}\simeq 10^5~\K$).
The star formation efficiency is set to $\epsilon_\star =0.05$.

Fig.~\ref{fig:time_Mdot} shows the time evolution of the accretion rate $\dot{M}_\bullet$ (solid curve).
As reference values, the Eddington accretion rate $\dot{M}_{\rm Edd}$ and the critical rate 
for the onset of hyper-Eddington accretion ($\simeq 500~\dot{M}_{\rm Edd}$; \citealt{IHO_2016})
are shown with dashed curves.
At the early stage of $t<1.6$ Myr, the mass accretion proceeds episodically.
While the accretion rate exceeds the Eddington value during those burst phases, the long-term average rate is 
$\simeq \dot{M}_{\rm Edd}$ due to radiative feedback associated with BH feeding.
At $t\simeq 1.6$ Myr, the mass accretion rate abruptly rises and reaches $\dot{M}_\bullet \simeq 10~\msunyr$,
which corresponds to $\simeq 2.5\times 10^3~\dot{M}_{\rm Edd}$.
After the peak, the rate gradually decreases with time but the super-Eddington accretion phase
lasts until $t\simeq 3$ Myr.
Note that the BH mass is $\simeq 6.3\times 10^5~\msun$ at the end of the simulation.
The overall behavior of the accretion rate is consistent with those found in previous RHD simulations
where the BH is not embedded in the external gravitational potential composed of stars and DM
\citep{IHO_2016, Takeo_2018,Takeo_2020,Park_2020,Toyouchi_2021}.

\begin{figure*}
\begin{center}
{\includegraphics[width=154mm]{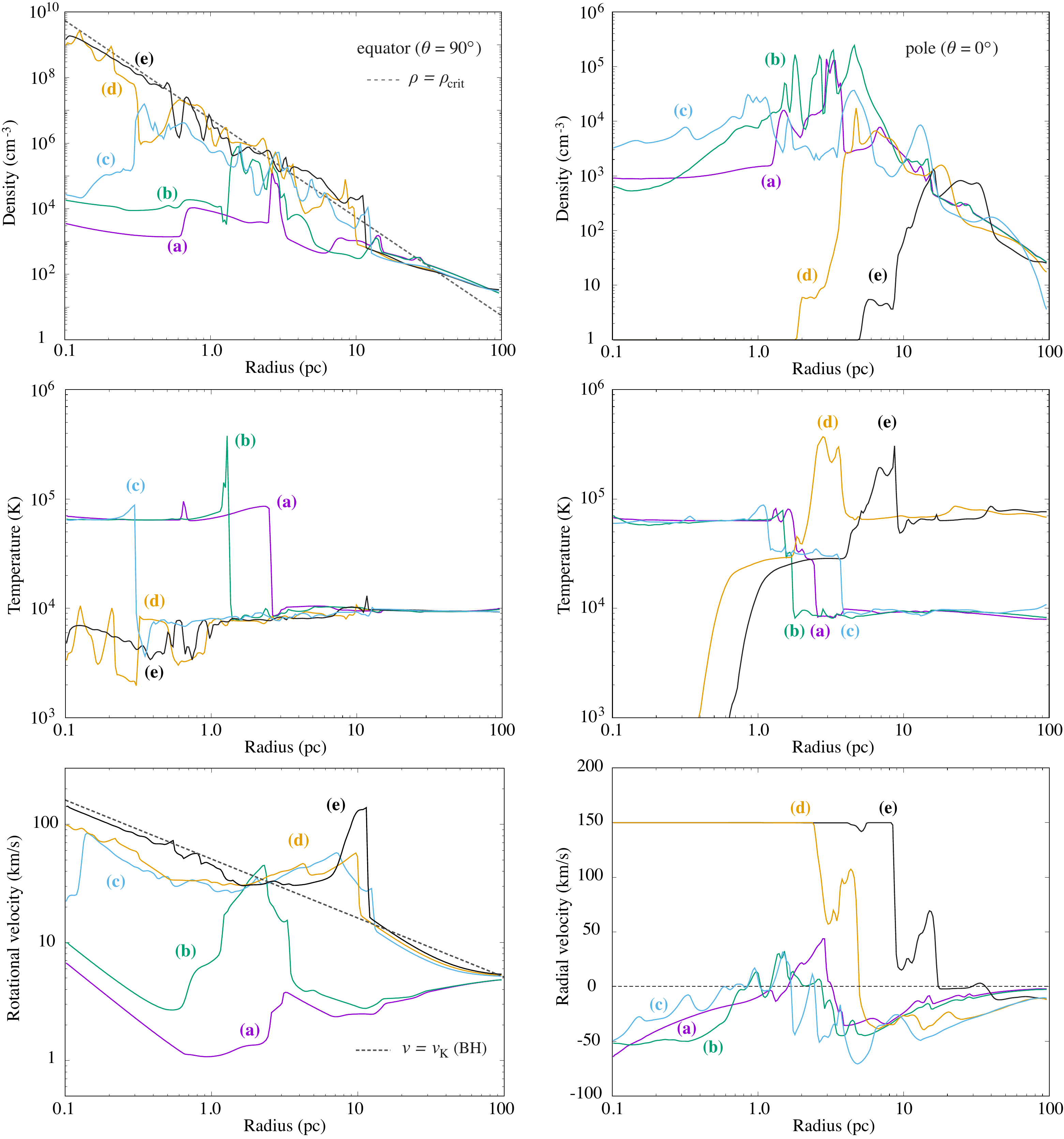}}
\caption{Radial structure of the gas density (top), temperature (middle), velocity (bottom) 
along the equator ($\theta = 90^\circ$; left-hand panels) and pole ($\theta = 0^\circ$; right-hand panels). 
The bottom panels present the rotational velocity along the equator and radial velocity along the polar direction, respectively.
In each panel, we show the profiles at different epochs during and after the transition: 
$t = 0.28~{\rm Myr}$ (phase $a$; purple), $0.35~{\rm Myr}$ (phase $b$; green), $1.5~{\rm Myr}$ (phase $c$; blue), 
$1.65~{\rm Myr}$ (phase $d$; yellow), and $1.8~{\rm Myr}$ (phase $e$; black). 
In the top-left and bottom-left panels, we overlay the critical density (see the viscous model below Eq.~\ref{eq:alpha}) and 
the Keplerian velocity for $M_\bullet =6\times 10^5~\msun$.
}
    \label{fig:rad_prof}
    \end{center}
\end{figure*}

Fig.~\ref{fig:rho_dist} presents the distribution of the gas density in the domain of $0\leq x \leq 8~\pc$ 
and $|z|\leq 6~\pc$ at six different elapsed times (phases $a$-$f$ of Fig.~\ref{fig:time_Mdot}) .
In each panel, the location of the ionization front where the neutral fraction is $x_{\rm HI} = 0.95$ (thin contours)
and the velocity vectors are overlaid.
Fig.~\ref{fig:rad_prof} presents the radial profiles of the gas density (top), temperature (middle), 
and velocity (bottom) along the equator ($\theta=90^\circ$; left panels) and the pole 
($\theta=0^\circ$; right panels), respectively.
In the bottom panels, we show the rotational velocity along the equator and radial velocity along the polar direction, respectively.
Note that the radial profiles at the phase ($f$) are not shown because they are similar to 
those at the phase ($e$) except that the shock front of the polar outflow moves forward. 
This indicates that the disk properties are in a quasi-steady state although the net accretion rate 
onto the BH gradually decreases.

At the early stage of $t<0.3$ Myr, the accreting BH emits nearly isotropic radiation
and creates an ionizing bubble.
The prolate shape of the bubble is caused by angular momentum of the inflowing gas with 
a lower density near the pole and by anisotropic radiation produced when the BH feeding rate 
exceeds the Eddington value in a short period.
While the size of the ionizing bubble is initially maximized at $r_{\rm ion}\sim 30~\pc$, 
it shrinks to $\sim 3~\pc$ by the phase ($a$) owing to efficient radiative recombination of the gas 
surrounding the Eddington-limited radiating BH.
As a result, the equatorial inflow of neutral gas penetrates into the interior of the BH gravitational 
influence radius of $r_{\rm B}\simeq 8~\pc$ for neutral gas with $T\simeq 8000~\K$ 
(see the left panels in Fig.~\ref{fig:rad_prof}).
The inflow with strong ram pressure leads to fragmentation of the shell of the ionized bubble.
Previous studies also found that collapse of an ionized region 
surrounding the accreting BH is an essential process to trigger the accretion transition
\citep{IHO_2016, Sakurai_2016,Toyouchi_2019, Toyouchi_2021}.
However, we note that gas rotation, anisotropic radiation, and dust obscuration moderate 
the propagation of ionizing radiation to the equatorial region and thus {\it do} promote the transition
\citep[see also][]{Takeo_2018,Takeo_2020}.

\begin{figure*}
\begin{center}
{\includegraphics[width=87mm]{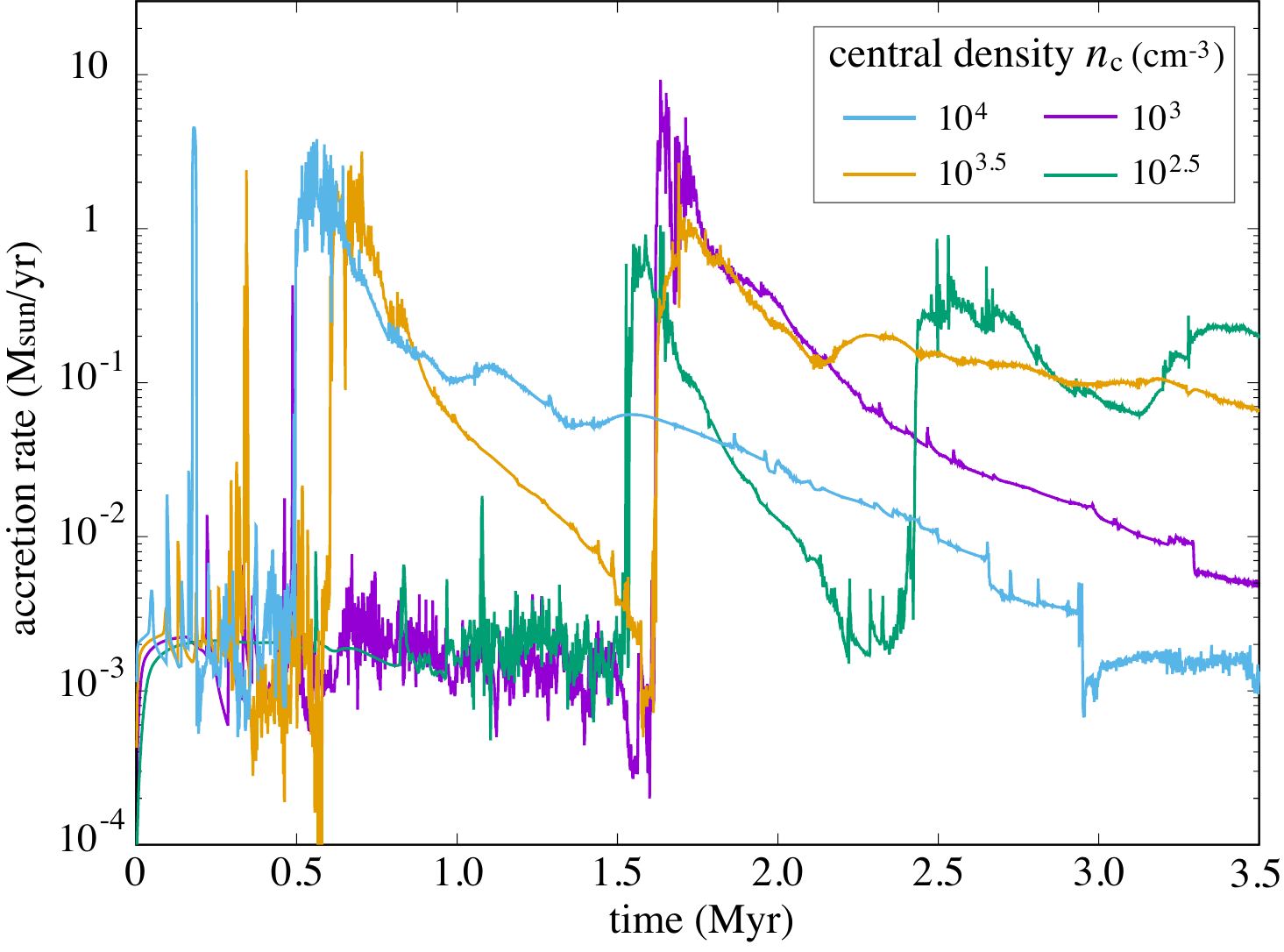}}\hspace{10pt}
{\includegraphics[width=87mm]{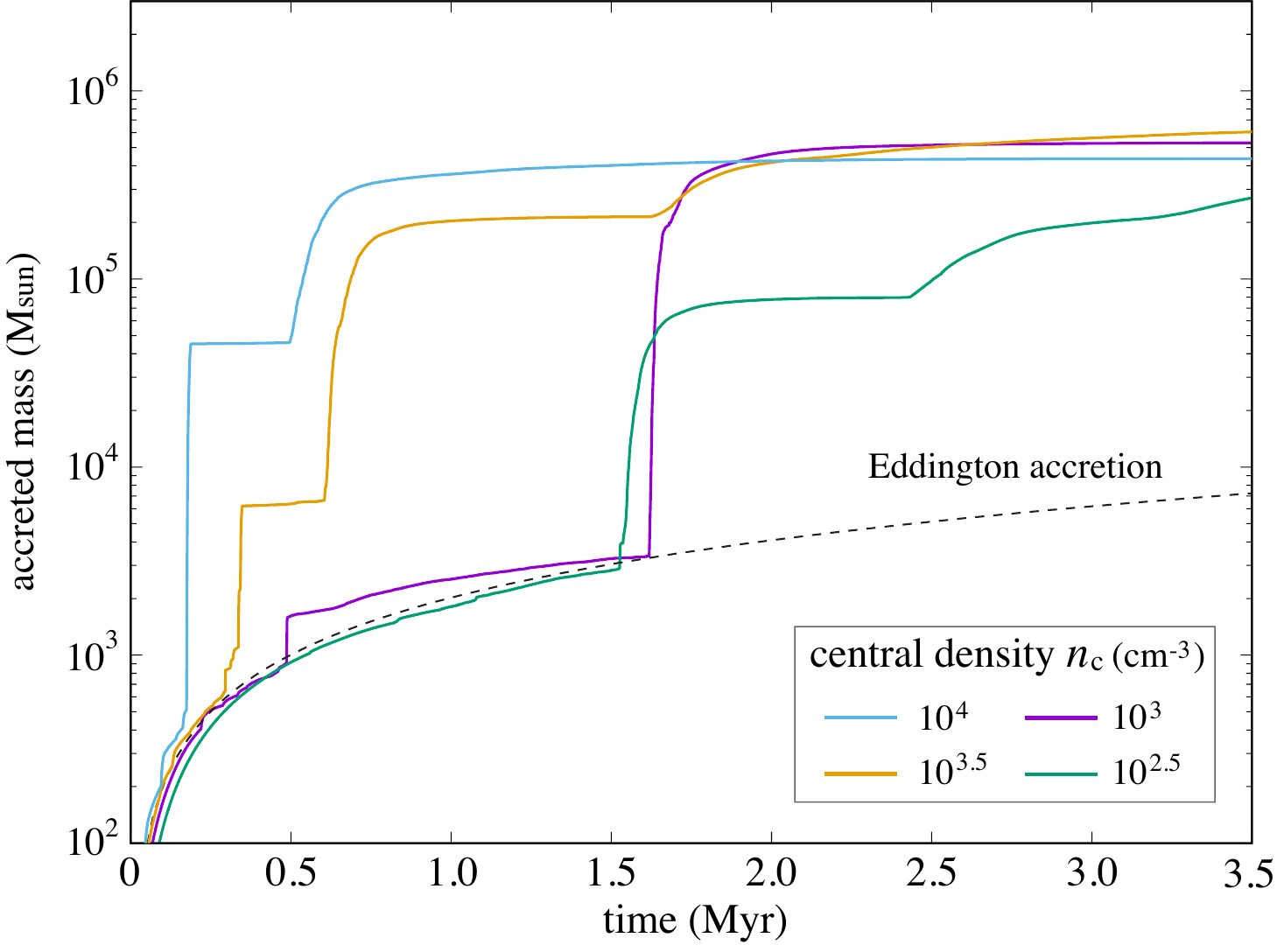}}
\caption{Time-evolution of the BH accretion rate (left) and 
the mass accreted onto the BH (right) with different values of the central density for the initial conditions; 
$n_{\rm c}=10^4~\cc$ (cyan), $10^{3.5}~\cc$ (orange), $10^3~\cc$ (purple; fiducial case), and $10^{2.5}~\cc$ (green).
The transition of rapid accretion tends to occur earlier with higher values of $n_{\rm c}$.
In spite of the stochastic nature of mass accretion, the mass accreted onto the BH reaches similar values of
$\Delta M_\bullet \simeq (3-6)\times 10^5~\msun$, which is substantially higher than that 
expected under the Eddington growth (dashed curve).}
\label{fig:nc}
\end{center}
\end{figure*}

At $t\simeq 0.35$ Myr (phase $b$), intense inflows of neutral gas feeds the nuclear scale within $\sim 2~\pc$,
but the BH feeding rate does not increase at the same moment.
This is primarily because the inflowing gas with angular momentum forms a rotationally-supported disk
and the rotational velocity exceeds the Keplerian value at $r\sim 2~\pc$ 
(see the left-bottom panel in Fig.~\ref{fig:rad_prof}), where the inflow speed through the disk slows down significantly.
In addition, the outward radiation force exerted through electron scattering, absorption of UV radiation
by atomic bound-free transitions and dust grain, and re-emission of IR radiation from heated dust
prevent the inflowing gas from feeding the BH at rates of $\ga \dot{M}_{\rm Edd}$ \citep{Toyouchi_2019}.
Meanwhile, the ionized regions become as small as $r_{\rm ion}\simeq 0.3~\pc$ and 
the inflow from larger radii accumulates mass in the nuclear region (phase $c$; 
see Fig.~\ref{fig:rho_dist} and the left panels in Fig.~\ref{fig:rad_prof}).

At the late stages of $t\ga 1.6$ Myr (phase $d$), the disk becomes opaque to UV by dust,
and the inner-edge of the dense dusty disk reaches the central cell with $r_{\rm min}=0.1~\pc$.
Meanwhile, the mid-plane density approaches the profiles of $\rho_{\rm crit}=\Omega_{\rm K}^2/(\pi G)$
(the dashed line in the top-left panel of Fig.~\ref{fig:rad_prof}),
where the Toomre's $Q$ parameter is nearly set to unity within the disk.
Since the disk feeds the BH at rates exceeding the Eddington value, the radiation flux from the center blows
the gas away and launches strong outflows toward the polar regions at velocities of 
$v_{\rm out}\simeq 150~\kms$, which is significantly faster than the escape velocity from the system (see Fig.~\ref{fig:rho_dist}).
In the phase ($e$), the radiation-driven outflow collides with the inflowing gas and creates 
bipolar low-density cavities surrounded by a dense shell
(see the right panels in Fig.~\ref{fig:rad_prof}).
As a result, the emergent UV radiation begins to heat the disk surface and drives thermally-driven outflows.
Although the radiation/mechanical feedback effect suppresses the mass supply from larger radii,
the dense accretion disk coexists with outflows and feeds the central BH at super-Eddington rates
$\ga \dot{M}_{\rm Edd}$ until $t\simeq 3.5$ Myr (phase $f$), when the simulation terminates.

\subsection{Parameter dependence}

Next, we investigate the dependence of the result on the model parameters:
(1) the central density $n_{\rm c}$, (2) the bulge mass $M_\star$, (3) the star formation efficiency
$\epsilon_\star$, and (4) the halo virial temperature $T_{\rm vir}$.

\subsubsection{Central density $n_{\rm c}$}\label{sec:nc}

In Fig.~\ref{fig:nc}, we present the time evolution of the BH accretion rate (left) and the 
mass accreted onto the BH (right) with different values of the central density for the initial conditions.
The transition to rapid accretion tends to occur earlier with higher values of $n_{\rm c}$.
This is because the size of the ionized region surrounding the BH becomes smaller 
due to efficient recombination and thus collapse of the compact ionized region triggers 
the BH feeding in a shorter dynamical timescale.
Although this trend holds for $n_{\rm c}\geq 10^3~\cc$, the accretion transition for the lowest
density case ($n_{\rm c}=10^{2.5}~\cc$) occurs earlier than that for the fiducial case due to the
stochastic nature of mass accretion through filamentary structures and a compact disk at $\sim 1~\pc$
(see the phase $b$ in Fig.~\ref{fig:rho_dist}).
Moreover, the two cases ($n_{\rm c}=10^{2.5}$ and $10^{3.5}~\cc$) show accretion bursts twice by the end of the simulations.
In those cases, the vertical oscillation of the accretion disk after its formation promotes mass loading into radiation-driven outflows
and the net accretion rate through the disk decreases quickly with time.
However, as a result of weak radiative feedback, continuous mass supply from large radii triggers the second accretion burst within $\sim 1$ Myr.
In spite of the stochastic nature of mass accretion, the gas mass accreted onto the BH reaches 
similar values of $\Delta M_\bullet \simeq (3-6)\times 10^5~\msun$, which is substantially higher than that expected 
under the Eddington growth (dashed curve).
We note that the accreted mass is comparable to that contained within the gaseous core in the initial state
($r\la r_0$).

It is worth mentioning that all the cases in Fig.~\ref{fig:nc} experience the transitions to rapid accretion
exceeding the Eddington rate.
As described in Fig.~\ref{fig:rad_prof}, the transition is induced by continuous accretion of neutral gas 
whose inward ram pressure pushes the ionization front (hereafter, I-front).
In \S\ref{sec:analytic}, we quantify the conditions required for the onset of rapid mass accretion.
On the contrary, when the central density is lower than $n_{\rm c}\sim 10^2~\cc$ (model: B7T5N2), 
the ionized region quickly reaches the core radius $r_0$.
Since the density follows $\rho \propto r^{-2}$ outside the core, the I-front further expands 
at speeds faster than the sound speed of ionized gas without disturbing the density structure 
(the so-called R-type I-front)
and the entire cloud is ionized \citep[e.g.,][]{Kitayama_2004,Whalen_Norman_2006}.
In this case, therefore the transition to rapid mass accretion does not occur.

\subsubsection{Bulge mass $M_\star$}\label{sec:Mb}

Fig.~\ref{fig:Mb} shows the time evolution of the mass accretion rate with four 
different bulge masses at $0 \leq M_\star /\msun \leq 10^7$.
In Fig.~\ref{fig:Mb_rho}, we also present the density distribution for the two cases of 
$M_\star =0$ (no bulge) and $10^{6.5}~\msun$ around the epochs when the simulations terminate,
respectively.

Without the bulge component, the accretion rate shows multiple bursts with a period of $\simeq 0.6$ Myr
and the time-averaged rate is significantly below the Eddington value; namely 
$\langle \dot{M}_\bullet \rangle \simeq 0.2~\dot{M}_{\rm Edd}$.
In this case, the I-front initially propagates up to $\sim 25~\pc$ 
and heats the ambient gas outside the BH influence radius $r_{\rm B} \simeq 5~\pc$. 
Within the hot ionized gas, the BH's gravity accelerates the inflow within a new sonic point in the hot region
($\simeq 0.4~\pc$ for $T\simeq 10^5~\K$), while the gas pressure (and partially the radiation pressure force) 
pushes the gas outwards at $r\ga 1~\pc$ and the mass accretion rate is suppressed.
In the quiescent phases, although the gas is in thermal-pressure equilibrium against the BH gravity, 
the mass depletion owing to BH feeding reduces the outward pressure gradient force.
As a result, the dense shell surrounding the ionized gas accretes to the BH and leads to burst-like 
accretion again.
The overall behavior of mass accretion is consistent with previous RHD simulations where the BH is 
assumed to be embedded in a uniform density distribution
\citep[e.g.,][]{Ciotti_Ostriker_2001,Milosavljevic_2009,PR_2011,PR_2012, IHO_2016,Park_2017}.
In our case, where the gas density decreases outward, the I-front continues to expand during 
the multiple episodes of burst-like accretion and reaches $\ga 40~\pc$ by $t\simeq 3.5$ Myr.

\begin{figure}
\begin{center}
{\includegraphics[width=85mm]{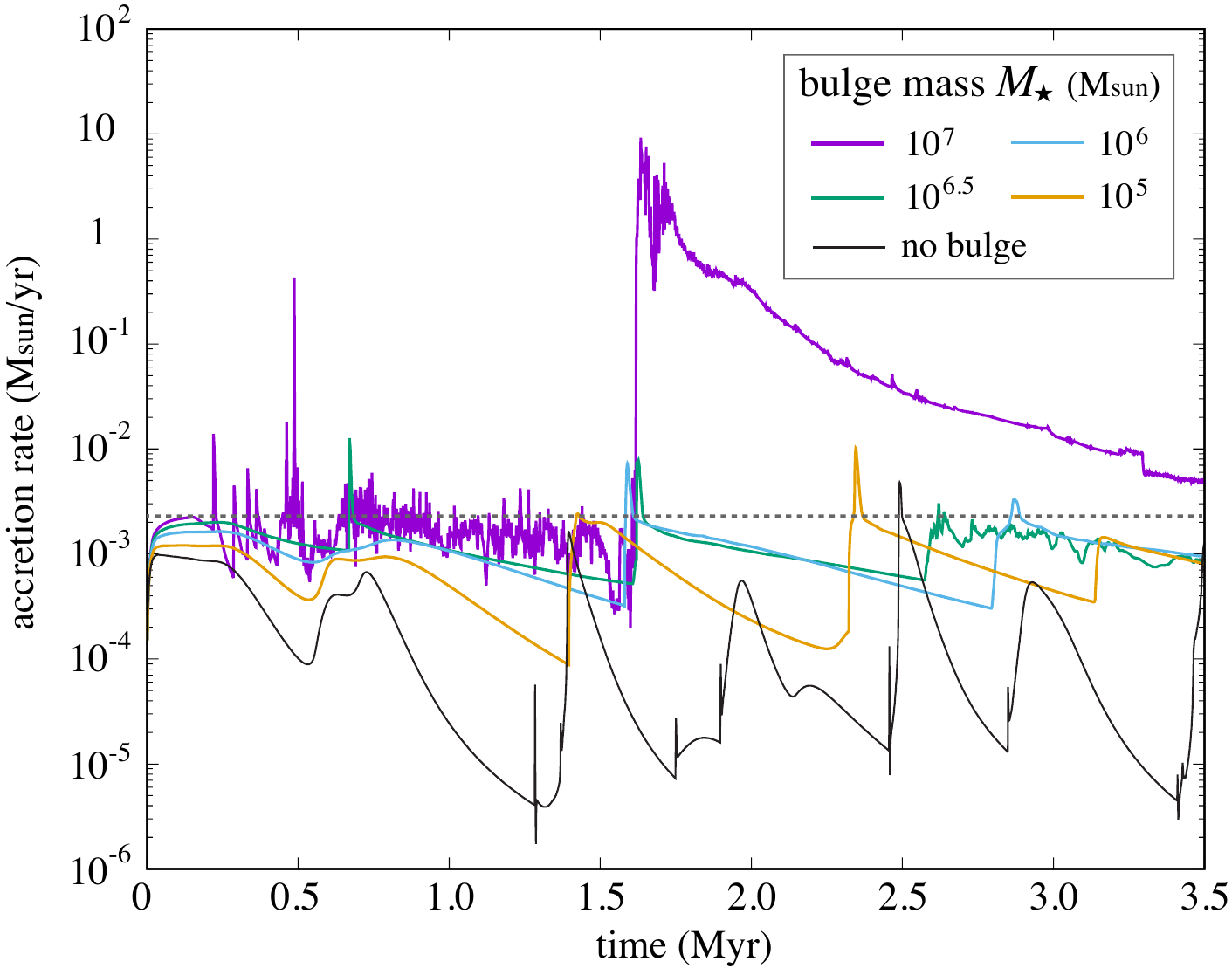}}
\caption{Time-evolution of the mass accretion rate onto the BH with 
different values of the bulge mass; $M_\star = 10^7~\msun$ (purple),
$10^{6.5}~\msun$ (green), $10^6~\msun$ (cyan), $10^5~\msun$ (orange), 
and no bulge (black).
The transition to rapid accretion phases occur when the bulge mass is as high as $M_\star =10^{7}~\msun$,
which corresponds to $\sim 100~M_\bullet$ (see also the discussion in \S\ref{sec:analytic}). 
As a reference, the Eddington accretion rate for $M_\bullet =10^5~\msun$ is overlaid (horizontal dashed line), 
above which anisotropic radiation is produced.
}
\label{fig:Mb}
\end{center}
\end{figure}

With the bulge mass increasing, the episodic behavior of mass accretion ceases and the time-average 
rate increases because the gas density and mass inflow rate are enhanced within the additional bulge gravitational potential.
When the bulge mass is lower than $M_\star < 10^7~\msun$, however, the ionized region
continues to expand and heats the surrounding gas in the same way as in the case without a bulge.
Note that this statement is still valid if the self-gravity of gas would be taken into account in our simulations,
since the total gas mass in the domain is lower than $10^7~\msun$.
As a result, the transition to rapid mass accretion does not occur by the end of the simulation.
Since the free-fall velocity at $r\simeq 50~\pc $ is $\sim 20~\kms$ (the enclosed mass is 
dominated by the bulge), the transition could not occur within another $\sim 2.5$ Myr 
even if all the radiative output turned off.

In summary, the size of the expanding I-front  determines the nature of mass accretion 
onto the BH, and the formation/acceleration of neutral-gas inflows from larger radii 
is a key process to trigger the accretion transition.
Based on those findings, we give a simple analytical argument for the conditions required
for the transition in \S\ref{sec:analytic}.

\begin{figure}
\begin{center}
{\includegraphics[width=85mm]{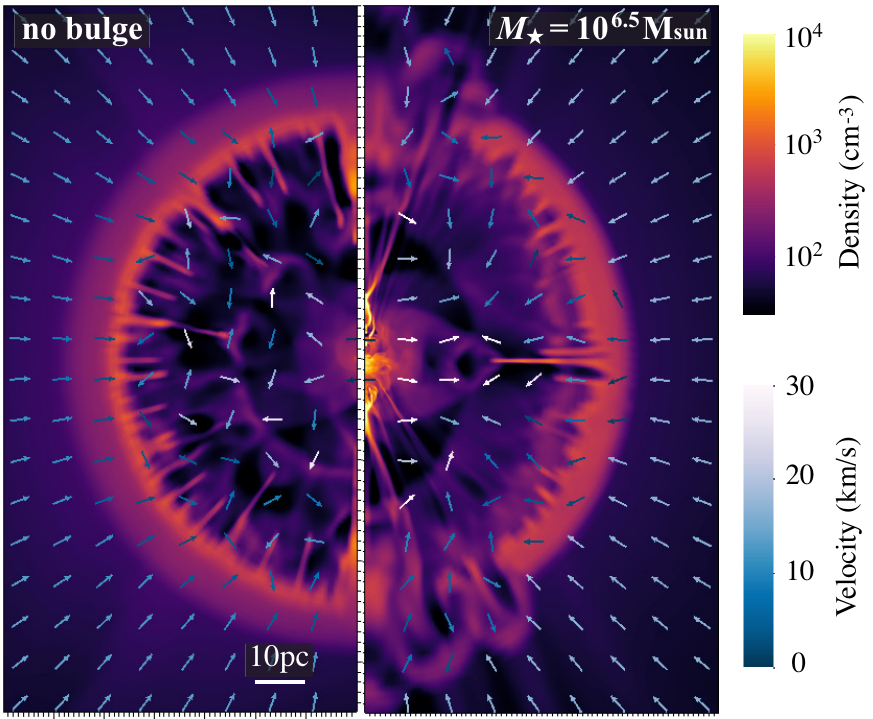}}
\caption{Distribution of the gas density for the case without bulge stars (left) and 
with the bulge of $M_\star =10^{6.5}~\msun$ (right) at the end of those simulations.
The arrows present the velocity vectors.
In both cases, the accreting BH is surrounded by an expanding ionized region and 
thus rapid inflows of neutral gas onto the central BH are not observed.
}
\label{fig:Mb_rho}
\end{center}
\end{figure}

\subsubsection{Star formation efficiency $\epsilon_\star$}\label{sec:highSFE}

Fig.~\ref{fig:MSFE} shows the time evolution of the mass accretion rate (solid), the Eddington accretion rate (dotted),
and the BH-to-bulge mass ratio of $M_\bullet /M_\star$ (dashed) for the case with a high SFE ($\epsilon_\star =0.5$; B7T5N3-highSFE).
The overall behavior of the accretion rate is consistent with the cases with the lower SFE as shown in Fig.~\ref{fig:nc}.
Compared to the fiducial case, the transition to rapid accretion occurs earlier because 
a larger amount of gas is accumulated within the gravitational influence radius of the fast growing bulge.
With the high SFR, intense stellar radiation heats the gas surrounding the BH but efficient radiative cooling 
keeps the gas neutral.
As a result, the attractive gravitational force by the massive bulge is more important than the negative 
feedback effect caused by the stellar irradiation.
In addition, the BH-to-bulge mass ratio evolves substantially within $\sim 3$ Myr and 
the ratio reaches $\sim 0.04$ at the end of the simulation.
This mass ratio is $\sim 10$ times higher than the BH-galaxy correlation seen in the local universe 
\citep{Kormendy_Ho_2013} and is consistent with those of bright quasars at $z>6$
(see more discussion in \S\ref{sec:Msigma}).

\subsubsection{Virial temperature of host DM halos $T_{\rm vir}$}\label{sec:Tv}

We examine three cases of BH accretion in a typical atomic-cooling halo with a virial temperature of 
$T_{\rm vir}=10^4~\K$ (see Table~\ref{tab:model}).
For all the cases, the I-front quickly expands and reaches the outer-most radius of the computational domain.
Since the I-front is R-type, the density profile does not change from the initial distribution except
within $1~\pc$, where the BH gravitational force pulls the ionized gas toward the center 
and increases the density as $\rho \propto r^{-3/2}$.
However, the accretion rate of the ionized gas is limited to the Eddington owing to the outward radiation 
force via electron scattering.
Due to the self-regulated nature of BH radiative outputs, the gas ejection from the halo
takes place in a relatively slow manner, unlike stellar irradiation and supernova feedback 
with given feedback energy \citep[e.g.,][]{Kitayama_2004, Kitayama_2005}.
Overall, the stunted growth of seed BHs in ``normal" atomic-cooling halos is consistent with 
previous studies that use cosmological hydrodynamical simulations \citep{Habouzit_2017,Latif_2018},
although our simulations focus on the early stage of the bulge formation without including
SN feedback.

\subsection{Radiative luminosity of rapidly accreting seed BHs}

\begin{figure}
\begin{center}
{\includegraphics[width=85mm]{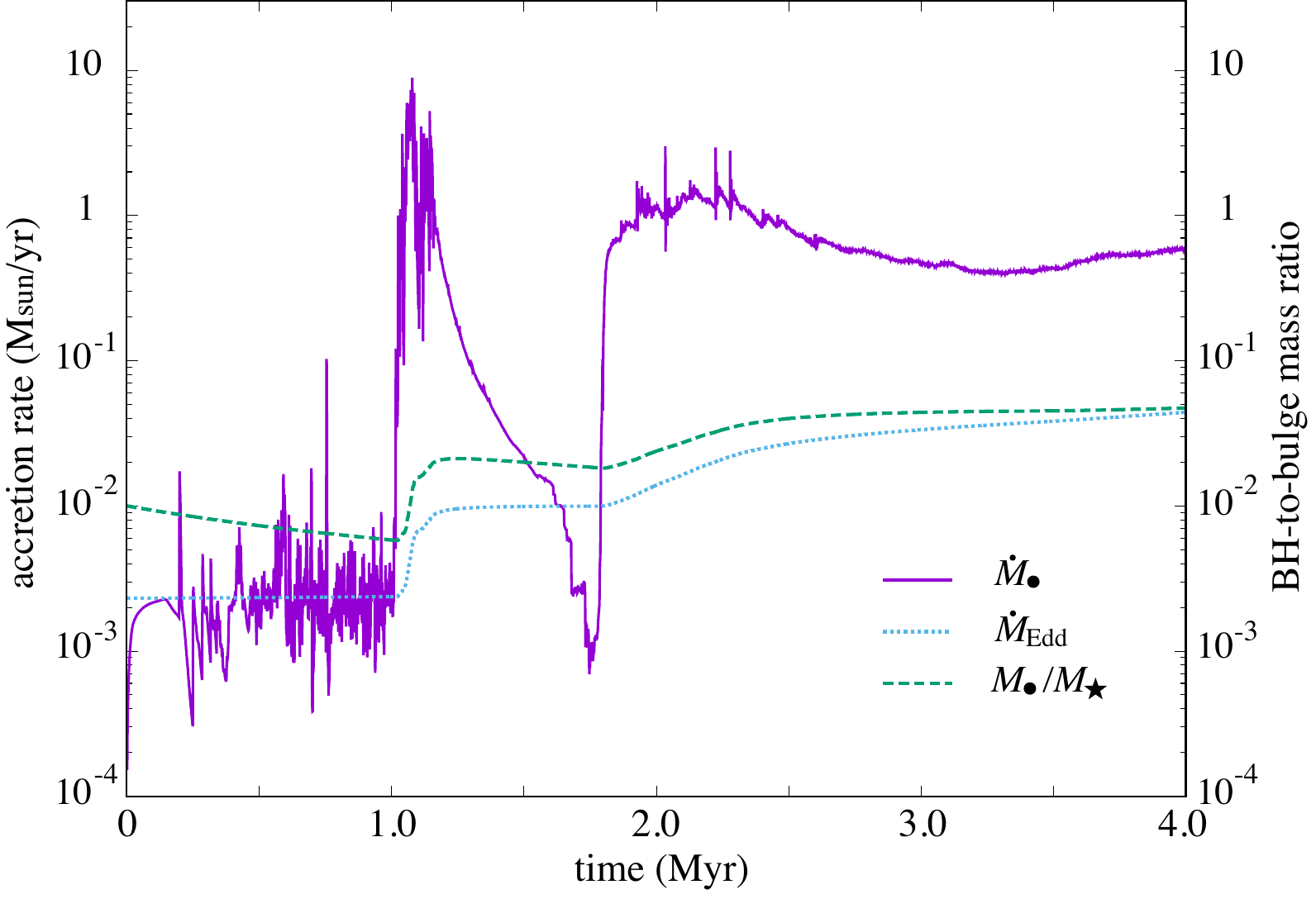}}
\caption{Time-evolution of the mass accretion rate onto a BH (solid),
the Eddington accretion rate (dotted), and the BH-to-bulge mass ratio (dashed) 
for the case with a high SFE ($\epsilon_\star =0.5$ and $f_0=4$).
In this case, both the BH and bulge evolve within the simulation time and the mass
ratio reaches $M_\bullet / M_\star \simeq 0.04$, which is consistent with those of bright quasars 
at $z>6$ (see also \S\ref{sec:Msigma}). 
}
\label{fig:MSFE}
\end{center}
\end{figure}

\begin{figure*}
\begin{center}
{\includegraphics[width=85mm]{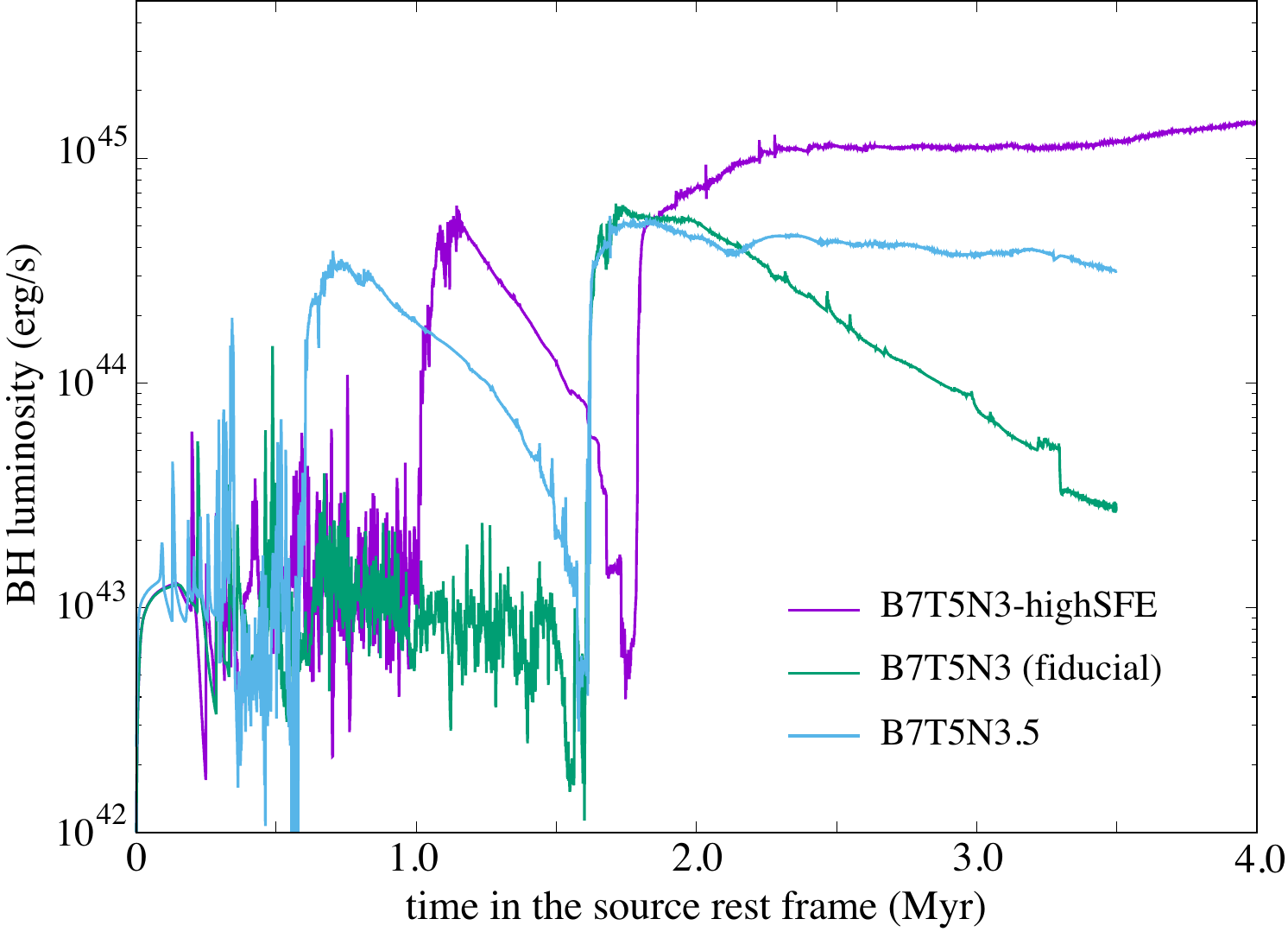}}\hspace{12pt}
{\includegraphics[width=90mm]{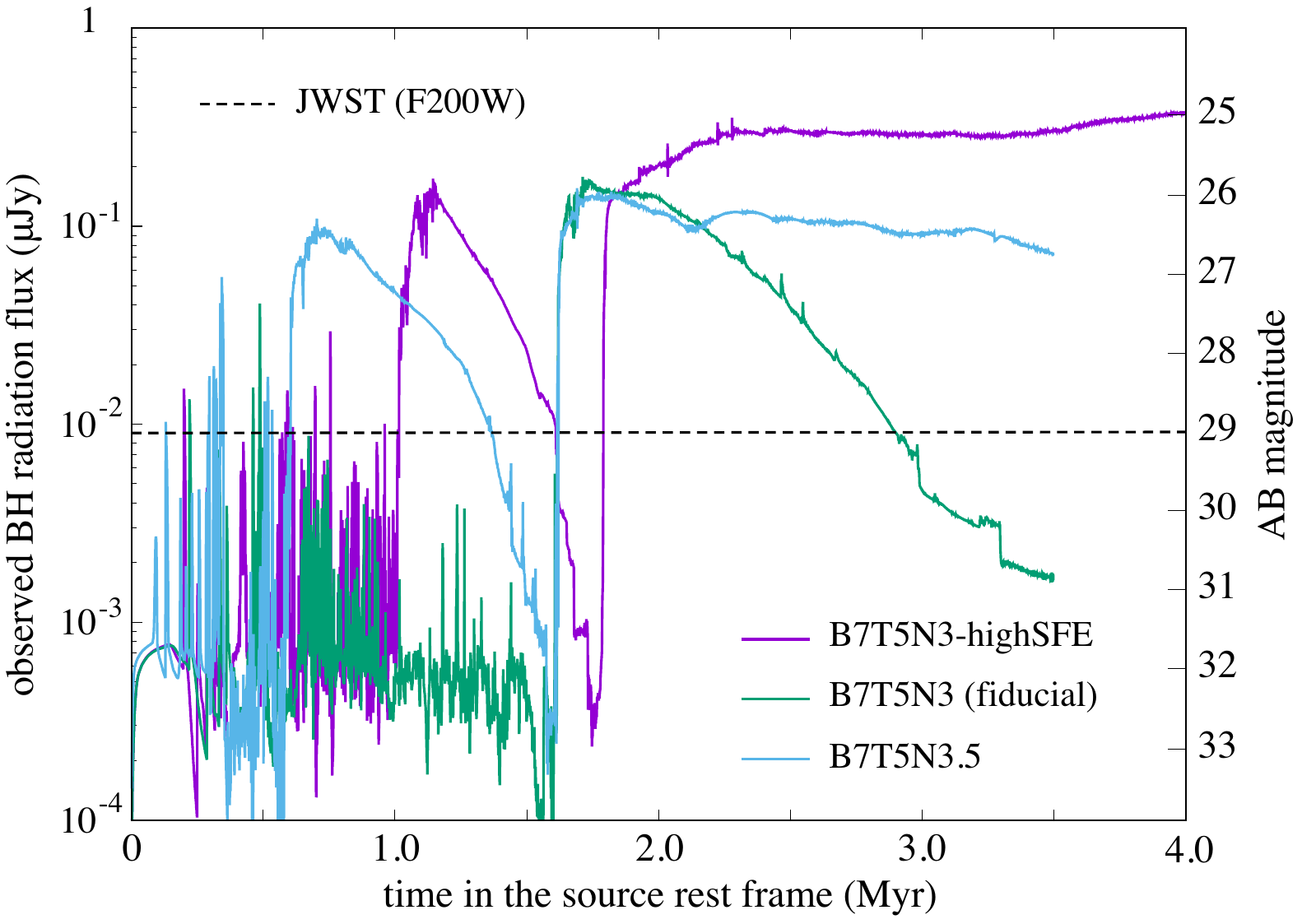}}
\caption{
{\t Left}: 
Radiative luminosity produced from a rapidly accreting BH for the high-SFE case (purple),
the fiducial case (green), and the high-density case with $n_{\rm c}=10^{3.5}~\cc$ (cyan).
During the growing phases from $M_{\bullet}=10^5~\msun$ to $\sim 10^6~\msun$, the luminosities 
becomes as high as $L_{\bullet} \simeq (0.3-1.2)\times 10^{45}~{\rm erg~s}^{-1}$.
{\it Right}:
Observed radiative flux at $\lambda_{\rm obs}=1.98~\mum$ emitted from an accreting BH 
in a protogalaxy at $z=15$ predicted for the three models. 
The viewing angle is set to $\theta_{\rm obs}=30^\circ$.
The horizontal, dashed line shows the $S/N=10$ detection limit of JWST/NIRCam imaging with the F200W filter
in a $10~{\rm ks}$ exposure time.
}
\label{fig:flux}
\end{center}
\end{figure*}

In the left panel of Fig.~\ref{fig:flux}, we present the time evolution of the radiative luminosity produced by
a rapidly accreting seed BH for the high-SFE case (purple),
the fiducial case (green), and the high-density case with $n_{\rm c}=10^{3.5}~\cc$ (cyan).
Here, the radiative luminosity is calculated by using Eq.~(\ref{eq:ledd}) as a function of the BH feeding rate.
We note that in general, this luminosity is not the one we observe but that injected through 
the inner-most cells of the simulation.
The two values are almost identical when and after the accretion transition occurs
because intense radiation ionizes the surrounding gas and further blows the gas away,
reducing the amount of absorbers along the line of sight.
However, in the early stage before the accretion transition, the injected radiation flux is attenuated 
by the surrounding neutral gas and dust to some degree.
When the mass accretion increases during the transition, the radiative luminosity also rises and reaches
$L_\bullet \simeq (0.3-1.2) \times 10^{45}~{\rm erg~s}^{-1}$, corresponding to $L_\bullet /L_{\rm Edd} \sim 4-16$.
We note that the luminosity of the accreting BH for each case substantially dominates that of the host galaxy; 
namely, the LW luminosity we assume as sources of stellar feedback is 
$1.9\times 10^{42}~{\rm erg~s}^{-1}$ (fiducial case) and $7.6\times 10^{43}~{\rm erg~s}^{-1}$ (high-SFE case), respectively.

With our spectral model for the injected radiation (see Eq.~\ref{eq:specin}), $\sim 55\%$ of 
the total radiation energy is emitted as photons with $h\nu< h\nu_0=10~{\rm eV}$ 
(the rest-frame wavelength is $\lambda_0=0.124~\mum$), 
which are not absorbed significantly by the intergalactic media.
The specific radiation luminosity at $\nu=\nu_0$ is calculated by $L_{\nu_0}=L_0=L_{\bullet}/(3.3\nu_0)$.
As an example, we consider that a seed BH is accreting in
a protogalaxy at $z\simeq 15$ (the luminosity distance is $D_{\rm L}\simeq 167~\gpc$).
In this case, the wavelength of the rest-frame $h\nu_0=10$ eV photons is redshifted to 
$\lambda_{\rm obs} \simeq 1.98~\mum ~[(1+z)/16]$, and the observed radiative flux is calculated via
\begin{equation}
F_{\nu_{\rm obs}}=\frac{(1+z)}{4\pi D_{\rm L}^2} \cdot \frac{L_{\rm Edd}+ \mathcal{F}(\theta_{\rm obs})\Delta L_\bullet}{3.3~\nu_0},
\end{equation}
where $\nu_{\rm obs}=\nu_0/(1+z) [=c/\lambda_{\rm obs}]$ is the observed frequency, 
$\Delta L_\bullet = L_\bullet-L_{\rm Edd}$, and the anisotropic degree of the flux is characterized 
by the function of $\mathcal{F}(\theta_{\rm obs})$ (see Eq.~\ref{eq:aniso}).
The right panel of Fig.~\ref{fig:flux} presents the radiative flux at $\lambda_{\rm obs}=1.98~\mum$ emitted from the accreting BH
in a protogalaxy at $z=15$, seen from a viewing angle of $\theta_{\rm obs}=30^\circ$.
We note that the x-axis is the time at the source rest-frame (the redshift effect is not considered).
After the transition of mass accretion, the observed flux becomes as high as $\sim 0.01-0.2~{\mu \rm Jy}$,
corresponding to $m_{\rm AB} \simeq 29-26$ mag.
For comparison, we overlay the $S/N=10$ detection limit of JWST/NIRCam imaging with the F200W filter 
at $1.755 \leq \lambda_{\rm obs}/\mum \leq 2.226$ in a $10~{\rm ks}$ exposure time ($m_{\rm AB}\simeq 29~{\rm mag}$
denoted by the dashed line; \citealt{Rieke_2019})\footnote{https://jwst-docs.stsci.edu/jwst-near-infrared-camera/nircam-instrumentation/nircam-filters}.
For all the cases, the radiation flux can be detectable (i.e., $m_{\rm AB}<29$ mag) 
in durations of $\Delta t \ga 2$ Myr at the source-rest frame.
The corresponding duty cycle is $f_{\rm duty}\sim 7\%$ of the cosmic time duration 
when the redshift changes by $\Delta z =\pm 0.5$.
Note that the host galaxy is as bright as $\sim 0.02~{\mu \rm Jy}$ ($m_{\rm AB}\ga 28~{\rm mag}$)
for the high-SFE case and would not be a serious contamination for hunting seed BHs but rather be 
a detectable extended source with JWST.

We finally estimate the number of rapidly accreting seed BHs detectable within a JWST/NIRCam field of view
(${\rm FoV}=9.7~{\rm arcmin}^2=0.0027~{\rm deg}^2$) as 
\begin{align}
N_{\rm det}^{\rm JWST} &=f_{\rm duty} \mathcal{N}_{\rm h,seed} \Delta V_{\rm c}(z) \frac{\Delta \Omega_{\rm obs}}{4\pi}\nonumber\\
&\simeq 1.1 \left(\frac{ f_{\rm duty}}{0.1}\right) \left(\frac{\Delta \Omega_{\rm obs}}{10~{\rm FoV}}\right)
 \left(\frac{\mathcal{N}_{\rm h,seed}}{10^{-4}~{\rm cMpc}^{-3}}\right),
\end{align}
where $\Delta \Omega_{\rm obs}$ is the observed solid angle and 
$\mathcal{N}_{\rm h,seed}$ is the comoving number density of DM halos that host growing seed BHs.
Here, we consider only the BH luminosity, because the seed BHs outshine their host galaxies.
The number density of DM halos with $M_{\rm h}\geq 2\times 10^9~\msun$ at $z=15$ is 
estimated as $\mathcal{N}_{\rm h}\simeq 2.5\times 10^{-3}~\mpc^{-3}$.
Therefore, if $\ga 4\%$ of those DM halos at $z=15$ contain seed BHs, we are able to detect 
one rapidly accreting seed BH by JWST in its 10 fields of view.
We also note that the detectability depends on the viewing angle $\theta_{\rm obs}$.
Namely, the direct component of the BH radiation flux is reduced by one order of magnitude for observers with 
$|\theta_{\rm obs}-90^\circ |\la 20^\circ$ (i.e., a nearly edge-on view), otherwise it is detectable with JWST.

Deep imaging surveys by the RST will also enable us to hunt for a larger number of rapidly accreting BHs in the early universe.
Assuming the survey area is as large as $\Delta \Omega_{\rm obs} \simeq 40~{\rm deg}^2$ and the limiting magnitude is 
$m_{\rm AB}\simeq 28-29$ with the F184 filter at $1.68 \leq \lambda_{\rm obs}/\mum \leq 2.00$ as planned in \cite{Akeson_2019},
the expected number of detected seed BHs would be
\begin{align}
N_{\rm det}^{\rm RST}\simeq 1.6\times 10^3 \left(\frac{ f_{\rm duty}}{0.1}\right) \left(\frac{\Delta \Omega_{\rm obs}}{40~{\rm deg}^2}\right)
 \left(\frac{\mathcal{N}_{\rm h,seed}}{10^{-4}~{\rm cMpc}^{-3}}\right).
\end{align}
We note that if the limiting magnitude is $m_{\rm AB}<27~{\rm mag}$, the observable duration becomes shorter 
and the number of detections is lowered.

The detailed modeling of their SED and the multi-band selection and spectroscopic diagnosis 
of accreting seed BH candidates will be discussed in future work.
Note that some previous studies modeled the radiation spectrum of an accreting seed BH
using a spherically-symmetric one-dimensional flow structure \citep{Pacucci_2015,Natarajan_2017,Valiante_2018}.


\section{Analytical derivation of the conditions for rapid accretion}
\label{sec:analytic}

We here derive the conditions required for the onset of rapid mass accretion onto 
a seed BH embedded within bulge stars.
Based on the RHD simulation results, the transition is triggered when (1) the I-front
is confined within the core region without propagating outward and (2) a sufficient amount of 
neutral gas is supplied from larger radii without being suppressed by radiative feedback.
In what follows, we quantify the two conditions with analytical expressions.

Let us consider that an accreting BH emits ionizing radiation at a rate of $Q_0$ and
the radiation propagates outward within a gas cloud with a distribution given by Eq.~(\ref{eq:nin}).
As a reference value, we estimate the size of the I-front in a uniform density with $n_{\rm c}$ as 
\begin{align}
r_{\rm ion} &=\left( \frac{3Q_0}{4\pi \alpha_{\rm rec,B} n_{\rm c}^2}\right)^{1/3},\nonumber\\
& \simeq 25.5~\lambda^{1/3} M_5^{1/3} n_{\rm c,3}^{-2/3}
\left(\frac{T_{\rm HII}}{7\times 10^4~\K}\right)^{0.28}~ \pc
\label{eq:rion}
\end{align}
where $\alpha_{\rm rec,B}$ is the case-B radiative recombination rate coefficient,
$T_{\rm HII}$ is the temperature of the ionized gas, 
the number flux of ionizing photons is given by $Q_0 \simeq 0.173~L_\bullet/(h\nu_{\rm Ly})$
for the radiation spectral model (Eq.~\ref{eq:specin}), and $\lambda (\equiv L_\bullet / L_{\rm Edd})$ is 
the Eddington ratio.
However, this estimate in Eq.~(\ref{eq:rion}) is no longer valid when the I-front size reaches the radius where
the density distribution decreases steeply.
With the gas remaining at rest, the critical power-law index of the density profile ($\rho \propto r^{-\beta}$)
for continuous expansion of the I-front is given by $\beta  = 3/2$ \citep{Franco_1990}.
Therefore, for the given initial density profile, one defines the critical radius, outside which 
the power-law index is steeper than the critical value and the I-front expansion is accelerated, 
by $r_{\rm crit} = \sqrt{3}~r_0$ or 
\begin{equation}
r_{\rm crit} =27.4~T_{\rm vir,5}^{1/2}~n_{\rm c,3}^{-1/2}~\pc .
\end{equation}
Thus, the first condition for the transition is given by $r_{\rm crit} \ga r_{\rm ion}$ or
\begin{equation}
M_\bullet \la 2\times 10^5 ~T_{\rm vir,5}^{3/2}~n_{\rm c,3}^{1/2} \left(\frac{\lambda}{0.5}\right)^{-1}~\msun,
\label{eq:cond1}
\end{equation}
where the Eddington ratio is set to $\lambda=0.5$, which is the typical value 
before the transition to rapid mass accretion.

Next, we consider the condition where radiative feedback does not affect mass inflows from larger radii.
Namely, this requires the effective gravitational influence radius to be larger than the size of 
the ionized region, i.e., $r_{\rm inf} \ga r_{\rm ion}$.
The effective gravitational influence radius is calculated with the balance between the total (BH+bulge+DM)
gravitational force  and the gas-pressure gradient force as
\begin{align}
\alpha \frac{c_{\rm s}^2}{r}
\simeq &
\frac{GM_\bullet}{r^2} + \frac{GM_\star}{(r+r_{\rm c})^2} 
-\frac{2k_{\rm B}T_{\rm vir}f(c_{\rm N})}{\mu m_{\rm p}r_{\rm s}} g(x)
\end{align}
where $x=r/r_{\rm s}$, $f(c_{\rm N})\simeq 5$ at $1\la c_{\rm N} \la 5$,
and $g(x)=[x-(1+x)\ln(1+x)]/[x^2(1+x)]$.
In the cases of interest, where $r_{\rm ion}\la r_{\rm crit} \ll r_{\rm s}$, the third term on the right-hand-side 
can be approximated as $\approx 5x k_{\rm B}T_{\rm vir} /(\mu m_{\rm p})$ and 
the first term is negligible compared to the second term.
Therefore, the force balance at $r_{\rm crit} \la r \ll r_{\rm s}$, where $\alpha \simeq 2$, 
is governed by the equation of 
\begin{equation}
2 c_{\rm s}^2 \simeq \frac{GM_\star r}{(r+r_{\rm c})^2}+ \frac{5k_{\rm B}T_{\rm vir}}{\mu m_{\rm p}}\left(\frac{r}{r_{\rm s}}\right).
\end{equation}
In addition, when $M_\star \ll 1.7\times 10^8~T_{\rm vir,5}n_{\rm c,3}^{-1}~\msun $, 
the bulge size is sufficiently small ($r \gg r_{\rm c}$), and the gravitational influence radius is 
expressed by\footnote{Another solution of $r \ga 0.4~(T_{\rm g}/T_{\rm vir})r_{\rm s}$
is not adopted.}.
\begin{equation}
r_{\rm inf} \simeq \frac{GM_\star}{2 c_{\rm s}^2}.
\label{eq:inf}
\end{equation}
Note that Eq.~(\ref{eq:inf}) is identical to the bulge-gravitational influence radius that approximates
the bulge to be a point gravitational source.
Therefore, the condition of $r_{\rm inf} \ga r_{\rm ion}$ is rewritten as 
\begin{equation}
\frac{M_\star}{M_\bullet} \ga 82~M_5^{-2/3}n_{\rm c,3}^{-2/3}
\left(\frac{\lambda}{0.5}\right)^{1/3}
\left(\frac{T_{\rm HII}}{7\times 10^4~\K}\right)^{1.28},
\label{eq:cond2}
\end{equation}
where the sound speed is evaluated with the temperature in the ionized region
as $c_{\rm s}\simeq 29.6~\kms (T_{\rm HII}/7\times 10^4~\K)^{1/2}$.
Indeed, for $M_\bullet =10^5~\msun$, the transient super-Eddington accretion mode
can be triggered when the bulge mass is as massive as $M_\star \ga 10^7~\msun$
(our fiducial case).
Finally, combining the two conditions given by Eqs.~(\ref{eq:cond1}) and (\ref{eq:cond2}),
we obtain the bulge-BH mass relation required to feed the BH efficiently as
\begin{equation}
\frac{M_\star}{M_\bullet} \ga 1.03\times 10^2~\lambda ~n_{\rm c,3}^{-1}~T_{\rm vir,5}^{-1},
\label{eq:cond3}
\end{equation}
where the dependence on $T_{\rm HII}$ is omitted.
For our cases with $M_\bullet =10^5~\msun$, the critical bulge mass for the onset of rapid accretion
is estimated from Eq.~(\ref{eq:cond3}) as $M_\star \simeq 10^7~\msun$.
This is consistent with the result of our parameter studies discussed in \S\ref{sec:Mb}.

In summary, for two given quantities to characterize the properties of gas in a DM halo ($T_{\rm vir}$ and $n_{\rm c}$),
we discuss whether seed BHs can grow via mass accretion at rates exceeding the Eddington value.
This transient growing phase takes place when the ionizing radiation produced by the BH does not break 
the gaseous core (see Eq.~\ref{eq:cond1}) {\it and} a massive bulge with $M_\star \ga 100~M_\bullet$ 
attracts a sufficient amount of gas within its gravitational influence radius (see Eq.~\ref{eq:cond3}).

Finally, we note that the critical bulge-to-BH mass ratio in Eq.~(\ref{eq:cond3})
is $\sim 10$ times higher than that obtained in the previous study by \cite{Park_2016},
where the gravitational effect of bulge stars is considered in spherically symmetric
one-dimensional RHD simulations for BH accretion.
Compared to their work, we further take into account the multi-dimensional effect
(e.g., gas angular momentum), irradiation by bulge stars, and metallicity and also adopt 
a different bulge model\footnote{
\cite{Park_2016} modeled it so that the mean stellar density within $r_{\rm c}$
is consistent with that of the Milky Way.
On the other hand, we model the bulge mass-size relation so that the observed 
$M_\star-\sigma_{\rm e}$ relation is reproduced (see \S\ref{sec:method}), 
yielding a more concentrated bulge.}.
However, the difference on the critical mass ratio is mainly caused by the initial density distribution.
Since \cite{Park_2016} adopted a uniform density distribution as their initial conditions,
the I-front always has a maximum size without continuous expansion.
On the other hand, in our cases where the initial density profile consists of the core and envelope,
it follows $n\propto r^{-2}$. 
Therefore, the condition of Eq.~(\ref{eq:cond1}) has to be required to trigger rapid BH accretion.
If we adopted a uniform density distribution, only the condition of Eq.~(\ref{eq:cond2}) would be 
considered.
For $M_\star = 10^6~\msun$, the critical mass ratio estimated from Eq.~(\ref{eq:cond2}) is 
$M_\star/M_\bullet \simeq 18$, which is consistent with that in \cite{Park_2016}.

\section{Discussion}\label{sec:discussion}

\subsection{Feasible conditions of rapid BH accretion in the galaxy assembly}\label{sec:cosmo}

In this section, we discuss whether seed BHs formed in high-redshift protogalaxies 
experience rapidly growing phases.
Let us first consider a BH with $M_\bullet =10^5~\msun$. 
From the condition of Eq.~(\ref{eq:cond1}), the host DM halo is required to be as massive as 
\begin{equation}
M_{\rm h} \gtrsim 2\times 10^9~M_{\bullet,5}~n_{\rm c,3}^{-1/2}\left(\frac{1+z}{16}\right)^{-3/2}~\msun,
\label{eq:halo}
\end{equation}
or equivalently, $T_{\rm vir}\gtrsim 10^5~M_{\bullet,5}^{2/3}n_{\rm c,3}^{-1/3}~\K$.
Now, we define the ratio of the galaxy stellar mass to the DM halo mass as 
$f_\star (\equiv M_\star/M_{\rm h})$ and assume that the bulge mass in the protogalaxy 
is comparable to the total stellar mass.
Then, using the condition of Eq.~(\ref{eq:cond3}), we obtain
\begin{equation}
f_\star \ga 5.2\times 10^{-3}~M_{\bullet,5}^{-2/3}~n_{\rm c,3}^{-1/6}\left(\frac{1+z}{16}\right)^{3/2}.
\label{eq:fstar}
\end{equation}
The required value of $f_\star$ corresponds to $\gtrsim 3\%$ of the conversion efficiency 
from gas into stars (i.e., the star formation efficiency $\epsilon_\star$), 
assuming the cosmic mean baryon fraction $f_{\rm b}=\Omega_{\rm b}/\Omega_{\rm m}\simeq 0.16$.
This value is consistent with those inferred by abundance matching and the observed UV luminosity 
function of galaxies at $z\simeq 6$ \citep{Bouwens_2015}.
Moreover, this choice of $\epsilon_\star \sim 0.03-0.05$ explains the cosmic reionization history
without violating the optical depth of the universe to electron scattering measured by the Planck
satellite \citep{Visbal_2015,Inayoshi_2021}\footnote{
An empirical model for linking  galaxy star formation rates to the properties 
of their host haloes ({\sc UniverseMachine}; \citealt{Behroozi_2020}) predicts the bulge-to-halo mass ratio as 
$f_\star \sim 3\times 10^{-4}(M_{\rm h}/10^{9}~\msun)^{0.75}$
at $10<z<15$ over $10^9 \la M_{\rm h}/~\msun \la 10^{10}$.
The error size with the $84\%$ confidence is $\sim \pm (0.3-0.5)$ dex at $z\simeq 10$
and $\sim \pm (0.7-1.0)$ dex at $z\simeq 14$, respectively.
Indeed, the upper envelope of the predicted values at the halo mass range is consistent with
$f_\star \sim 5\times 10^{-3}$.}.
Adopting a single value of $\epsilon_\star =0.05$, therefore the condition of Eq.~(\ref{eq:fstar})
is rewritten as 
\begin{equation}
M_\bullet \ga 5.2\times 10^4~n_{\rm c,3}^{-1/4}
\left(\frac{\epsilon_\star}{0.05}\right)^{-3/2}
\left(\frac{1+z}{16}\right)^{9/4} ~\msun,
\label{eq:fstar1}
\end{equation}
which holds for $M_\bullet =10^5~\msun$ seed BHs at $z<20$.

\begin{figure}
\begin{center}
{\includegraphics[width=85mm]{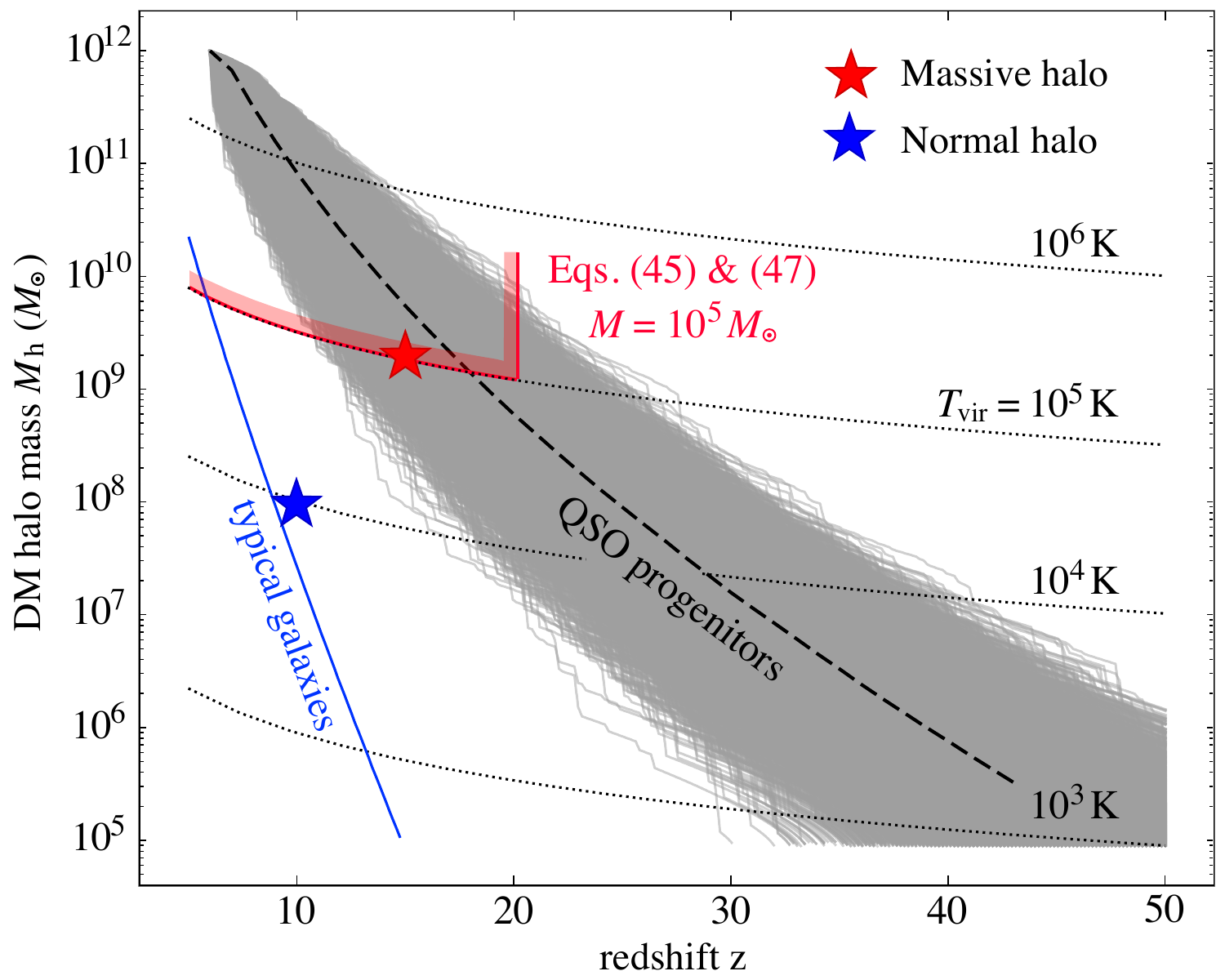}}
\caption{Summary of the analytical argument in \S\ref{sec:cosmo} on the redshift-halo mass plane.
In the region enclosed with the two conditions of Eqs.~(\ref{eq:halo}) and (\ref{eq:fstar1}),
rapid mass accretion onto seed BHs with $M_\bullet =10^5~\msun$ can take place (red curve).
The star symbols represent the mass and redshift of the DM halos studied in this paper; Massive halo (red)
and Normal halo (blue), respectively (see Table \ref{tab:model}).
The gray curves show $10^4$ merger trees of the high-redshift quasar main 
progenitors, corresponding to the $3-4\sigma$ mass variance, and the black dashed curve presents 
the median halo mass. 
For comparison, the blue curve is the assembly history of the typical first galaxies, 
corresponding to the $2\sigma$ mass variance.
The dotted curves indicate constant virial temperatures, the values of which are denoted by numbers in the figure.
}
\label{fig:diagram}
\end{center}
\end{figure}

\begin{figure*}
\begin{center}
{\includegraphics[width=130mm]{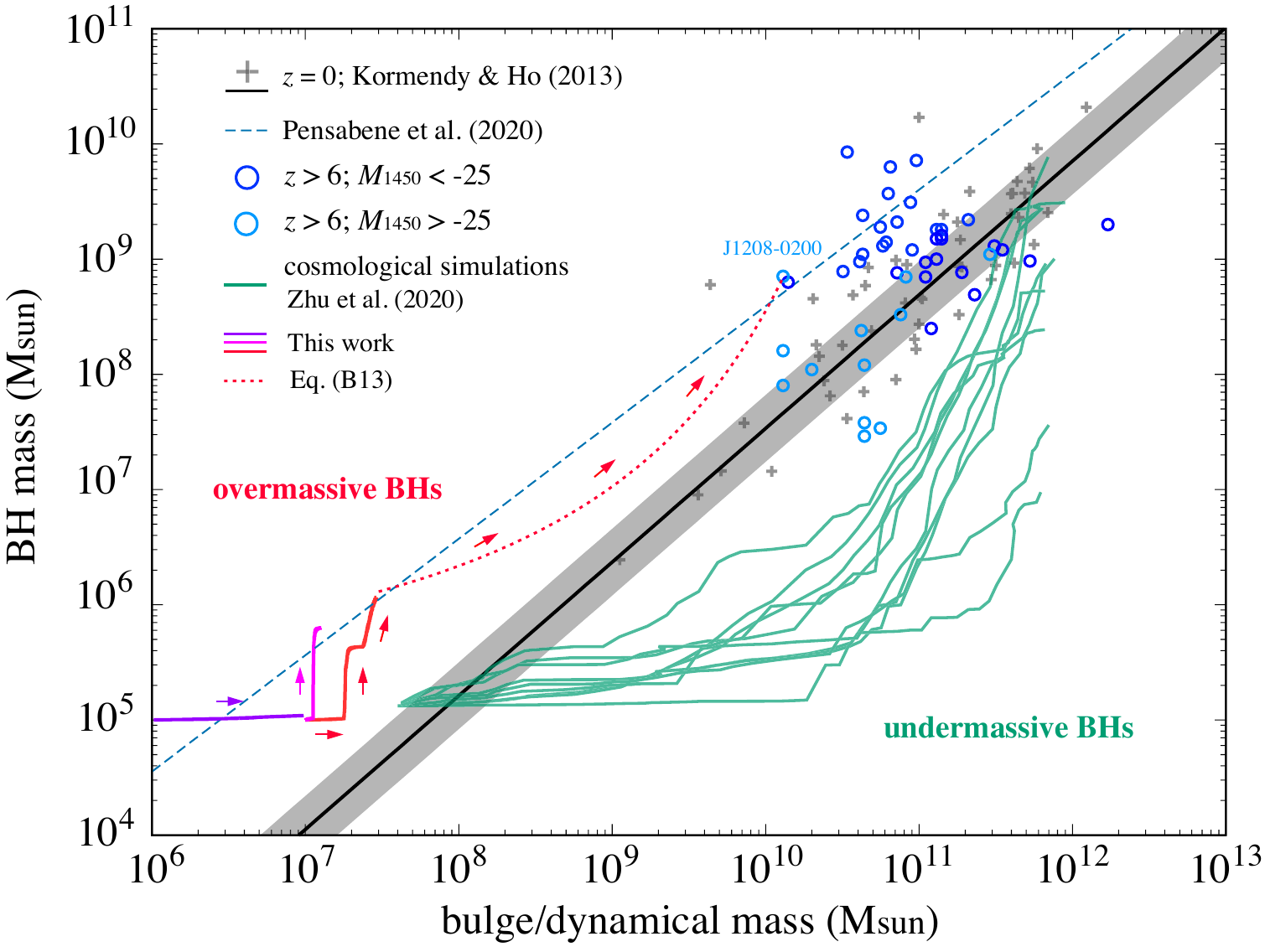}}
\caption{The $M_\bullet-M_\star$ relation for the three simulations: B7T5N3-highSFE (red), 
B7T5N3 (magenta), and B6T5N3 (purple).
Circle symbols show the $z>6$ quasar samples compiled by \cite{Izumi_2019,Izumi_2021}; brighter ones 
with $M_{1450}<-25$ (blue) and fainter ones with $M_{1450}>-25$ (cyan).
The cross symbols are the observational samples in the local universe provided by \cite{Kormendy_Ho_2013}.
Each line represents the local relation of $M_\bullet/M_\star$ (black solid, with $\sim 1\sigma$ errors), 
and the best-fit relations for the brightest $z>6$ quasars by \cite{Pensabene_2020}.
The red dotted curve presents the evolutionary track of the $M_\bullet / M_\star$ ratio after rapid accretion phases
predicted by a phenomenological model (see Appendix \ref{sec:appb}).
As a reference, the evolutionary tracks of the $M_\bullet/M_\star$ ratio obtained by 
cosmological simulations \citep{Zhu_2020} are overlaid (green curves).
}
\label{fig:Msigma}
\end{center}
\end{figure*}

To see the feasibility of rapid BH accretion in the hierarchical structure formation paradigm, 
we consider two different assembly histories of DM halos.
One is the evolution of the high-redshift quasar main progenitors, i.e., the most massive halos 
at each epoch among halos that end up in $M_{\rm h} = 10^{12}~\msun$ at $z = 6$.
In Fig.~\ref{fig:diagram}, we present $10^4$ merger trees of the high-redshift quasar main 
progenitors (gray curves), corresponding to the $3-4\sigma$ mass variance, and 
the median halo mass is shown by the dashed curve.
Along with the merger history of the high-redshift quasar hosts, there is a parameter region where
the two conditions given by Eqs.~(\ref{eq:halo}) and (\ref{eq:fstar1}) are satisfied
and the seed BH can experience transient super-Eddington accretion phases (red curve).
The other one is the assembly history of the typical first galaxies that form in DM halos with 
$M_{\rm h}\ga 10^7~\msun$ at $z\simeq 10$, corresponding to the $2\sigma$ mass variance
(blue curve).
Along with this evolutionary track, seed BHs with $M_\bullet =10^5~\msun$ could grow via rapid accretion 
at $z<6$ (i.e., $T_{\rm vir}\ga 10^5~\K$).
It is also worth noting that strong suppression of BH growth from $M_\bullet \sim 10^5~\msun$
in typical galaxies with $T_{\rm vir}\simeq 10^4~\K$ is commonly seen in cosmological 
hydrodynamical simulations.
For instance, \cite{Latif_2018} showed that the early growth of a $10^5~\msun$ BH is totally quenched 
in an atomic-cooling halo with the $2\sigma$ mass variance.

Next, we briefly discuss the cases with different BH masses.
When the seed mass is $M_\bullet =10^4~\msun$, the halo condition yields $T_{\rm vir} \ga 2\times 10^4~\K$, 
but the high value of $f_\star $ can be achieved at lower redshift of $1+z \la 7.8~(\epsilon_\star/0.05)^{2/3}$.
On the other hand, when the seed mass is $M_\bullet =10^6~\msun$,
the halo condition is satisfied in massive DM halos with $M_{\rm h}\ga 10^{11}~\msun$ 
(or $T_{\rm vir} \ga 4.6\times 10^5~\K$),
where all the high-redshift quasar progenitor halos are expected to have sufficiently massive bulges.
However, the host galaxies formed in those massive halos would be already polluted by heavy elements.
While our simulation results hold for lower metallicity environments with $Z\simeq 0.01~\zsun$, 
the radiative feedback effect caused by the accreting BH would be stronger in more metal-enriched environments.
This would quench their efficient growth of BHs and change the accretion mode to the Eddington-limited one.
Exploration of the critical metallicity to terminate super-Eddington accretion and 
the impact of cosmological metal-enrichment processes is left for future work.

\subsection{Early coevolution of seed BHs with host galaxies}\label{sec:Msigma}

The empirical relation between the mass of SMBHs and the properties of their host galaxies
are considered to be one of the most important outcomes caused by their coevolution over
the cosmic timescale \citep[e.g.,][]{Magorrian_1998,Ferrarese_2000,Kormendy_Ho_2013}.
Theoretical models for explaining the tight correlations have been proposed but the origin is still unclear.
To understand the nature of these correlations, it is critically important to study them beyond the local universe,
characterizing how and when the relations have been established and evolved until now.
So far, a large number of observational studies have extensively investigated the redshift dependence of 
the BH-to-bulge mass ratio of $M_\bullet/M_\star$ and overall suggested its positive redshift dependence,
i.e., the ratio increases with redshift \citep{Bennert_2011,Schramm_Silverman_2013,Ding_2020}.
Beyond $z\sim 6$, ALMA is a powerful tool to measure the dynamical mass of gas in quasar host galaxies
and allows us to explore the early stage of the BH/galaxy correlation 
\citep[e.g.,][]{Wang_2010,Wang_2013,Venemans_2017}.
In addition, observations with the Subaru HSC provide low-luminosity and less massive BH samples,
which are unique populations to determine the $M_\bullet/M_\star$ ratio at $z>6$ \citep{Izumi_2019,Izumi_2021}.
Fig.~\ref{fig:Msigma} shows the distribution of $z>6$ quasars compiled in \cite{Izumi_2021}, 
together with those in the local universe \citep{Kormendy_Ho_2013}.
First, the brightest $z>6$ quasars with $M_{1450}<-25$ mag tend to have $M_\bullet/M_\star$
ratios higher than those seen in the local universe.
Namely, the mass ratio for those brightest objects is boosted by a factor of $\sim 10$ (blue dashed line; \citealt{Pensabene_2020}).
On the other hand, the fainter quasars with $M_{1450}>-25$ mag appear to follow the local relation,
although those BHs are considered to grow at rates of $\ga 0.05~{\rm SFR}$ and will be overmassive at lower redshifts\footnote{
In this paper, overmassive BHs are referred to as a BH population with a BH-to-galaxy mass ratio higher 
than that observed in the local universe; $M_\bullet/M_\star \ga 4.9^{+0.6}_{-0.5}\times 10^{-3}$ \citep{Kormendy_Ho_2013}.
We employ this terminology to clearly contrast the difference between the overmassive and undermassive BH population
with respect to the local value
(see Fig.~\ref{fig:Msigma}).
Note that a previous study by \cite{Agarwal_2013} used a term of ``obese BH", which refers to a BH population dominating
over the stellar mass of its host galaxy at least in the initial growing stage (i.e., $M_\bullet/M_\star >1$).
}.
We note that for all the $z>6$ samples, the values of the x-axis are not the bulge mass of their host galaxies 
but the dynamical mass measured by [CII] 158 $\mum$ lines.
In general, the dynamical mass is considered to be higher than the true bulge mass.
With a high-resolution ALMA observation, \cite{Izumi_2021} found that the gas dynamics of the core component 
of a low-luminosity quasar at $z=7.07$ (HSC J1243+0100) is 
governed by rotation associated with a compact bulge and estimated its mass as $\sim 50\%$ of the [CII]-based dynamical mass.
Therefore, the correlation at $z>6$ might be shifted to the left if the conversion factor from the dynamical mass to the bulge mass
is taken into account.

At the left-bottom corner of Fig.~\ref{fig:Msigma}, we present the evolution tracks of the $M_\bullet/M_\star$
ratio obtained from our simulations.
When the bulge mass is $M_\star <10^7~\msun$ (purple curve), the seed BH hardly grows in mass owing to radiative feedback
and the $M_\bullet/M_\star$ moves to the right, i.e., only the bulge mass grows.
When the bulge is as massive as $M_\star \geq 10^7~\msun$ (magenta and red curves), the efficient mass accretion phases are triggered 
and thus the BH mass abruptly increases within $\sim$ a few Myrs.
As a result, the $M_\bullet/M_\star$ ratio becomes as high as $\sim 0.03-0.05$, which is consistent with those 
of $z>6$ (see also the line obtained by \citealt{Pensabene_2020}).
In the high-SEF model ($\epsilon_\star =0.5$), where the bulge mass reaches $M_\star \simeq 3\times 10^7~\msun$ 
within $\sim 3$ Myr, the BH-to-bulge mass ratio is $M_\bullet/M_\star \simeq 0.03$
at the end of the simulation\footnote{
A semi-analytical study by \cite{Agarwal_2013} proposed a pathway yielding a more extreme BH-to-galaxy mass ratio of
$M_\bullet/M_\star > 1$ in atomic-cooling halos with $T_{\rm vir}\simeq 10^4~\K$. 
Based on their model, the number density of such obese BHs with $M_\bullet > 10^7~\msun$ is estimated as 
$\sim 0.03~{\rm cMpc}^{-3}$ at $z\sim 6$.
However, the cumulative mass density of those BHs overshoots the local mass density of SMBHs \citep[e.g.,][]{Shankar_2009}
and a constraint from the unresolved cosmic X-ray background \citep[e.g.,][]{Salvaterra_2012}.}.
After the rapid BH growing phase terminates, both the BH and host galaxy evolve in mass and reach the observed values as
high-redshift quasars. 
As an example, we present the subsequent BH/galaxy evolutionary track to a $z>6$ HSC quasar \citep{Izumi_2019}
predicted by a phenomenological model (see more details in Appendix \ref{sec:appb}).
In summary, seed BHs formed in high-redshift quasar hosts can be substantially overmassive during the early 
bulge assembly at $z\ga 15$ and the mass ratio would approach the ratio of $M_\bullet/M_\star$ seen at $z\sim 6$ 
via the subsequent growth.

It is worth noting that most cosmological simulations studying the BH-galaxy coevolution 
concluded that seed BHs hardly grow in mass via gas accretion because dense, cold gas 
is expelled by energetic SN feedback associated with star formation 
\citep[e.g.,][]{Sijacki_2009,Habouzit_2017,Lupi_2019}.
As a result of SN feedback, BH growth is strongly quenched until the host galaxies become
heavier than $M_\star \ga 10^9~\msun$, whose critical stellar mass depends on the sub-grid models
for star formation, SN feedback, and AGN feedback.
There is a model parameter set for which the most massive BHs in their simulations can reach 
$M_\bullet \simeq 10^9~\msun$ by $z\sim 6$.
However, the predicted shape of the $M_\bullet-M_\star$ relation shows that 
most of the BHs are {\it undermassive}, compared to the local relation 
\citep[e.g.,][]{Zhu_2020,Valentini_2021}.
As a reference, the evolutionary tracks of the $M_\bullet/M_\star$ ratio obtained by \cite{Zhu_2020}\footnote{
\cite{Zhu_2020} have extensively investigated the effect of feedback, BH seeding, and accretion models
on the BH growth. Among their simulation results, we show the cases where the initial BH mass is 
$M_\bullet =10^5~\msun$ for comparison.} are overlaid in Fig.~\ref{fig:Msigma} (green curves).

As discussed in \cite{Inayoshi_ARAA_2020}, there are important limitations of current 
numerical simulations.
First, most large-scale cosmological simulations resolve the dynamics of DM/gas/stars on 
galactic scales at $\sim O({\rm kpc})$, but do not resolve the BH gravitational influence radius. 
Secondly, owing to simplified star formation models where gas particles denser than a threshold 
are replaced with stars, dense clouds in the nuclear region would be disrupted and thus the BH growth
could be quenched as seen in many cosmological simulation studies.
In contrast, as shown in our simulations that resolve sub-pc scales, a fraction of seed BHs that
were born in highly biased regions of the universe with mass variance of $\ga 3-4~\sigma$ could 
be fed through dense, cold accretion flows \citep[see also][]{Li_2021}.
Additionally, the existence of such {\it overmassive} BHs in protogalaxies will provide us with a unique 
opportunity to detect highly accreting seed BHs in the very early universe at $z>10$, 
unlike the undermassive-BH scenarios supported by cosmological simulations.

\subsection{Young quasars with low radiative efficiencies}

Our RHD simulations suggest the existence of high-redshift quasars that accrete at super-Eddington rates 
and fade out in a timescale of $\sim {\rm a~few}$ Myr.
The duration of such rapid accretion is generally consistent with lifetimes of $z \ga 6$ quasars ($t_{\rm Q}\la 1-10~{\rm Myr}$),
which are estimated by the measurement of the physical extents of hydrogen Ly$\alpha$ proximity zones observed in the 
rest-frame UV spectra \citep[e.g.,][]{Eilers_2018,Davies_2019,Eilers_2021}.
Since the inferred quasar lifetimes are substantially shorter than the $e$-folding timescale assuming Eddington-limited accretion, 
$t_{\rm Edd}~(\equiv M_\bullet /\dot{M}_{\rm Edd}) \simeq 45~{\rm Myr}$, some of those $z\ga 6$ quasars are expected to 
undergo radiatively inefficient super-Eddington accretion to grow up to $M_\bullet \sim 10^9~\msun$ in such a short duration.
For instance, \cite{Davies_2019} proposed that two $z>7$ quasars (ULAS J1120+0641 and ULAS J1342+0928) would have 
a small radiative efficiency significantly below $\eta_0 \simeq 0.1$.
Based on the above, mupltiple episodes of rapidly accretion would be an intriguing counterpart of such young high-redshift quasars, 
although the observed quasars are hosted in substantially heavier halos and exhibit star formation rates higher than we consider. 
Moreover, the episodic nature of quasar activity produces a complex structure of ionized and neutral gas surrounding the growing BH,
unlike the situation where continuous output of radiation from the BH is assumed (the so-called ``light-bulb” light curve model). 
The variable light-curve effect on the formation of proximity zones needs to be explored in more detail \citep[see e.g.,][]{Davies_2020}. 
To perform RT calculations of Ly$\alpha$ lines under a realistic density profile obtained from
simulations for BH accretion is left for future investigations.

\section{Summary}\label{sec:summary}

We study the early growth of massive seed BHs via accretion at the centers of protogalaxies 
where the stellar bulge component is assembled, performing axisymmetric two-dimensional RHD simulations.
We find that when a seed BH with $M_\bullet \sim 10^5~\msun$ is embedded 
in dense metal-poor gas ($Z=0.01~\zsun$) with a density of $n_{\rm c} \ga 100~\cc$ and bulge stars with a total mass of 
$M_\star \ga 100~M_\bullet$, a massive gaseous disk feeds the BH efficiently at super-Eddington rates 
of $\ga 0.3-1~\msunyr$.
The peak rate of $\dot{M}_\bullet \sim 10~\msunyr$ corresponds to $\sim 2.5\times 10^3~\dot{M}_{\rm Edd}$.
This rapid accretion phase lasts until a good fraction of the gas bounded within the bulge accretes onto the BH,
although the feeding rate is regulated owing to strong outflows driven by ionizing radiation
emitted from the accreting BH.
As a result, the BH mass increases to $\sim 10^6~\msun $ within $\sim 2$ Myr
after the onset of accretion bursts.
In contrast, when the seed BH is surrounded by diffuse gaseous media and/or less massive bulges,
the BH accretion proceeds episodically due to radiative feedback and thus the average accretion 
rate is limited below the Eddington rate.

We provide the analytical expressions of the conditions required for the onset of super-Eddington 
accreting phases of seed BHs embedded within growing bulge stars.
This transient growing phase takes place when (1) the ionizing radiation produced by the BH does not break 
the gaseous core (see Eq.~\ref{eq:cond1}) and (2) a massive bulge with $M_\star \ga 100~M_\bullet$ 
attracts a sufficient amount of gas within its gravitational influence radius (see Eq.~\ref{eq:cond3}).
In the framework of the cosmological hierarchical structure formation, those conditions can be achieved
for seed BHs formed in massive DM halos with masses of $\ga 10^9~\msun$ at $z\sim 15-20$
(the virial temperature is $T_{\rm vir}\simeq 10^5~\K$).
The host halos are heavier and rarer than those of typical first galaxies with $T_{\rm vir}\simeq 10^4~\K$, 
but are more likely to end up in quasar hosts by $z\simeq 6$.

The rapid accretion mechanism found with our simulations that resolve sub-pc scales naturally 
yields a high BH-to-bulge mass ratio of $M_\bullet/M_\star \simeq 0.03-0.05$.
The ratio is significantly higher than that seen in the local universe, but is consistent with
those of the brightest high-redshift quasars with $M_{1450}<-25$ mag.
The rarity of those overmassive BHs in the brightest quasars can also be explained by the fact that 
the transient rapid growth would take place in highly biased regions of the universe with mass variance 
of $3-4\sigma$.
In contrast, most large-scale cosmological simulations that resolve the dynamics of DM/gas/stars on 
galactic scales at $\sim O({\rm kpc})$ predict that SMBHs (or their seeds) tend to be undermassive
at high redshifts compared to the local BH-galaxy correlation.
Both observational and numerical studies are further required to better understand the early development
of the BH-galaxy correlation.

Moreover, the existence of such overmassive BHs provides us with a unique opportunity 
to detect highly accreting seed BHs by the upcoming observations by JWST.
In fact, seed BHs accreting at super-Eddington rates produce radiative luminosities of 
$L_\bullet \simeq (0.3-1.2)\times 10^{45}~{\rm erg~s}^{-1}$, corresponding to 
$L_\bullet /L_{\rm Edd}\sim 4-16$.
If those BHs are in $z\simeq 15$ protogalaxies, the radiation flux at $\lambda_{\rm obs} \simeq 2~\mum$
($h\nu=10$ eV in the rest frame) is estimated as $\sim 0.01-0.2~{\rm \mu Jy}$,
which corresponds to $m_{\rm AB}=26-29$ mag and can be detectable with the $S/N=10$ detection
limit of JWST NIRCam imaging with the F200W filter in a 10 ks exposure time.
If $\ga 4\%$ of the DM halos where the transition conditions are satisfied contain seed BHs, 
we are able to detect one rapidly accreting seed BH by JWST in its 10 fields of view.\\

\acknowledgments

We greatly thank Luis C. Ho, Takuma Izumi, and Mingyang Zhuang for constructive discussions. 
K.~I. acknowledges support from the National Natural Science Foundation
of China (12073003, 12003003, 11721303, 11991052, 11950410493), the National Key R\&D Program of China
(2016YFA0400702), and the China Manned Space Project with NO. CMS-CSST-2021-A06.
R.~N. is supported by the Special Postdoctoral Researcher (SPDR) Program at RIKEN and by a Grant-in-Aid for Research Activity Start-up (19K23469).
T.~H. acknowledges support from the JSPS KAKENHI grant Nos. 17H06360, 17H01102, and 19H01934. 
R.~K. acknowledges financial support via the Emmy Noether and Heisenberg Research Grants funded by the German Research Foundation (DFG) under grant no.~KU 2849/3 and 2849/9.
R.~K. also acknowledges financial support via the JSPS Invitational Fellowship for Research in Japan under the Fellowship ID S20156.
The numerical simulations were performed with the Cray XC50 at the Center for Computational Astrophysics (CfCA) 
of the National Astronomical Observatory of Japan and with the High-performance Computing Platform of 
Peking University.

\appendix
\section{bulge size-mass relation and formation}\label{sec:appA}
We consider the gravitational effect of bulge stars on the thermal dynamics of gas.
In our simulations, we adopt a Hernquist density profile \citep{Hernquist_1990}
\begin{equation}
\rho(r) = \frac{M_\star}{2\pi}\frac{r_{\rm c}}{r(r+r_{\rm c})^3},
\label{eq:rho}
\end{equation}
where $M_\star$ is the total bulge mass and $r_{\rm c}$ is the core radius.
Integrating the density profile, the enclosed mass is given by
\begin{equation}
M(r) = M_\star\frac{r^2}{(r+r_{\rm c})^2},
\end{equation}
and thus the half-mass radius is calculated by
\begin{equation}
R_{\rm e} = (1+\sqrt{2})r_{\rm c}.
\end{equation}
Assuming isotropic motion of stars, the velocity dispersion is analytically calculated as
\begin{equation}
\sigma^2(r) = \frac{GM_\star}{r_{\rm c}}
\left[
xy^3\ln \left(\frac{y}{x}\right) - \frac{x}{y}\left(\frac{1}{4}+\frac{y}{3}+\frac{y^2}{2}+y^3\right)
\right],
\end{equation}
where $x=r/r_{\rm c}$ and $y=1+x$.
The velocity dispersion at the half-mass radius is 
\begin{equation}
\sigma_e^2 \simeq 0.055~\frac{GM_\star}{r_{\rm c}}.
\end{equation}
In our paper, we quantify the core radius of $r_{\rm c}$ so that the correlation between $M_\star$ and $\sigma_{\rm }$ 
becomes consistent with that observed in the local universe \citep{Kormendy_Ho_2013}, as shown in Eq.~(\ref{eq:rbulge}).

As described in \S\ref{sec:BC}, mass growth of the bulge is calculated with the SFR given by the DM halo properties
(see Eq. \ref{eq:SFRmodel}) and the spherical mass distribution is imposed without solving stellar dynamics.
Here, we briefly discuss the dynamical evolution of newly-forming stars and show that a quasi-spherical stellar distribution
can be achieved in a short timescale of $\sim0.1$~Myr.
Let us consider an extreme case where star formation is suppressed by anisotropic irradiation by an accreting BH
but stars form in the equatorial region owing to the shadow effect.
In this case, most of the stars are initially aligned with the disk plane.
The relaxation timescale of the inclination is 
estimated as \citep{Stewart_Ida_2000, Kocsis_Loeb_2011}
\begin{align}
t_{\rm relax,disk} &\simeq \frac{0.44 \pi  \langle e^2\rangle ^2 M_\bullet^2}{\Omega \langle m_\star \rangle M_\star \ln \Lambda} 
\simeq \frac{7.8\times 10^4~{\rm yr}}{(\ln \Lambda/7)}
\left(\frac{r}{10~{\rm pc}}\right)^{3/2}
\left(\frac{M_\bullet}{10^5~\msun}\right)^2
\left(\frac{M_\star}{10^7~\msun}\right)^{-3/2}
\left(\frac{\langle m_\star \rangle}{3~\msun}\right)^{-1},
\end{align}
where $\langle e^{2}\rangle^{1/2} =0.3$ is the mean eccentricity, $\langle m_\star \rangle$ is 
the mean stellar mass for the Salpeter IMF at $1-100~\msun$, $\Omega^{-1} = \sqrt{GM(<r)/r^3}$,
and $\Lambda = \langle e^{2}\rangle^{3/2}M_\bullet / \langle m_\star \rangle$.
Therefore, the disk stars will evolve into a quasi-spherical stellar cluster within $\la 0.1$ Myr, 
which is substantially shorter than the timescale of interest in our simulations.
Note that this argument leads to a consistent result for stellar cluster formation around a massive heavy seed BH
in an atomic-cooling halo \citep[e.g.,][]{Kashiyama_Inayoshi_2016}.

\vspace{5mm}

\section{Phenomenological model for the $M_\bullet/M_\star $ ratio}\label{sec:appb}

We provide a phenomenological model for calculating the redshift-dependent $M_\bullet/M_\star $ ratio.
First, we characterize the mass growth of a BH using an exponential function of time $t$,
\begin{equation}
\frac{M_\bullet }{M_{\bullet i}} =\exp \left[ \langle A \rangle ~ \frac{t-t_i}{t_{\rm Edd}} \right],
\end{equation}
where $t_{\rm Edd}=45$ Myr is the Salpeter timescale, $M_{\bullet i}$ is the BH mass at $t=t_i$,
$A$ is the product of the Eddington ratio and duty cycle of the BH active phase, and $\langle \cdot \rangle $
means the time-averaged value over the cosmic time duration of interest.
Next, we express the mass growth of the host galaxy with a functional form of $M_\star \propto e^{-Bz}$,
which is well-known to nicely reproduce the mass growth of DM halos $M_{\rm h}$ 
\citep[e.g.,][]{Wechsler_2002,Neistein_Dekel_2008,Fakhouri_2010}.
In fact, this redshift dependence leads to the halo-mass growth rate of $d(\ln M_{\rm h})/dt \propto (1+z)^{5/2}$,
which has been understood based on the extended Press-Schechter formalism and also derived by a fit to merger trees 
from cosmological N-body simulations \citep{Dekel_2013}.
Despite the complex nature of galaxy formation processes, cosmological hydrodynamical simulations have shown that 
the galaxy mass assembly history appears to follow a similar function form \citep[e.g., see Fig.~7 in][]{Zhu_2020}.
Note that the value of $B$ for galaxy assembly is different form that for halo assembly.
Motivated by this fact, we express the mass growth of the host galaxy as
\begin{equation}
\frac{M_\star}{M_{\star i}} =\exp \left[\langle B\rangle  \left( 1-\frac{1+z}{1+z_i}\right)\right] ,
\end{equation}
where $M_{\star i}$ is the galaxy mass at $z=z_i$ and $\langle B\rangle $ is a parameter to characterize the growth speed of the galaxy.
Combining the two equations, we calculate the mass ratio of $M_\bullet/M_\star $ as
\begin{align}
\frac{M_\bullet}{M_\star} &= \frac{M_{\bullet i}}{M_{\star i}} ~
\exp \left[\langle A \rangle~ \frac{t-t_i}{t_{\rm Edd}} - \langle B \rangle \left( 1-\frac{1+z}{1+z_i}\right)\right],
\nonumber\\[10pt]
&\simeq \frac{M_{\bullet i}}{M_{\star i}} ~
\exp \left[\langle A\rangle (D^{3/2}-1)\frac{t_i}{t_{\rm Edd}} - \langle B \rangle  (1-D^{-1})\right],
\end{align}
where $D(z)\equiv (1+z_i)/(1+z)$ and $t_{\rm H}(z_i)= (2/3H_0)[\Omega_{\rm m}(1+z_i)^3]^{-1/2}$ is the Hubble time at $z=z_i$,
which is a good approximation in the matter dominant universe ($z>1$).
Finally, we determine $\langle A \rangle $ and $\langle B\rangle $ by
\begin{equation}
\langle A \rangle  = \frac{t_{\rm Edd}}{t_i}\frac{\ln (M_{\bullet f}/M_{\bullet i})}{D_f^{3/2}-1} 
~~~~~{\rm and}~~~~~
\langle B\rangle  = \frac{\ln (M_{\star f}/M_{\star i})}{1-D_f^{-1}},
\end{equation}
where $M_{\bullet f}$ and $M_{\star f}$ are the BH and galaxy mass at $z=z_f$ and $D_f = D(z_f)$.

As an example, we adopt a high-redshift HSC quasar (J1208-0200).
For this quasar, the BH mass, dynamical mass, and redshift are measured as $M_\bullet =7.1\times 10^8~\msun$,
$M_{\rm dyn}=1.3\times 10^{10}~\msun$, and $z=6.2$, respectively \citep{Izumi_2019}.
For simplicity, the galaxy mass is assumed to be equal to the dynamical mass (i.e., $M_\star = M_{\rm dyn}$).
Let us consider the BH after the transient rapid accretion phases ($M_{\bullet i} = 1.3\times 10^6~\msun$, 
$M_{\star i}=3\times 10^7~\msun$, and $z_i=15$) to be its seed (see the red curve in Fig.~\ref{fig:Msigma}). 
Following the method above, one obtain $\langle A \rangle = 0.4765$ and $\langle B \rangle = 11.039$.
The evolution track of the $M_\bullet/M_\star$ ratio is shown in Fig.~\ref{fig:Msigma} (red dashed curve).
The mass ratio decreases initially, reaches a minimum value of $\sim 0.01$ at $z\simeq 10$, and increases
to the observed value.

\bibliography{ref}{}

\begin{thebibliography}{}
\expandafter\ifx\csname natexlab\endcsname\relax\def\natexlab#1{#1}\fi
\providecommand{\url}[1]{\href{#1}{#1}}
\providecommand{\dodoi}[1]{doi:~\href{http://doi.org/#1}{\nolinkurl{#1}}}
\providecommand{\doeprint}[1]{\href{http://ascl.net/#1}{\nolinkurl{http://ascl.net/#1}}}
\providecommand{\doarXiv}[1]{\href{https://arxiv.org/abs/#1}{\nolinkurl{https://arxiv.org/abs/#1}}}

\bibitem[{{Abel} {et~al.}(2002){Abel}, {Bryan}, \& {Norman}}]{Abel_2002}
{Abel}, T., {Bryan}, G.~L., \& {Norman}, M.~L. 2002, Science, 295, 93,
  \dodoi{10.1126/science.295.5552.93}

\bibitem[{{Abramowicz} {et~al.}(1988){Abramowicz}, {Czerny}, {Lasota}, \&
  {Szuszkiewicz}}]{Abramowicz_1988}
{Abramowicz}, M.~A., {Czerny}, B., {Lasota}, J.~P., \& {Szuszkiewicz}, E. 1988,
  \apj, 332, 646, \dodoi{10.1086/166683}

\bibitem[{{Agarwal} {et~al.}(2013){Agarwal}, {Davis}, {Khochfar}, {Natarajan},
  \& {Dunlop}}]{Agarwal_2013}
{Agarwal}, B., {Davis}, A.~J., {Khochfar}, S., {Natarajan}, P., \& {Dunlop},
  J.~S. 2013, \mnras, 432, 3438, \dodoi{10.1093/mnras/stt696}

\bibitem[{{Akeson} {et~al.}(2019){Akeson}, {Armus}, {Bachelet}, {Bailey},
  {Bartusek}, {Bellini}, {Benford}, {Bennett}, {Bhattacharya}, {Bohlin},
  {Boyer}, {Bozza}, {Bryden}, {Calchi Novati}, {Carpenter}, {Casertano},
  {Choi}, {Content}, {Dayal}, {Dressler}, {Dor{\'e}}, {Fall}, {Fan}, {Fang},
  {Filippenko}, {Finkelstein}, {Foley}, {Furlanetto}, {Kalirai}, {Gaudi},
  {Gilbert}, {Girard}, {Grady}, {Greene}, {Guhathakurta}, {Heinrich},
  {Hemmati}, {Hendel}, {Henderson}, {Henning}, {Hirata}, {Ho}, {Huff},
  {Hutter}, {Jansen}, {Jha}, {Johnson}, {Jones}, {Kasdin}, {Kelly}, {Kirshner},
  {Koekemoer}, {Kruk}, {Lewis}, {Macintosh}, {Madau}, {Malhotra}, {Mandel},
  {Massara}, {Masters}, {McEnery}, {McQuinn}, {Melchior}, {Melton},
  {Mennesson}, {Peeples}, {Penny}, {Perlmutter}, {Pisani}, {Plazas}, {Poleski},
  {Postman}, {Ranc}, {Rauscher}, {Rest}, {Roberge}, {Robertson}, {Rodney},
  {Rhoads}, {Rhodes}, {Ryan}, {Sahu}, {Sand}, {Scolnic}, {Seth}, {Shvartzvald},
  {Siellez}, {Smith}, {Spergel}, {Stassun}, {Street}, {Strolger}, {Szalay},
  {Trauger}, {Troxel}, {Turnbull}, {van der Marel}, {von der Linden}, {Wang},
  {Weinberg}, {Williams}, {Windhorst}, {Wollack}, {Wu}, {Yee}, \&
  {Zimmerman}}]{Akeson_2019}
{Akeson}, R., {Armus}, L., {Bachelet}, E., {et~al.} 2019, arXiv e-prints,
  arXiv:1902.05569.
\newblock \doarXiv{1902.05569}

\bibitem[{{Alexander} \& {Natarajan}(2014)}]{Alexander_Natarajan_2014}
{Alexander}, T., \& {Natarajan}, P. 2014, Science, 345, 1330,
  \dodoi{10.1126/science.1251053}

\bibitem[{{Angl{\'e}s-Alc{\'a}zar} {et~al.}(2017){Angl{\'e}s-Alc{\'a}zar},
  {Faucher-Gigu{\`e}re}, {Quataert}, {Hopkins}, {Feldmann}, {Torrey}, {Wetzel},
  \& {Kere{\v{s}}}}]{Angles-Alcazar_2017}
{Angl{\'e}s-Alc{\'a}zar}, D., {Faucher-Gigu{\`e}re}, C.-A., {Quataert}, E.,
  {et~al.} 2017, \mnras, 472, L109, \dodoi{10.1093/mnrasl/slx161}

\bibitem[{{Ba{\~n}ados} {et~al.}(2018){Ba{\~n}ados}, {Venemans},
  {Mazzucchelli}, {Farina}, {Walter}, {Wang}, {Decarli}, {Stern}, {Fan},
  {Davies}, {Hennawi}, {Simcoe}, {Turner}, {Rix}, {Yang}, {Kelson}, {Rudie}, \&
  {Winters}}]{Banados_2018}
{Ba{\~n}ados}, E., {Venemans}, B.~P., {Mazzucchelli}, C., {et~al.} 2018, \nat,
  553, 473, \dodoi{10.1038/nature25180}

\bibitem[{{Bai}(2011)}]{Bai_2011}
{Bai}, X.-N. 2011, \apj, 739, 50, \dodoi{10.1088/0004-637X/739/1/50}

\bibitem[{{Balbus} \& {Hawley}(1998)}]{Balbus_Hawley_1998}
{Balbus}, S.~A., \& {Hawley}, J.~F. 1998, Reviews of Modern Physics, 70, 1,
  \dodoi{10.1103/RevModPhys.70.1}

\bibitem[{{Becerra} {et~al.}(2015){Becerra}, {Greif}, {Springel}, \&
  {Hernquist}}]{Becerra_2015}
{Becerra}, F., {Greif}, T.~H., {Springel}, V., \& {Hernquist}, L.~E. 2015,
  \mnras, 446, 2380, \dodoi{10.1093/mnras/stu2284}

\bibitem[{{Behroozi} {et~al.}(2020){Behroozi}, {Conroy}, {Wechsler}, {Hearin},
  {Williams}, {Moster}, {Yung}, {Somerville}, {Gottl{\"o}ber}, {Yepes}, \&
  {Endsley}}]{Behroozi_2020}
{Behroozi}, P., {Conroy}, C., {Wechsler}, R.~H., {et~al.} 2020, \mnras, 499,
  5702, \dodoi{10.1093/mnras/staa3164}

\bibitem[{{Bennert} {et~al.}(2011){Bennert}, {Auger}, {Treu}, {Woo}, \&
  {Malkan}}]{Bennert_2011}
{Bennert}, V.~N., {Auger}, M.~W., {Treu}, T., {Woo}, J.-H., \& {Malkan}, M.~A.
  2011, \apj, 742, 107, \dodoi{10.1088/0004-637X/742/2/107}

\bibitem[{{Bouwens} {et~al.}(2015){Bouwens}, {Illingworth}, {Oesch}, {Trenti},
  {Labb{\'e}}, {Bradley}, {Carollo}, {van Dokkum}, {Gonzalez}, {Holwerda},
  {Franx}, {Spitler}, {Smit}, \& {Magee}}]{Bouwens_2015}
{Bouwens}, R.~J., {Illingworth}, G.~D., {Oesch}, P.~A., {et~al.} 2015, \apj,
  803, 34, \dodoi{10.1088/0004-637X/803/1/34}

\bibitem[{{Bromm} \& {Loeb}(2003)}]{Bromm_Loeb_2003}
{Bromm}, V., \& {Loeb}, A. 2003, \apj, 596, 34, \dodoi{10.1086/377529}

\bibitem[{{Bromm} \& {Yoshida}(2011)}]{Bromm_Yoshida_2011}
{Bromm}, V., \& {Yoshida}, N. 2011, \araa, 49, 373,
  \dodoi{10.1146/annurev-astro-081710-102608}

\bibitem[{{Bullock} {et~al.}(2001){Bullock}, {Kolatt}, {Sigad}, {Somerville},
  {Kravtsov}, {Klypin}, {Primack}, \& {Dekel}}]{Bullock_2001}
{Bullock}, J.~S., {Kolatt}, T.~S., {Sigad}, Y., {et~al.} 2001, \mnras, 321,
  559, \dodoi{10.1046/j.1365-8711.2001.04068.x}

\bibitem[{{Chon} {et~al.}(2018){Chon}, {Hosokawa}, \& {Yoshida}}]{Chon_2018}
{Chon}, S., {Hosokawa}, T., \& {Yoshida}, N. 2018, \mnras, 475, 4104,
  \dodoi{10.1093/mnras/sty086}

\bibitem[{{Chon} \& {Omukai}(2020)}]{Chon_Omukai_2020}
{Chon}, S., \& {Omukai}, K. 2020, \mnras, 494, 2851,
  \dodoi{10.1093/mnras/staa863}

\bibitem[{{Ciotti} \& {Ostriker}(2001)}]{Ciotti_Ostriker_2001}
{Ciotti}, L., \& {Ostriker}, J.~P. 2001, \apj, 551, 131, \dodoi{10.1086/320053}

\bibitem[{{Davies} {et~al.}(2019){Davies}, {Hennawi}, \&
  {Eilers}}]{Davies_2019}
{Davies}, F.~B., {Hennawi}, J.~F., \& {Eilers}, A.-C. 2019, \apjl, 884, L19,
  \dodoi{10.3847/2041-8213/ab42e3}

\bibitem[{{Davies} {et~al.}(2020){Davies}, {Hennawi}, \&
  {Eilers}}]{Davies_2020}
---. 2020, \mnras, 493, 1330, \dodoi{10.1093/mnras/stz3303}

\bibitem[{{Dekel} \& {Birnboim}(2006)}]{Dekel_2006}
{Dekel}, A., \& {Birnboim}, Y. 2006, \mnras, 368, 2,
  \dodoi{10.1111/j.1365-2966.2006.10145.x}

\bibitem[{{Dekel} {et~al.}(2013){Dekel}, {Zolotov}, {Tweed}, {Cacciato},
  {Ceverino}, \& {Primack}}]{Dekel_2013}
{Dekel}, A., {Zolotov}, A., {Tweed}, D., {et~al.} 2013, \mnras, 435, 999,
  \dodoi{10.1093/mnras/stt1338}

\bibitem[{{Devecchi} \& {Volonteri}(2009)}]{Devecchi_2009}
{Devecchi}, B., \& {Volonteri}, M. 2009, \apj, 694, 302,
  \dodoi{10.1088/0004-637X/694/1/302}

\bibitem[{{Di Matteo} {et~al.}(2012){Di Matteo}, {Khandai}, {DeGraf}, {Feng},
  {Croft}, {Lopez}, \& {Springel}}]{DiMatteo_2012}
{Di Matteo}, T., {Khandai}, N., {DeGraf}, C., {et~al.} 2012, \apjl, 745, L29,
  \dodoi{10.1088/2041-8205/745/2/L29}

\bibitem[{{Ding} {et~al.}(2020){Ding}, {Silverman}, {Treu}, {Schulze},
  {Schramm}, {Birrer}, {Park}, {Jahnke}, {Bennert}, {Kartaltepe}, {Koekemoer},
  {Malkan}, \& {Sanders}}]{Ding_2020}
{Ding}, X., {Silverman}, J., {Treu}, T., {et~al.} 2020, \apj, 888, 37,
  \dodoi{10.3847/1538-4357/ab5b90}

\bibitem[{{Draine} \& {Lee}(1984)}]{Draine_Lee_1984}
{Draine}, B.~T., \& {Lee}, H.~M. 1984, \apj, 285, 89, \dodoi{10.1086/162480}

\bibitem[{{Dubois} {et~al.}(2013){Dubois}, {Pichon}, {Devriendt}, {Silk},
  {Haehnelt}, {Kimm}, \& {Slyz}}]{Dubois_2013}
{Dubois}, Y., {Pichon}, C., {Devriendt}, J., {et~al.} 2013, \mnras, 428, 2885,
  \dodoi{10.1093/mnras/sts224}

\bibitem[{{Eilers} {et~al.}(2018){Eilers}, {Hennawi}, \&
  {Davies}}]{Eilers_2018}
{Eilers}, A.-C., {Hennawi}, J.~F., \& {Davies}, F.~B. 2018, \apj, 867, 30,
  \dodoi{10.3847/1538-4357/aae081}

\bibitem[{{Eilers} {et~al.}(2021){Eilers}, {Hennawi}, {Davies}, \&
  {Simcoe}}]{Eilers_2021}
{Eilers}, A.-C., {Hennawi}, J.~F., {Davies}, F.~B., \& {Simcoe}, R.~A. 2021,
  \apj, 917, 38, \dodoi{10.3847/1538-4357/ac0a76}

\bibitem[{{Fakhouri} {et~al.}(2010){Fakhouri}, {Ma}, \&
  {Boylan-Kolchin}}]{Fakhouri_2010}
{Fakhouri}, O., {Ma}, C.-P., \& {Boylan-Kolchin}, M. 2010, \mnras, 406, 2267,
  \dodoi{10.1111/j.1365-2966.2010.16859.x}

\bibitem[{{Fan}(2006)}]{Fan_2006}
{Fan}, X. 2006, \nar, 50, 665, \dodoi{10.1016/j.newar.2006.06.077}

\bibitem[{{Fern{\'a}ndez} \& {Metzger}(2013)}]{Fernandez_2013}
{Fern{\'a}ndez}, R., \& {Metzger}, B.~D. 2013, \apj, 763, 108,
  \dodoi{10.1088/0004-637X/763/2/108}

\bibitem[{{Ferrarese} \& {Merritt}(2000)}]{Ferrarese_2000}
{Ferrarese}, L., \& {Merritt}, D. 2000, \apjl, 539, L9, \dodoi{10.1086/312838}

\bibitem[{{Franco} {et~al.}(1990){Franco}, {Tenorio-Tagle}, \&
  {Bodenheimer}}]{Franco_1990}
{Franco}, J., {Tenorio-Tagle}, G., \& {Bodenheimer}, P. 1990, \apj, 349, 126,
  \dodoi{10.1086/168300}

\bibitem[{{Fukushima} \& {Yajima}(2021)}]{Fukushima_2021}
{Fukushima}, H., \& {Yajima}, H. 2021, \mnras, 506, 5512,
  \dodoi{10.1093/mnras/stab2099}

\bibitem[{{Fukushima} {et~al.}(2020){Fukushima}, {Yajima}, {Sugimura},
  {Hosokawa}, {Omukai}, \& {Matsumoto}}]{Fukushima_2020}
{Fukushima}, H., {Yajima}, H., {Sugimura}, K., {et~al.} 2020, \mnras, 497,
  3830, \dodoi{10.1093/mnras/staa2062}

\bibitem[{{Galli} \& {Palla}(1998)}]{Galli_Palla_1998}
{Galli}, D., \& {Palla}, F. 1998, \aap, 335, 403

\bibitem[{{Glover} \& {Jappsen}(2007)}]{Glover_Jappsen_2007}
{Glover}, S.~C.~O., \& {Jappsen}, A.-K. 2007, \apj, 666, 1,
  \dodoi{10.1086/519445}

\bibitem[{{Grimm} {et~al.}(2003){Grimm}, {Gilfanov}, \& {Sunyaev}}]{Grimm_2003}
{Grimm}, H.-J., {Gilfanov}, M., \& {Sunyaev}, R. 2003, \mnras, 339, 793,
  \dodoi{10.1046/j.1365-8711.2003.06224.x}

\bibitem[{{Habouzit} {et~al.}(2017){Habouzit}, {Volonteri}, \&
  {Dubois}}]{Habouzit_2017}
{Habouzit}, M., {Volonteri}, M., \& {Dubois}, Y. 2017, \mnras, 468, 3935,
  \dodoi{10.1093/mnras/stx666}

\bibitem[{{Haiman}(2013)}]{Haiman_2013}
{Haiman}, Z. 2013, in Astrophysics and Space Science Library, Vol. 396,
  Astrophysics and Space Science Library, ed. T.~{Wiklind}, B.~{Mobasher}, \&
  V.~{Bromm}, 293, \dodoi{10.1007/978-3-642-32362-1_6}

\bibitem[{{Hernquist}(1990)}]{Hernquist_1990}
{Hernquist}, L. 1990, \apj, 356, 359, \dodoi{10.1086/168845}

\bibitem[{{Hirano} {et~al.}(2014){Hirano}, {Hosokawa}, {Yoshida}, {Umeda},
  {Omukai}, {Chiaki}, \& {Yorke}}]{Hirano_2014}
{Hirano}, S., {Hosokawa}, T., {Yoshida}, N., {et~al.} 2014, \apj, 781, 60,
  \dodoi{10.1088/0004-637X/781/2/60}

\bibitem[{{Hollenbach} \& {McKee}(1989)}]{Hollenbach_McKee_1989}
{Hollenbach}, D., \& {McKee}, C.~F. 1989, \apj, 342, 306,
  \dodoi{10.1086/167595}

\bibitem[{{Hosokawa} {et~al.}(2011){Hosokawa}, {Omukai}, {Yoshida}, \&
  {Yorke}}]{Hosokawa_2011}
{Hosokawa}, T., {Omukai}, K., {Yoshida}, N., \& {Yorke}, H.~W. 2011, Science,
  334, 1250, \dodoi{10.1126/science.1207433}

\bibitem[{{Iliev} \& {Shapiro}(2001)}]{Iliev_2001}
{Iliev}, I.~T., \& {Shapiro}, P.~R. 2001, \mnras, 325, 468,
  \dodoi{10.1046/j.1365-8711.2001.04422.x}

\bibitem[{{Inayoshi} {et~al.}(2016){Inayoshi}, {Haiman}, \&
  {Ostriker}}]{IHO_2016}
{Inayoshi}, K., {Haiman}, Z., \& {Ostriker}, J.~P. 2016, \mnras, 459, 3738,
  \dodoi{10.1093/mnras/stw836}

\bibitem[{{Inayoshi} {et~al.}(2019){Inayoshi}, {Ichikawa}, {Ostriker}, \&
  {Kuiper}}]{Inayoshi_2019}
{Inayoshi}, K., {Ichikawa}, K., {Ostriker}, J.~P., \& {Kuiper}, R. 2019,
  \mnras, 486, 5377, \dodoi{10.1093/mnras/stz1189}

\bibitem[{{Inayoshi} {et~al.}(2021){Inayoshi}, {Kashiyama}, {Visbal}, \&
  {Haiman}}]{Inayoshi_2021}
{Inayoshi}, K., {Kashiyama}, K., {Visbal}, E., \& {Haiman}, Z. 2021, arXiv
  e-prints, arXiv:2103.12755.
\newblock \doarXiv{2103.12755}

\bibitem[{{Inayoshi} \& {Omukai}(2011)}]{IO11}
{Inayoshi}, K., \& {Omukai}, K. 2011, \mnras, 416, 2748,
  \dodoi{10.1111/j.1365-2966.2011.19229.x}

\bibitem[{{Inayoshi} {et~al.}(2014){Inayoshi}, {Omukai}, \&
  {Tasker}}]{IOT_2014}
{Inayoshi}, K., {Omukai}, K., \& {Tasker}, E. 2014, \mnras, 445, L109,
  \dodoi{10.1093/mnrasl/slu151}

\bibitem[{{Inayoshi} \& {Tanaka}(2015)}]{Inayoshi_Tanaka_2015}
{Inayoshi}, K., \& {Tanaka}, T.~L. 2015, \mnras, 450, 4350,
  \dodoi{10.1093/mnras/stv871}

\bibitem[{{Inayoshi} {et~al.}(2020){Inayoshi}, {Visbal}, \&
  {Haiman}}]{Inayoshi_ARAA_2020}
{Inayoshi}, K., {Visbal}, E., \& {Haiman}, Z. 2020, \araa, 58, 27,
  \dodoi{10.1146/annurev-astro-120419-014455}

\bibitem[{{Inoue}(2011)}]{Inoue_2011}
{Inoue}, A.~K. 2011, \mnras, 415, 2920,
  \dodoi{10.1111/j.1365-2966.2011.18906.x}

\bibitem[{{Izumi} {et~al.}(2019){Izumi}, {Onoue}, {Matsuoka}, {Nagao},
  {Strauss}, {Imanishi}, {Kashikawa}, {Fujimoto}, {Kohno}, {Toba}, {Umehata},
  {Goto}, {Ueda}, {Shirakata}, {Silverman}, {Greene}, {Harikane}, {Hashimoto},
  {Ikarashi}, {Iono}, {Iwasawa}, {Lee}, {Minezaki}, {Nakanishi}, {Tamura},
  {Tang}, \& {Taniguchi}}]{Izumi_2019}
{Izumi}, T., {Onoue}, M., {Matsuoka}, Y., {et~al.} 2019, \pasj, 71, 111,
  \dodoi{10.1093/pasj/psz096}

\bibitem[{{Izumi} {et~al.}(2021){Izumi}, {Matsuoka}, {Fujimoto}, {Onoue},
  {Strauss}, {Umehata}, {Imanishi}, {Kohno}, {Kawaguchi}, {Kawamuro}, {Baba},
  {Nagao}, {Toba}, {Inayoshi}, {Silverman}, {Inoue}, {Ikarashi}, {Iwasawa},
  {Kashikawa}, {Hashimoto}, {Nakanishi}, {Ueda}, {Schramm}, {Lee}, \&
  {Suh}}]{Izumi_2021}
{Izumi}, T., {Matsuoka}, Y., {Fujimoto}, S., {et~al.} 2021, \apj, 914, 36,
  \dodoi{10.3847/1538-4357/abf6dc}

\bibitem[{{Jeon} {et~al.}(2012){Jeon}, {Pawlik}, {Greif}, {Glover}, {Bromm},
  {Milosavljevi{\'c}}, \& {Klessen}}]{Jeon_2012}
{Jeon}, M., {Pawlik}, A.~H., {Greif}, T.~H., {et~al.} 2012, \apj, 754, 34,
  \dodoi{10.1088/0004-637X/754/1/34}

\bibitem[{{Jiang} {et~al.}(2014){Jiang}, {Stone}, \& {Davis}}]{Jiang_2014}
{Jiang}, Y.-F., {Stone}, J.~M., \& {Davis}, S.~W. 2014, \apj, 796, 106,
  \dodoi{10.1088/0004-637X/796/2/106}

\bibitem[{{Johnson} {et~al.}(2011){Johnson}, {Khochfar}, {Greif}, \&
  {Durier}}]{Johnson_2011}
{Johnson}, J.~L., {Khochfar}, S., {Greif}, T.~H., \& {Durier}, F. 2011, \mnras,
  410, 919, \dodoi{10.1111/j.1365-2966.2010.17491.x}

\bibitem[{{Kashiyama} \& {Inayoshi}(2016)}]{Kashiyama_Inayoshi_2016}
{Kashiyama}, K., \& {Inayoshi}, K. 2016, \apj, 826, 80,
  \dodoi{10.3847/0004-637X/826/1/80}

\bibitem[{{Kawinwanichakij} {et~al.}(2021){Kawinwanichakij}, {Silverman},
  {Ding}, {George}, {Damjanov}, {Sawicki}, {Tanaka}, {Taranu}, {Birrer},
  {Huang}, {Li}, {Onodera}, {Shibuya}, \& {Yasuda}}]{Kawinwanichakij_2021}
{Kawinwanichakij}, L., {Silverman}, J.~D., {Ding}, X., {et~al.} 2021, arXiv
  e-prints, arXiv:2109.09766.
\newblock \doarXiv{2109.09766}

\bibitem[{{Kitayama} \& {Yoshida}(2005)}]{Kitayama_2005}
{Kitayama}, T., \& {Yoshida}, N. 2005, \apj, 630, 675, \dodoi{10.1086/432114}

\bibitem[{{Kitayama} {et~al.}(2004){Kitayama}, {Yoshida}, {Susa}, \&
  {Umemura}}]{Kitayama_2004}
{Kitayama}, T., {Yoshida}, N., {Susa}, H., \& {Umemura}, M. 2004, \apj, 613,
  631, \dodoi{10.1086/423313}

\bibitem[{{Kocsis} \& {Tremaine}(2011)}]{Kocsis_Loeb_2011}
{Kocsis}, B., \& {Tremaine}, S. 2011, \mnras, 412, 187,
  \dodoi{10.1111/j.1365-2966.2010.17897.x}

\bibitem[{{Kormendy} \& {Ho}(2013)}]{Kormendy_Ho_2013}
{Kormendy}, J., \& {Ho}, L.~C. 2013, \araa, 51, 511,
  \dodoi{10.1146/annurev-astro-082708-101811}

\bibitem[{{Kratter} \& {Lodato}(2016)}]{Kratter_2016}
{Kratter}, K., \& {Lodato}, G. 2016, \araa, 54, 271,
  \dodoi{10.1146/annurev-astro-081915-023307}

\bibitem[{{Kuiper} {et~al.}(2010){Kuiper}, {Klahr}, {Beuther}, \&
  {Henning}}]{Kuiper_2010}
{Kuiper}, R., {Klahr}, H., {Beuther}, H., \& {Henning}, T. 2010, \apj, 722,
  1556, \dodoi{10.1088/0004-637X/722/2/1556}

\bibitem[{{Kuiper} {et~al.}(2011){Kuiper}, {Klahr}, {Beuther}, \&
  {Henning}}]{Kuiper_2011}
---. 2011, \apj, 732, 20, \dodoi{10.1088/0004-637X/732/1/20}

\bibitem[{{Kuiper} {et~al.}(2020){Kuiper}, {Yorke}, \& {Mignone}}]{Kuiper_2020}
{Kuiper}, R., {Yorke}, H.~W., \& {Mignone}, A. 2020, \apjs, 250, 13,
  \dodoi{10.3847/1538-4365/ab9a36}

\bibitem[{{Latif} {et~al.}(2013){Latif}, {Schleicher}, {Schmidt}, \&
  {Niemeyer}}]{Latif_2013}
{Latif}, M.~A., {Schleicher}, D.~R.~G., {Schmidt}, W., \& {Niemeyer}, J. 2013,
  \mnras, 433, 1607, \dodoi{10.1093/mnras/stt834}

\bibitem[{{Latif} {et~al.}(2018){Latif}, {Volonteri}, \& {Wise}}]{Latif_2018}
{Latif}, M.~A., {Volonteri}, M., \& {Wise}, J.~H. 2018, \mnras, 476, 5016,
  \dodoi{10.1093/mnras/sty622}

\bibitem[{{Levermore} \& {Pomraning}(1981)}]{Levermore_1981}
{Levermore}, C.~D., \& {Pomraning}, G.~C. 1981, \apj, 248, 321,
  \dodoi{10.1086/159157}

\bibitem[{{Li} {et~al.}(2021{\natexlab{a}}){Li}, {Silverman}, {Ding},
  {Strauss}, {Goulding}, {Birrer}, {Yesuf}, {Xue}, {Kawinwanichakij},
  {Matsuoka}, {Toba}, {Nagao}, {Schramm}, \& {Inayoshi}}]{Li_size_2021}
{Li}, J., {Silverman}, J.~D., {Ding}, X., {et~al.} 2021{\natexlab{a}}, \apj,
  918, 22, \dodoi{10.3847/1538-4357/ac06a8}

\bibitem[{{Li} {et~al.}(2021{\natexlab{b}}){Li}, {Inayoshi}, \&
  {Qiu}}]{Li_2021}
{Li}, W., {Inayoshi}, K., \& {Qiu}, Y. 2021{\natexlab{b}}, arXiv e-prints,
  arXiv:2105.12637.
\newblock \doarXiv{2105.12637}

\bibitem[{{Li} {et~al.}(2007){Li}, {Hernquist}, {Robertson}, {Cox}, {Hopkins},
  {Springel}, {Gao}, {Di Matteo}, {Zentner}, {Jenkins}, \& {Yoshida}}]{Li_2007}
{Li}, Y., {Hernquist}, L., {Robertson}, B., {et~al.} 2007, \apj, 665, 187,
  \dodoi{10.1086/519297}

\bibitem[{{Lodato} \& {Natarajan}(2006)}]{Lodato_Natarajan_2006}
{Lodato}, G., \& {Natarajan}, P. 2006, \mnras, 371, 1813,
  \dodoi{10.1111/j.1365-2966.2006.10801.x}

\bibitem[{{Lupi} {et~al.}(2021){Lupi}, {Haiman}, \& {Volonteri}}]{Lupi_2021}
{Lupi}, A., {Haiman}, Z., \& {Volonteri}, M. 2021, \mnras, 503, 5046,
  \dodoi{10.1093/mnras/stab692}

\bibitem[{{Lupi} {et~al.}(2019){Lupi}, {Volonteri}, {Decarli}, {Bovino},
  {Silk}, \& {Bergeron}}]{Lupi_2019}
{Lupi}, A., {Volonteri}, M., {Decarli}, R., {et~al.} 2019, \mnras, 488, 4004,
  \dodoi{10.1093/mnras/stz1959}

\bibitem[{{Lusso} {et~al.}(2015){Lusso}, {Worseck}, {Hennawi}, {Prochaska},
  {Vignali}, {Stern}, \& {O'Meara}}]{Lusso_2015}
{Lusso}, E., {Worseck}, G., {Hennawi}, J.~F., {et~al.} 2015, \mnras, 449, 4204,
  \dodoi{10.1093/mnras/stv516}

\bibitem[{{Magorrian} {et~al.}(1998){Magorrian}, {Tremaine}, {Richstone},
  {Bender}, {Bower}, {Dressler}, {Faber}, {Gebhardt}, {Green}, {Grillmair},
  {Kormendy}, \& {Lauer}}]{Magorrian_1998}
{Magorrian}, J., {Tremaine}, S., {Richstone}, D., {et~al.} 1998, \aj, 115,
  2285, \dodoi{10.1086/300353}

\bibitem[{{Matsuoka} {et~al.}(2016){Matsuoka}, {Onoue}, {Kashikawa}, {Iwasawa},
  {Strauss}, {Nagao}, {Imanishi}, {Niida}, {Toba}, {Akiyama}, {Asami}, {Bosch},
  {Foucaud}, {Furusawa}, {Goto}, {Gunn}, {Harikane}, {Ikeda}, {Kawaguchi},
  {Kikuta}, {Komiyama}, {Lupton}, {Minezaki}, {Miyazaki}, {Morokuma},
  {Murayama}, {Nishizawa}, {Ono}, {Ouchi}, {Price}, {Sameshima}, {Silverman},
  {Sugiyama}, {Tait}, {Takada}, {Takata}, {Tanaka}, {Tang}, \&
  {Utsumi}}]{Matsuoka_2016}
{Matsuoka}, Y., {Onoue}, M., {Kashikawa}, N., {et~al.} 2016, \apj, 828, 26,
  \dodoi{10.3847/0004-637X/828/1/26}

\bibitem[{{Matsuoka} {et~al.}(2018){Matsuoka}, {Strauss}, {Kashikawa}, {Onoue},
  {Iwasawa}, {Tang}, {Lee}, {Imanishi}, {Nagao}, {Akiyama}, {Asami}, {Bosch},
  {Furusawa}, {Goto}, {Gunn}, {Harikane}, {Ikeda}, {Izumi}, {Kawaguchi},
  {Kato}, {Kikuta}, {Kohno}, {Komiyama}, {Lupton}, {Minezaki}, {Miyazaki},
  {Murayama}, {Niida}, {Nishizawa}, {Noboriguchi}, {Oguri}, {Ono}, {Ouchi},
  {Price}, {Sameshima}, {Schulze}, {Shirakata}, {Silverman}, {Sugiyama},
  {Tait}, {Takada}, {Takata}, {Tanaka}, {Toba}, {Utsumi}, {Wang}, \&
  {Yamashita}}]{Matsuoka_2018}
{Matsuoka}, Y., {Strauss}, M.~A., {Kashikawa}, N., {et~al.} 2018, \apj, 869,
  150, \dodoi{10.3847/1538-4357/aaee7a}

\bibitem[{{McKinney} {et~al.}(2015){McKinney}, {Dai}, \&
  {Avara}}]{McKinney_2015}
{McKinney}, J.~C., {Dai}, L., \& {Avara}, M.~J. 2015, \mnras, 454, L6,
  \dodoi{10.1093/mnrasl/slv115}

\bibitem[{{McKinney} \& {Gammie}(2004)}]{McKinney_&_Gammie_2004}
{McKinney}, J.~C., \& {Gammie}, C.~F. 2004, \apj, 611, 977,
  \dodoi{10.1086/422244}

\bibitem[{{Mignone} {et~al.}(2007){Mignone}, {Bodo}, {Massaglia}, {Matsakos},
  {Tesileanu}, {Zanni}, \& {Ferrari}}]{Mignone_2007}
{Mignone}, A., {Bodo}, G., {Massaglia}, S., {et~al.} 2007, \apjs, 170, 228,
  \dodoi{10.1086/513316}

\bibitem[{{Milosavljevi{\'c}} {et~al.}(2009){Milosavljevi{\'c}}, {Bromm},
  {Couch}, \& {Oh}}]{Milosavljevic_2009}
{Milosavljevi{\'c}}, M., {Bromm}, V., {Couch}, S.~M., \& {Oh}, S.~P. 2009,
  \apj, 698, 766, \dodoi{10.1088/0004-637X/698/1/766}

\bibitem[{{Mineo} {et~al.}(2014){Mineo}, {Gilfanov}, {Lehmer}, {Morrison}, \&
  {Sunyaev}}]{Mineo_2014}
{Mineo}, S., {Gilfanov}, M., {Lehmer}, B.~D., {Morrison}, G.~E., \& {Sunyaev},
  R. 2014, \mnras, 437, 1698, \dodoi{10.1093/mnras/stt1999}

\bibitem[{{Mo} {et~al.}(1998){Mo}, {Mao}, \& {White}}]{Mo_1998}
{Mo}, H.~J., {Mao}, S., \& {White}, S. D.~M. 1998, \mnras, 295, 319,
  \dodoi{10.1046/j.1365-8711.1998.01227.x}

\bibitem[{{Mortlock} {et~al.}(2011){Mortlock}, {Warren}, {Venemans}, {Patel},
  {Hewett}, {McMahon}, {Simpson}, {Theuns}, {Gonz{\'a}les-Solares}, {Adamson},
  {Dye}, {Hambly}, {Hirst}, {Irwin}, {Kuiper}, {Lawrence}, \&
  {R{\"o}ttgering}}]{Mortlock_2011}
{Mortlock}, D.~J., {Warren}, S.~J., {Venemans}, B.~P., {et~al.} 2011, \nat,
  474, 616, \dodoi{10.1038/nature10159}

\bibitem[{{Mowla} {et~al.}(2019){Mowla}, {van der Wel}, {van Dokkum}, \&
  {Miller}}]{Mowla_2019}
{Mowla}, L., {van der Wel}, A., {van Dokkum}, P., \& {Miller}, T.~B. 2019,
  \apjl, 872, L13, \dodoi{10.3847/2041-8213/ab0379}

\bibitem[{{Nakatani} {et~al.}(2018{\natexlab{a}}){Nakatani}, {Hosokawa},
  {Yoshida}, {Nomura}, \& {Kuiper}}]{Nakatani_2018a}
{Nakatani}, R., {Hosokawa}, T., {Yoshida}, N., {Nomura}, H., \& {Kuiper}, R.
  2018{\natexlab{a}}, \apj, 857, 57, \dodoi{10.3847/1538-4357/aab70b}

\bibitem[{{Nakatani} {et~al.}(2018{\natexlab{b}}){Nakatani}, {Hosokawa},
  {Yoshida}, {Nomura}, \& {Kuiper}}]{Nakatani_2018b}
---. 2018{\natexlab{b}}, \apj, 865, 75, \dodoi{10.3847/1538-4357/aad9fd}

\bibitem[{{Narayan} {et~al.}(2012){Narayan}, {S\c{a}dowski}, {Penna}, \&
  {Kulkarni}}]{Narayan_2012}
{Narayan}, R., {S\c{a}dowski}, A., {Penna}, R.~F., \& {Kulkarni}, A.~K. 2012,
  \mnras, 426, 3241, \dodoi{10.1111/j.1365-2966.2012.22002.x}

\bibitem[{{Natarajan}(2021)}]{Natarajan_2021}
{Natarajan}, P. 2021, \mnras, 501, 1413, \dodoi{10.1093/mnras/staa3724}

\bibitem[{{Natarajan} {et~al.}(2017){Natarajan}, {Pacucci}, {Ferrara},
  {Agarwal}, {Ricarte}, {Zackrisson}, \& {Cappelluti}}]{Natarajan_2017}
{Natarajan}, P., {Pacucci}, F., {Ferrara}, A., {et~al.} 2017, \apj, 838, 117,
  \dodoi{10.3847/1538-4357/aa6330}

\bibitem[{{Navarro} {et~al.}(1997){Navarro}, {Frenk}, \& {White}}]{NFW_1997}
{Navarro}, J.~F., {Frenk}, C.~S., \& {White}, S. D.~M. 1997, \apj, 490, 493,
  \dodoi{10.1086/304888}

\bibitem[{{Neistein} \& {Dekel}(2008)}]{Neistein_Dekel_2008}
{Neistein}, E., \& {Dekel}, A. 2008, \mnras, 388, 1792,
  \dodoi{10.1111/j.1365-2966.2008.13525.x}

\bibitem[{{Ohsuga} {et~al.}(2005){Ohsuga}, {Mori}, {Nakamoto}, \&
  {Mineshige}}]{Ohsuga_2005}
{Ohsuga}, K., {Mori}, M., {Nakamoto}, T., \& {Mineshige}, S. 2005, \apj, 628,
  368, \dodoi{10.1086/430728}

\bibitem[{{Omukai}(2001)}]{O01}
{Omukai}, K. 2001, \apj, 546, 635, \dodoi{10.1086/318296}

\bibitem[{{Omukai} \& {Nishi}(1998)}]{Omukai_Nishi_1998}
{Omukai}, K., \& {Nishi}, R. 1998, \apj, 508, 141, \dodoi{10.1086/306395}

\bibitem[{{Onoue} {et~al.}(2019){Onoue}, {Kashikawa}, {Matsuoka}, {Kato},
  {Izumi}, {Nagao}, {Strauss}, {Harikane}, {Imanishi}, {Ito}, {Iwasawa},
  {Kawaguchi}, {Lee}, {Noboriguchi}, {Suh}, {Tanaka}, \& {Toba}}]{Onoue_2019}
{Onoue}, M., {Kashikawa}, N., {Matsuoka}, Y., {et~al.} 2019, \apj, 880, 77,
  \dodoi{10.3847/1538-4357/ab29e9}

\bibitem[{{Pacucci} {et~al.}(2015){Pacucci}, {Ferrara}, {Volonteri}, \&
  {Dubus}}]{Pacucci_2015}
{Pacucci}, F., {Ferrara}, A., {Volonteri}, M., \& {Dubus}, G. 2015, \mnras,
  454, 3771, \dodoi{10.1093/mnras/stv2196}

\bibitem[{{Park} \& {Ricotti}(2011)}]{PR_2011}
{Park}, K., \& {Ricotti}, M. 2011, \apj, 739, 2,
  \dodoi{10.1088/0004-637X/739/1/2}

\bibitem[{{Park} \& {Ricotti}(2012)}]{PR_2012}
---. 2012, \apj, 747, 9, \dodoi{10.1088/0004-637X/747/1/9}

\bibitem[{{Park} {et~al.}(2014){Park}, {Ricotti}, {Di Matteo}, \&
  {Reynolds}}]{Park_2014}
{Park}, K., {Ricotti}, M., {Di Matteo}, T., \& {Reynolds}, C.~S. 2014, \mnras,
  437, 2856, \dodoi{10.1093/mnras/stt2096}

\bibitem[{{Park} {et~al.}(2016){Park}, {Ricotti}, {Natarajan},
  {Bogdanovi{\'c}}, \& {Wise}}]{Park_2016}
{Park}, K., {Ricotti}, M., {Natarajan}, P., {Bogdanovi{\'c}}, T., \& {Wise},
  J.~H. 2016, \apj, 818, 184, \dodoi{10.3847/0004-637X/818/2/184}

\bibitem[{{Park} {et~al.}(2017){Park}, {Wise}, \& {Bogdanovi{\'c}}}]{Park_2017}
{Park}, K., {Wise}, J.~H., \& {Bogdanovi{\'c}}, T. 2017, \apj, 847, 70,
  \dodoi{10.3847/1538-4357/aa8729}

\bibitem[{{Park} {et~al.}(2020){Park}, {Wise}, {Bogdanovi{\'c}}, \&
  {Ricotti}}]{Park_2020}
{Park}, K., {Wise}, J.~H., {Bogdanovi{\'c}}, T., \& {Ricotti}, M. 2020, \apj,
  905, 92, \dodoi{10.3847/1538-4357/abc336}

\bibitem[{{Pensabene} {et~al.}(2020){Pensabene}, {Carniani}, {Perna}, {Cresci},
  {Decarli}, {Maiolino}, \& {Marconi}}]{Pensabene_2020}
{Pensabene}, A., {Carniani}, S., {Perna}, M., {et~al.} 2020, \aap, 637, A84,
  \dodoi{10.1051/0004-6361/201936634}

\bibitem[{{Prieto} \& {Escala}(2016)}]{Prieto_2016}
{Prieto}, J., \& {Escala}, A. 2016, \mnras, 460, 4018,
  \dodoi{10.1093/mnras/stw1285}

\bibitem[{{Regan} {et~al.}(2014){Regan}, {Johansson}, \&
  {Haehnelt}}]{Regan_2014}
{Regan}, J.~A., {Johansson}, P.~H., \& {Haehnelt}, M.~G. 2014, \mnras, 439,
  1160, \dodoi{10.1093/mnras/stu068}

\bibitem[{{Regan} {et~al.}(2020){Regan}, {Wise}, {Woods}, {Downes}, {O'Shea},
  \& {Norman}}]{Regan_2020}
{Regan}, J.~A., {Wise}, J.~H., {Woods}, T.~E., {et~al.} 2020, The Open Journal
  of Astrophysics, 3, 15, \dodoi{10.21105/astro.2008.08090}

\bibitem[{{Rieke} {et~al.}(2019){Rieke}, {Arribas}, {Bunker}, {Charlot},
  {Finkelstein}, {Maiolino}, {Robertson}, {Willott}, {Windhorst}, {Eisenstein},
  {Nelson}, {Tacchella}, {Egami}, {Endsley}, {Frye}, {Hainline}, {Hviding},
  {Rieke}, {Williams}, {Willmer}, \& {Woodrum}}]{Rieke_2019}
{Rieke}, M., {Arribas}, S., {Bunker}, A., {et~al.} 2019, \baas, 51, 45

\bibitem[{{Sakurai} {et~al.}(2016){Sakurai}, {Inayoshi}, \&
  {Haiman}}]{Sakurai_2016}
{Sakurai}, Y., {Inayoshi}, K., \& {Haiman}, Z. 2016, \mnras, 461, 4496,
  \dodoi{10.1093/mnras/stw1652}

\bibitem[{{Sakurai} {et~al.}(2017){Sakurai}, {Yoshida}, {Fujii}, \&
  {Hirano}}]{Sakurai_2017}
{Sakurai}, Y., {Yoshida}, N., {Fujii}, M.~S., \& {Hirano}, S. 2017, \mnras,
  472, 1677, \dodoi{10.1093/mnras/stx2044}

\bibitem[{{Salvaterra} {et~al.}(2012){Salvaterra}, {Haardt}, {Volonteri}, \&
  {Moretti}}]{Salvaterra_2012}
{Salvaterra}, R., {Haardt}, F., {Volonteri}, M., \& {Moretti}, A. 2012, \aap,
  545, L6, \dodoi{10.1051/0004-6361/201219965}

\bibitem[{{Sassano} {et~al.}(2021){Sassano}, {Schneider}, {Valiante},
  {Inayoshi}, {Chon}, {Omukai}, {Mayer}, \& {Capelo}}]{Sassano_2021}
{Sassano}, F., {Schneider}, R., {Valiante}, R., {et~al.} 2021, \mnras, 506,
  613, \dodoi{10.1093/mnras/stab1737}

\bibitem[{{Sazonov} {et~al.}(2004){Sazonov}, {Ostriker}, \&
  {Sunyaev}}]{Sazonov_2004}
{Sazonov}, S.~Y., {Ostriker}, J.~P., \& {Sunyaev}, R.~A. 2004, \mnras, 347,
  144, \dodoi{10.1111/j.1365-2966.2004.07184.x}

\bibitem[{{S{\c a}dowski} {et~al.}(2015){S{\c a}dowski}, {Narayan},
  {Tchekhovskoy}, {Abarca}, {Zhu}, \& {McKinney}}]{Sadowski_2015}
{S{\c a}dowski}, A., {Narayan}, R., {Tchekhovskoy}, A., {et~al.} 2015, \mnras,
  447, 49, \dodoi{10.1093/mnras/stu2387}

\bibitem[{{Schaerer}(2003)}]{Schaerer_2003}
{Schaerer}, D. 2003, \aap, 397, 527, \dodoi{10.1051/0004-6361:20021525}

\bibitem[{{Schramm} \& {Silverman}(2013)}]{Schramm_Silverman_2013}
{Schramm}, M., \& {Silverman}, J.~D. 2013, \apj, 767, 13,
  \dodoi{10.1088/0004-637X/767/1/13}

\bibitem[{{Shakura} \& {Sunyaev}(1973)}]{SS_1973}
{Shakura}, N.~I., \& {Sunyaev}, R.~A. 1973, \aap, 24, 337

\bibitem[{{Shang} {et~al.}(2010){Shang}, {Bryan}, \& {Haiman}}]{Shang_2010}
{Shang}, C., {Bryan}, G.~L., \& {Haiman}, Z. 2010, \mnras, 402, 1249,
  \dodoi{10.1111/j.1365-2966.2009.15960.x}

\bibitem[{{Shankar} {et~al.}(2009){Shankar}, {Weinberg}, \&
  {Miralda-Escud{\'e}}}]{Shankar_2009}
{Shankar}, F., {Weinberg}, D.~H., \& {Miralda-Escud{\'e}}, J. 2009, \apj, 690,
  20, \dodoi{10.1088/0004-637X/690/1/20}

\bibitem[{{Shapiro} {et~al.}(1999){Shapiro}, {Iliev}, \& {Raga}}]{Shapiro_1999}
{Shapiro}, P.~R., {Iliev}, I.~T., \& {Raga}, A.~C. 1999, \mnras, 307, 203,
  \dodoi{10.1046/j.1365-8711.1999.02609.x}

\bibitem[{{Shen} {et~al.}(2003){Shen}, {Mo}, {White}, {Blanton}, {Kauffmann},
  {Voges}, {Brinkmann}, \& {Csabai}}]{Shen_2003}
{Shen}, S., {Mo}, H.~J., {White}, S. D.~M., {et~al.} 2003, \mnras, 343, 978,
  \dodoi{10.1046/j.1365-8711.2003.06740.x}

\bibitem[{{Shull} \& {van Steenberg}(1985)}]{Shull_1985}
{Shull}, J.~M., \& {van Steenberg}, M.~E. 1985, \apj, 298, 268,
  \dodoi{10.1086/163605}

\bibitem[{{Sijacki} {et~al.}(2009){Sijacki}, {Springel}, \&
  {Haehnelt}}]{Sijacki_2009}
{Sijacki}, D., {Springel}, V., \& {Haehnelt}, M.~G. 2009, \mnras, 400, 100,
  \dodoi{10.1111/j.1365-2966.2009.15452.x}

\bibitem[{{Smith} {et~al.}(2017){Smith}, {Becerra}, {Bromm}, \&
  {Hernquist}}]{Smith_2017}
{Smith}, A., {Becerra}, F., {Bromm}, V., \& {Hernquist}, L. 2017, \mnras, 472,
  205, \dodoi{10.1093/mnras/stx1993}

\bibitem[{{Stewart} \& {Ida}(2000)}]{Stewart_Ida_2000}
{Stewart}, G.~R., \& {Ida}, S. 2000, \icarus, 143, 28,
  \dodoi{10.1006/icar.1999.6242}

\bibitem[{{Stone} \& {Pringle}(2001)}]{Stone_Pringle_2001}
{Stone}, J.~M., \& {Pringle}, J.~E. 2001, \mnras, 322, 461,
  \dodoi{10.1046/j.1365-8711.2001.04138.x}

\bibitem[{{Stone} {et~al.}(1999){Stone}, {Pringle}, \& {Begelman}}]{Stone_1999}
{Stone}, J.~M., {Pringle}, J.~E., \& {Begelman}, M.~C. 1999, \mnras, 310, 1002,
  \dodoi{10.1046/j.1365-8711.1999.03024.x}

\bibitem[{{Sugimura} {et~al.}(2018){Sugimura}, {Hosokawa}, {Yajima},
  {Inayoshi}, \& {Omukai}}]{Sugimura_2018}
{Sugimura}, K., {Hosokawa}, T., {Yajima}, H., {Inayoshi}, K., \& {Omukai}, K.
  2018, \mnras, 478, 3961, \dodoi{10.1093/mnras/sty1298}

\bibitem[{{Sugimura} {et~al.}(2017){Sugimura}, {Hosokawa}, {Yajima}, \&
  {Omukai}}]{Sugimura_2017}
{Sugimura}, K., {Hosokawa}, T., {Yajima}, H., \& {Omukai}, K. 2017, \mnras,
  469, 62, \dodoi{10.1093/mnras/stx769}

\bibitem[{{Tagawa} {et~al.}(2020){Tagawa}, {Haiman}, \& {Kocsis}}]{Tagawa_2020}
{Tagawa}, H., {Haiman}, Z., \& {Kocsis}, B. 2020, \apj, 892, 36,
  \dodoi{10.3847/1538-4357/ab7922}

\bibitem[{{Takahashi} {et~al.}(2013){Takahashi}, {Inutsuka}, \&
  {Machida}}]{Takahashi_2013}
{Takahashi}, S.~Z., {Inutsuka}, S.-i., \& {Machida}, M.~N. 2013, \apj, 770, 71,
  \dodoi{10.1088/0004-637X/770/1/71}

\bibitem[{{Takasao} {et~al.}(2018){Takasao}, {Tomida}, {Iwasaki}, \&
  {Suzuki}}]{Takasao_2018}
{Takasao}, S., {Tomida}, K., {Iwasaki}, K., \& {Suzuki}, T.~K. 2018, \apj, 857,
  4, \dodoi{10.3847/1538-4357/aab5b3}

\bibitem[{{Takeo} {et~al.}(2020){Takeo}, {Inayoshi}, \&
  {Mineshige}}]{Takeo_2020}
{Takeo}, E., {Inayoshi}, K., \& {Mineshige}, S. 2020, \mnras, 497, 302,
  \dodoi{10.1093/mnras/staa1906}

\bibitem[{{Takeo} {et~al.}(2018){Takeo}, {Inayoshi}, {Ohsuga}, {Takahashi}, \&
  {Mineshige}}]{Takeo_2018}
{Takeo}, E., {Inayoshi}, K., {Ohsuga}, K., {Takahashi}, H.~R., \& {Mineshige},
  S. 2018, \mnras, 476, 673, \dodoi{10.1093/mnras/sty264}

\bibitem[{{Toomre}(1964)}]{Toomre_1964}
{Toomre}, A. 1964, \apj, 139, 1217, \dodoi{10.1086/147861}

\bibitem[{{Toyouchi} {et~al.}(2019){Toyouchi}, {Hosokawa}, {Sugimura},
  {Nakatani}, \& {Kuiper}}]{Toyouchi_2019}
{Toyouchi}, D., {Hosokawa}, T., {Sugimura}, K., {Nakatani}, R., \& {Kuiper}, R.
  2019, \mnras, 483, 2031, \dodoi{10.1093/mnras/sty3012}

\bibitem[{{Toyouchi} {et~al.}(2021){Toyouchi}, {Inayoshi}, {Hosokawa}, \&
  {Kuiper}}]{Toyouchi_2021}
{Toyouchi}, D., {Inayoshi}, K., {Hosokawa}, T., \& {Kuiper}, R. 2021, \apj,
  907, 74, \dodoi{10.3847/1538-4357/abcfc2}

\bibitem[{{Valentini} {et~al.}(2021){Valentini}, {Gallerani}, \&
  {Ferrara}}]{Valentini_2021}
{Valentini}, M., {Gallerani}, S., \& {Ferrara}, A. 2021, \mnras, 507, 1,
  \dodoi{10.1093/mnras/stab1992}

\bibitem[{{Valiante} {et~al.}(2018){Valiante}, {Schneider}, {Zappacosta},
  {Graziani}, {Pezzulli}, \& {Volonteri}}]{Valiante_2018}
{Valiante}, R., {Schneider}, R., {Zappacosta}, L., {et~al.} 2018, \mnras, 476,
  407, \dodoi{10.1093/mnras/sty213}

\bibitem[{{Venemans} {et~al.}(2017){Venemans}, {Walter}, {Decarli},
  {Ba{\~n}ados}, {Hodge}, {Hewett}, {McMahon}, {Mortlock}, \&
  {Simpson}}]{Venemans_2017}
{Venemans}, B.~P., {Walter}, F., {Decarli}, R., {et~al.} 2017, \apj, 837, 146,
  \dodoi{10.3847/1538-4357/aa62ac}

\bibitem[{{Verner} {et~al.}(1996){Verner}, {Ferland}, {Korista}, \&
  {Yakovlev}}]{Verner_1996}
{Verner}, D.~A., {Ferland}, G.~J., {Korista}, K.~T., \& {Yakovlev}, D.~G. 1996,
  \apj, 465, 487, \dodoi{10.1086/177435}

\bibitem[{{Visbal} {et~al.}(2015){Visbal}, {Haiman}, \& {Bryan}}]{Visbal_2015}
{Visbal}, E., {Haiman}, Z., \& {Bryan}, G.~L. 2015, \mnras, 453, 4456,
  \dodoi{10.1093/mnras/stv1941}

\bibitem[{{Volonteri}(2012)}]{Volonteri_2012}
{Volonteri}, M. 2012, Science, 337, 544, \dodoi{10.1126/science.1220843}

\bibitem[{{Volonteri} \& {Rees}(2005)}]{Volonteri_Rees_2005}
{Volonteri}, M., \& {Rees}, M.~J. 2005, \apj, 633, 624, \dodoi{10.1086/466521}

\bibitem[{{Wang} {et~al.}(2021){Wang}, {Yang}, {Fan}, {Hennawi}, {Barth},
  {Banados}, {Bian}, {Boutsia}, {Connor}, {Davies}, {Decarli}, {Eilers},
  {Farina}, {Green}, {Jiang}, {Li}, {Mazzucchelli}, {Nanni}, {Schindler},
  {Venemans}, {Walter}, {Wu}, \& {Yue}}]{Wang_2021}
{Wang}, F., {Yang}, J., {Fan}, X., {et~al.} 2021, \apjl, 907, L1,
  \dodoi{10.3847/2041-8213/abd8c6}

\bibitem[{{Wang} {et~al.}(2010){Wang}, {Carilli}, {Neri}, {Riechers}, {Wagg},
  {Walter}, {Bertoldi}, {Menten}, {Omont}, {Cox}, \& {Fan}}]{Wang_2010}
{Wang}, R., {Carilli}, C.~L., {Neri}, R., {et~al.} 2010, \apj, 714, 699,
  \dodoi{10.1088/0004-637X/714/1/699}

\bibitem[{{Wang} {et~al.}(2013){Wang}, {Wagg}, {Carilli}, {Walter}, {Lentati},
  {Fan}, {Riechers}, {Bertoldi}, {Narayanan}, {Strauss}, {Cox}, {Omont},
  {Menten}, {Knudsen}, {Neri}, \& {Jiang}}]{Wang_2013}
{Wang}, R., {Wagg}, J., {Carilli}, C.~L., {et~al.} 2013, \apj, 773, 44,
  \dodoi{10.1088/0004-637X/773/1/44}

\bibitem[{{Watarai} {et~al.}(2000){Watarai}, {Fukue}, {Takeuchi}, \&
  {Mineshige}}]{Watarai_2000}
{Watarai}, K.-y., {Fukue}, J., {Takeuchi}, M., \& {Mineshige}, S. 2000, \pasj,
  52, 133, \dodoi{10.1093/pasj/52.1.133}

\bibitem[{{Wechsler} {et~al.}(2002){Wechsler}, {Bullock}, {Primack},
  {Kravtsov}, \& {Dekel}}]{Wechsler_2002}
{Wechsler}, R.~H., {Bullock}, J.~S., {Primack}, J.~R., {Kravtsov}, A.~V., \&
  {Dekel}, A. 2002, \apj, 568, 52, \dodoi{10.1086/338765}

\bibitem[{{Whalen} \& {Norman}(2006)}]{Whalen_Norman_2006}
{Whalen}, D., \& {Norman}, M.~L. 2006, \apjs, 162, 281, \dodoi{10.1086/499072}

\bibitem[{{Willott} {et~al.}(2010){Willott}, {Delorme}, {Reyl{\'e}}, {Albert},
  {Bergeron}, {Crampton}, {Delfosse}, {Forveille}, {Hutchings}, {McLure},
  {Omont}, \& {Schade}}]{Willott_2010}
{Willott}, C.~J., {Delorme}, P., {Reyl{\'e}}, C., {et~al.} 2010, \aj, 139, 906,
  \dodoi{10.1088/0004-6256/139/3/906}

\bibitem[{{Wise} {et~al.}(2019){Wise}, {Regan}, {O'Shea}, {Norman}, {Downes},
  \& {Xu}}]{Wise_2019}
{Wise}, J.~H., {Regan}, J.~A., {O'Shea}, B.~W., {et~al.} 2019, \nat, 566, 85,
  \dodoi{10.1038/s41586-019-0873-4}

\bibitem[{{Wise} {et~al.}(2008){Wise}, {Turk}, \& {Abel}}]{Wise_2008}
{Wise}, J.~H., {Turk}, M.~J., \& {Abel}, T. 2008, \apj, 682, 745,
  \dodoi{10.1086/588209}

\bibitem[{{Wu} {et~al.}(2015){Wu}, {Wang}, {Fan}, {Yi}, {Zuo}, {Bian}, {Jiang},
  {McGreer}, {Wang}, {Yang}, {Yang}, {Thompson}, \& {Beletsky}}]{Wu_2015}
{Wu}, X.-B., {Wang}, F., {Fan}, X., {et~al.} 2015, \nat, 518, 512,
  \dodoi{10.1038/nature14241}

\bibitem[{{Yajima} {et~al.}(2017){Yajima}, {Ricotti}, {Park}, \&
  {Sugimura}}]{Yajima_2017}
{Yajima}, H., {Ricotti}, M., {Park}, K., \& {Sugimura}, K. 2017, \apj, 846, 3,
  \dodoi{10.3847/1538-4357/aa8269}

\bibitem[{{Yan} {et~al.}(1998){Yan}, {Sadeghpour}, \& {Dalgarno}}]{Yan_1998}
{Yan}, M., {Sadeghpour}, H.~R., \& {Dalgarno}, A. 1998, \apj, 496, 1044,
  \dodoi{10.1086/305420}

\bibitem[{{Yang} {et~al.}(2021){Yang}, {Wang}, {Fan}, {Barth}, {Hennawi},
  {Nanni}, {Bian}, {Davies}, {Farina}, {Schindler}, {Banados}, {Decarli},
  {Eilers}, {Green}, {Guo}, {Jiang}, {Li}, {Venemans}, {Walter}, {Wu}, \&
  {Yue}}]{Yang_2021}
{Yang}, J., {Wang}, F., {Fan}, X., {et~al.} 2021, arXiv e-prints,
  arXiv:2109.13942.
\newblock \doarXiv{2109.13942}

\bibitem[{{Yoshida} {et~al.}(2008){Yoshida}, {Omukai}, \&
  {Hernquist}}]{Yoshida_2008}
{Yoshida}, N., {Omukai}, K., \& {Hernquist}, L. 2008, Science, 321, 669,
  \dodoi{10.1126/science.1160259}

\bibitem[{{Zhu} {et~al.}(2020){Zhu}, {Li}, {Li}, {Maji}, {Yajima}, {Schneider},
  \& {Hernquist}}]{Zhu_2020}
{Zhu}, Q., {Li}, Y., {Li}, Y., {et~al.} 2020, arXiv e-prints, arXiv:2012.01458.
\newblock \doarXiv{2012.01458}

\bibitem[{{Zhu} {et~al.}(2009){Zhu}, {Hartmann}, {Gammie}, \&
  {McKinney}}]{Zhu_2009}
{Zhu}, Z., {Hartmann}, L., {Gammie}, C., \& {McKinney}, J.~C. 2009, \apj, 701,
  620, \dodoi{10.1088/0004-637X/701/1/620}

\bibitem[{{Zhu} \& {Stone}(2018)}]{Zhu_Stone_2018}
{Zhu}, Z., \& {Stone}, J.~M. 2018, \apj, 857, 34,
  \dodoi{10.3847/1538-4357/aaafc9}

\end{thebibliography}
\bibliographystyle{aasjournal}

\end{document}